\documentclass[11pt]{article}
\usepackage[T1]{fontenc}
\usepackage{lmodern,amsmath,amssymb,amsthm,mathtools,microtype}
\usepackage[margin=1in]{geometry}
\usepackage[colorlinks=true,allcolors=blue]{hyperref}
\usepackage{cleveref}
\usepackage{enumitem}
\usepackage{mdframed}
\usepackage{comment}
\usepackage{xcolor}
\usepackage{lipsum}
\usepackage{enumitem}
\usepackage[color]{changebar}

\makeatletter
\newif\if@inbar
\@inbarfalse
\newcommand{\@currentbarcolor}{none}

\newcommand{\startmarkcolor}[1]{%
  \if@inbar
    \def\@tempa{#1}%
    \ifx\@tempa\@currentbarcolor\relax
    \else
      \cbend
      \cbcolor{#1}%
      \def\@currentbarcolor{#1}%
      \cbstart
    \fi
  \else
    \cbcolor{#1}%
    \def\@currentbarcolor{#1}%
    \cbstart
    \@inbartrue
  \fi
}

\newcommand{\stopmark}{%
  \if@inbar
    \cbend
    \@inbarfalse
    \def\@currentbarcolor{none}%
  \fi
}

\AtEndDocument{\stopmark}
\makeatother

\setlist{itemsep=3pt,topsep=5pt}
\newtheorem{theorem}{Theorem}
\newtheorem{lemma}[theorem]{Lemma}

\theoremstyle{definition}

\newtheorem{remark}[theorem]{Remark}
\newtheorem{definition}[theorem]{Definition}
\newtheorem{observation}[theorem]{Observation}
\theoremstyle{plain}
\newtheorem*{thmrandomgraphHC}{\Cref{thm:randomgraphHC}}
\newtheorem*{thmrandomgraphRC}{\Cref{thm:randomgraphRC}}
\newtheorem*{lemZbound}{\Cref{lem:ZboundsHC}}
\newtheorem*{lemZboundsRC}{\Cref{lem:ZboundsRC}}
\newtheorem*{lemstatistics}{\Cref{lem:statistics}}
\newtheorem*{lemstatisticsRC}{\Cref{lem:statisticsRC}}
\newtheorem*{leminit}{\Cref{lem:init}}
\newtheorem*{leminitRC}{\Cref{lem:initRC}}
\newtheorem*{leminitIS}{\Cref{lem:initIS}}

\newcommand{\pl}{{\textup{\texttt{+}}}}
\newcommand{\mi}{{\textup{\texttt{-}}}}

\newcommand{\plm}{{\textup{\texttt{\textpm}}}}
\newcommand{\E}{\mathbb E}

\newcommand{\Var}{\operatorname{Var}}

\newcommand{\TV}{\mathrm{TV}}

\newcommand{\inact}{{\mathrm{inact}}}
\newcommand{\edge}{{\mathrm{edge}}}
\newcommand{\ver}{{\mathrm{ver}}}
\newcommand{\cI}{\mathcal I}
\newcommand{\cE}{\mathcal E}
\newcommand{\emm}{\mathrm e}

\newcommand{\cA}{\mathcal A}
\newcommand{\cB}{\mathcal B}
\newcommand{\cG}{\mathcal G}

\newcommand{\cH}{\mathcal H}
\newcommand{\cF}{\mathcal F}

\newcommand{\cC}{\mathcal C}
\newcommand{\CM}{\mathsf{CM}}
\newcommand{\bF}{\boldsymbol{F}}

\newcommand{\bG}{\boldsymbol{G}}
\newcommand{\bGp}{\hat{\bG}{}^{\pl}}

\newcommand{\bsigma}{\boldsymbol{\sigma}}
\newcommand{\alphab}{\boldsymbol{\alpha}}
\newcommand{\betab}{\boldsymbol{\beta}}
\newcommand{\omegab}{\boldsymbol{\omega}}
\newcommand{\xb}{\boldsymbol{x}}
\newcommand{\rb}{\mathbf{r}}
\newcommand{\cb}{\mathbf{c}}
\newcommand{\Bb}{\mathbf{B}}
\newcommand{\gammab}{\boldsymbol{\gamma}}
\newcommand{\deltab}{\boldsymbol{\delta}}
\newcommand{\yb}{\mathbf{y}}
\newcommand{\bone}{{\boldsymbol{1}}}
\newcommand{\btwo}{{\boldsymbol{2}}}

\let\eps\varepsilon

\title{Low-Temperature Sampling via Equilibrium Disagreements}
\author{Andreas Galanis\thanks{Department of Computer Science, University of Oxford.
  Email: \texttt{andreas.galanis@cs.ox.ac.uk}.}
  \and Daniel \v{S}tefankovi\v{c}\thanks{Department of Computer Science, University of Rochester.
  Email: \texttt{stefanko@cs.rochester.edu}.}}
\date{5 November 2026}
\begin{document}
\maketitle
\vspace{-1.8em}
\begin{abstract}
We develop a new framework for sampling at low temperatures using Glauber dynamics. In this regime,
the coexistence of competing modes (commonly referred to as phases)  creates bottlenecks that can make
worst-case mixing exponentially slow. The main new ingredient in our  framework is to relate 
convergence within a phase to the geometry of disagreements
between two independent configurations drawn from that phase. 

As applications of our approach, we consider  three canonical models on random
regular graphs that have been central to low-temperature sampling, and establish fast convergence from natural
initialisations of Glauber dynamics for every fixed degree. We show:
\begin{itemize}
    \item For the hard-core model on random bipartite
    $d$-regular graphs, Glauber dynamics initialised from
    either one-sided independent set mixes within the
    corresponding phase in $n^{3/2+o(1)}$ steps
    throughout the non-uniqueness regime
    $\lambda>\lambda_c(d)$.

    \item For the random-cluster model on random
    $d$-regular graphs, for every real $q>2$,
    Glauber dynamics  initialised from the all-in 
     edge set mixes within the ordered
    phase in $n^{3/2+o(1)}$ steps for all $\beta\geq \beta_c$. Similarly,  Glauber dynamics  initialised from the all-out  
     edge set mixes within the disordered phase in $n^{3/2+o(1)}$ steps for $\beta\leq \beta_c$.

    \item For the ferromagnetic Ising model on random
    $d$-regular graphs, Glauber dynamics initialised from
    all-plus or all-minus mixes within the corresponding
    phase in $n^{3/2+o(1)}$ steps throughout the
    low-temperature regime $\beta>\beta_c(d)$.
\end{itemize}
We also give an $n^{2+o(1)}$-time algorithm that combines
phase samples into samples from the full distribution, avoiding in particular time-costly approximations of the corresponding partition functions to the target sampling accuracy.  These results resolve previously open parts of  the low-temperature regime that either required sufficiently strong interactions or large degree $d$.

Our main contribution is a simple combinatorial
route to weak Poincar\'e inequalities, motivated by the
sampling framework of Huang, Mohanty, Rajaraman, and Wu
(STOC 2025). Their approach to such inequalities is analytical, relying on spectral-norm
bounds under typical pinnings or along localization
processes.
Our approach instead works directly with the geometry
of disagreements between two independent samples from
the target phase distribution.
These disagreements typically split into small,
tree-like components of size $O(\log n)$. We then convert this geometric understanding of the disagreements  
into a weak Poincar\'e
inequality  via an elementary canonical-path argument between typical pairs, with exceptional pairs of configurations contributing only to
the additive error in the resulting weak Poincar\'e inequality. This geometric criterion is amenable to techniques from
the analysis of random CSPs, and we verify the required
conditions using double-planted models. Combined with simulated
annealing and monotonicity, the resulting inequalities
yield fast convergence of Glauber dynamics from natural initialisations. \let\thefootnote\relax\footnotetext{For the purpose of Open Access, the authors have applied a CC BY copyright licence to any Author Accepted Manuscript version arising from this submission. All data is provided in full in the results section of this paper.}
\end{abstract}

\section{Introduction}\label{sec:main}
A central obstacle in sampling from high dimensional distributions is the presence of multiple modes in the  underlying distribution, causing strong correlations and  making standard sampling methods hard to analyse or even apply. Such multimodalities arise in various well-studied problems, canonical examples in computer science include the problem of sampling independent sets on bipartite graphs (hard core model with parameter $\lambda$) where the vast majority of independent sets are correlated with one of the two sides of the bipartition for large $\lambda$, or the ferromagnetic Ising model with large $\beta$ where 
typical spin configurations have a nontrivial
bias towards the all-plus or all-minus configuration. 

It is a standard fact by now that the presence of multiple modes typically creates bottlenecks
that make the standard Glauber dynamics mix exponentially slowly from worst-case configurations. More generally, multimodalities and the underlying correlations pose a rather formidable obstruction that frequently obliterate standard analysis tools that we have at our disposal (including spectral independence and correlation decay algorithms).  In fact, the presence of multiple phases frequently underpins various strong inapproximability results at low temperatures \cite{Sly,SlySun}.

The emergence of these competing modes is a  central object of study in the theory of phase transitions and has been analysed in great detail for graph classes of particular interest, examples include random graphs and lattices.  Still, from the sampling perspective, the picture is not well understood despite substantial progress that has been made in recent years in designing
low-temperature sampling algorithms.  We review in detail progress on this front but in summary, the state-of-the-art even for canonical models (such as the Ising and hard-core models) leaves open tantalizing parts of the parameter space, typically close to the corresponding phase-transition thresholds. In particular,  
existing approaches for sampling at low temperatures typically require extra conditions such as sufficiently strong
correlations or additional structural conditions. 
However, it is widely conjectured that low-temperature sampling should not depend so heavily on the proximity to the corresponding phase transition threshold, so the key unresolved question is to identify those properties that will allow for a more universal approach.

Our main contribution in this work is to make a major step towards this goal, by developing a new framework for low-temperature sampling
based on the structure of disagreements between two
independent equilibrium configurations drawn from the
same phase; we then show a generic approach on how to lift this structural understanding  to fast sampling using canonical paths. The main feature of our approach is that for key models of interest it works throughout the low-temperature regime and applying the method is fairly intuitive/mechanical.

To demonstrate this, as main applications of our new approach, we apply the framework to three extensively
studied models on random graphs: 
 We give the precise statements in the upcoming sections, but here are our main results in a nutshell.\footnote{Below, we use $\tilde{O}(g(n))$ to denote functions of the form $g(n) n^{o(1)}$}
\begin{itemize}
    \item For the hard-core model on random bipartite
    $d$-regular graphs, Glauber dynamics initialised from
    either one-sided independent set mixes within the
    corresponding phase in $\widetilde O(n^{3/2})$ steps
    throughout the non-uniqueness regime
    $\lambda>\lambda_c(d)$.

    \item For the random-cluster model on random
    $d$-regular graphs, for every real $q>2$,
    Glauber dynamics  initialised from the all-in 
     edge set mixes within the ordered
    phase in $\widetilde O(n^{3/2})$ steps for all $\beta\geq \beta_c$. Similarly for the disordered phase for $\beta\leq \beta_c$. Our results for both phases actually extend in a window around $\beta_c$ allowing us to handle the full range of $\beta$.

    \item For the ferromagnetic Ising model on random
    $d$-regular graphs, Glauber dynamics initialised from
    all-plus or all-minus mixes within the corresponding
    phase in $\widetilde O(n^{3/2})$ steps throughout the
    low-temperature regime $\beta>\beta_c(d)$.
\end{itemize}
Another contribution of this work is  also to show a fast $\tilde{O}(n^2)$ algorithm that combines the phase samples into an $\eps$-sample from the full distribution, avoiding in particular time-costly approximations of the corresponding partition functions. We remark that, in the case of the Ising model, this is fairly easy to achieve by spin-flip symmetry, so this is more important for the other two models which do not enjoy such nice symmetry. 

Note that our hard-core and ferro Ising results apply to every fixed
$\lambda>\lambda_c(d)$ and $\beta>\beta_c(d)$, with constants that may
depend on the distance from the threshold.
For the random-cluster model with $q>2$, our guarantees
remain uniform in a neighbourhood of $\beta_c$,
including at $\beta_c$ itself.
This distinction reflects the different nature of the
phase transitions: the hard-core phases merge at the
uniqueness threshold, whereas the ordered and disordered
random-cluster phases remain distinct (and remain stable)
at the threshold $\beta_c$.
Extending our hard-core guarantees to activities whose
distance from $\lambda_c(d)$ vanishes with $n$
remains an interesting open direction.

Our framework connects the geometry of equilibrium configurations
to the convergence of Glauber dynamics. Although two independent
samples from the same phase may differ on a linear number of
vertices, their disagreements typically break into small,
tree-like components. We use this structure to construct local
paths and derive weak Poincar\'e inequalities, which imply rapid
convergence from warm starts while allowing a small exceptional
set of configuration pairs. Simulated annealing and monotonicity
then turn these guarantees into convergence from natural
deterministic initialisations. We explain the framework and its
main ingredients in Section~\ref{sec:framework}.

\subsection{Our results}

\subsubsection{Results for bipartite independent sets}
For real  $\lambda>0$,  the hard-core (HC) distribution on a graph $G=(V,E)$ is over the set of independent sets $\cI_G$ of the graph, given by $\mu_G(I)=\lambda^{|I|}/Z_G$ where $Z_G=\sum_{I\in \cI_G}\lambda^{|I|}$ is the partition function. 
For an integer $d\geq 3$, let\footnote{Throughout we use the $\sqcup$ symbol to denote the union of two disjoint sets.} $G=(V^\pl\sqcup V^\mi,E)$ be a bipartite $d$-regular graph with $|V^\plm|=n$. We let $\cG_{n,d}^{\mathrm{bip}}$ be the uniform distribution over such bipartite graphs, and use $G\sim \cG_{n,d}^{\mathrm{bip}}$ to denote a random graph $G$ selected from $\cG_{n,d}^{\mathrm{bip}}$ uniformly at random.\footnote{\label{foo:configmodel}A standard way to sample $G\sim \cG_{n,d}^{\mathrm{bip}}$ is via the configuration model where a random multigraph $\bG$ is sampled by assigning every vertex in $V^\pl,V^\mi$ a collection of $d$ labelled-half edges and then choosing uniformly a random pairing of the half-edges between the two sides. Conditioned on no cycles of length 2 (which happens with probability $\Omega(1)$), $\bG$ is uniform on $\cG_{n,d}^{\mathrm{bip}}$.}

It is by now a classical result \cite{MWw09,DFJ02,SlySun,GSV16} that, whp over $G\sim \cG_{n,d}^{\mathrm{bip}}$, for $\lambda>\lambda_c(d)$ the distribution $\mu_G$ undergoes a phase transition at $\lambda=\lambda_c(d)$, where $\lambda_c(d)=\frac{(d-1)^{d-1}}{(d-2)^d}$ is known as the tree uniqueness threshold. Namely, for $\lambda>\lambda_c(d)$, the distribution $\mu_G$ exhibits a bimodality where the two modes are non-trivially correlated with one of the two sides of the graph (i.e., one side has a constant fraction more vertices in the independent set than the other). This bimodality causes bottlenecks and makes Glauber dynamics mix exponentially slowly from worst-case configurations; in fact, this bimodality on the random bipartite graph  is behind strong inapproximability results for $\lambda>\lambda_c(d)$, see \cite{Sly,SlySun,GSV16}.

To state our phase sampling results, let $\cI^\pl_G$ denote the independent sets with at least as many  vertices occupied on the $\pl$ side versus on the $\mi$ side. Let  $\cI^\mi_G=\cI_G\backslash \cI^\pl_G$  denote the independent sets that favor the $\mi$ side instead. We consider the conditional distributions 
\[\mu^\pl_G(\cdot)=\mu_G(\cdot\mid \cI^\pl_G), \qquad  \mu^\mi_G(\cdot)=\mu_G(\cdot\mid \cI^\mi_G) \]
which focus on independent sets that favor one of the two-sides.  We will refer to $\mu^{\plm}_G$ as the phase distributions, whereas to $\mu_G$ as the full distribution. We let $Z_G^\plm=\sum_{I\in \cI_G^\plm}\lambda^{|I|}$ denote the corresponding partition functions.

Our main result for the hard-core model on random bipartite graphs is as follows. At each step, the hard-core Glauber dynamics $(X_t)_{t\geq 0}$ chooses a vertex u.a.r. and resamples whether to include it in the current independent set from $\mu_G$ conditional on the states of its neighbours.  For a bipartite  graph~$G$, we use  $(X_t^\pl)_{t\geq 0}$ to denote Glauber dynamics initialised at the one-side independent set $X_0^\pl=V^\pl$, and define similarly  $(X_t^\mi)_{t\geq 0}$. Below, we use $\tilde{O}$ to denote asymptotics up to an $n^{o(1)}$ factor.
\begin{theorem}\label{thm:samplerHC1}
Let $d\geq 3$ be an integer and $\lambda>\lambda_c(d)$. For any $K>0$, the following holds whp over $G\sim \cG^{\mathrm{bip}}_{n,d}$. Let $\eps=1/n^K$. Then, 
\begin{enumerate}
    \item   $(X_t^\pl)_{t\geq 0}$ on $G$ is within TV-distance $\eps$ from $\mu^\pl_G$ in $t=\tilde{O}(n^{3/2})$ steps. Similarly for $(X_t^\mi)_{t\geq 0}$.
    \item  There is an algorithm that outputs in time $\tilde{O}(n^{2})$ an $\eps$-sample from the full distribution $\mu_G$.
\end{enumerate}
\end{theorem}
We give the proof of the first part of Theorem~\ref{thm:samplerHC1} in Section~\ref{sec:GlauberHC} (where it is stated as Lemma~\ref{lem:samplerbHC1}). We complete the proof of~\Cref{thm:samplerHC1} in Section~\ref{sec:finalcombo}, after developing the general framework to combine the individual phase samplers in the appropriate mixture.

\subsubsection{Results for the Random-Cluster model}
For real parameters $\beta>0$, $q>1$ the random-cluster (RC) distribution on a graph $G=(V,E)$ is over all edge subsets $F\subseteq E$ given by
\[
 \pi_G(F)\propto(\emm^\beta-1)^{|F|}q^{\kappa(F)} \qquad (\forall F\subseteq E),
\]
where $\kappa(F)$ is the number of connected components
of $(V,F)$, including isolated vertices. For integer $q\geq 3$, the random-cluster model is an edge-representation of the well-known Potts model (with $q=2$ being the Ising case).

For real $q>2$ and every degree $d\geq 3$, \cite{bencs2022random} showed that on the random $d$-regular graph, the model undergoes a phase transition from ``disordered'' to ``ordered'' at $\beta=\beta_c(q,d)$ where 
$\beta_c(q,d)=\log\left(\frac{q-2}{(q-1)^{1-2/d}-1}\right)$, 
see also \cite{BIShard,RCM-Helmuth2020,BDS25,DZ25}. Roughly, in the disordered regime $\beta<\beta_c$, the components in a typical configuration have size $O(\log n)$ whereas in the ordered regime $\beta>\beta_c$ there is a giant component of size $\Omega(n)$; at $\beta=\beta_c$ the two types of configurations have inverse-polynomially large mass.

The critical point $\beta_c$ lies strictly inside a so-called  metastability
window $(\beta_u,\beta_s)$, where $\beta_u$ and
$\beta_s=\log(1+q/(d-2))$ are the lower and upper uniqueness thresholds
of the RC model on the wired $d$-regular tree.
For integer $q\geq 3$, it has been established that the ordered and disordered phases are both
present throughout this window as ``locally stable'' modes, although only one dominates
away from $\beta_c$. For every fixed $\beta$ in this window, the
metastability result of Coja-Oghlan et al.~\cite{coja2023} implies
that the worst-case mixing time of RC Glauber dynamics is
$\exp(\Omega(n))$ whp on the random $d$-regular graph.

To state our mixing results more formally, for $\mathfrak{m}_c=\mathfrak{m}_c(q,d)\in (0,1)$ to be specified\footnote{See Remark~\ref{rem:pcvalue} for the explicit definition using the densities of the expected number of edges at the critical threshold $\beta_c$.}, we consider the following subsets of configurations which will be most relevant
\begin{equation}\label{eq:phasesRCdef}
\begin{aligned}\text{minus/disordered phase: }\quad & \cE^\mi_G=\{F\subseteq E\mid |F|\leq \tfrac{dn}{2}\mathfrak{m}_c\},\\ \text{plus/ordered phase: }\quad &\cE^\pl_G=\{F\subseteq E\mid |F|> \tfrac{dn}{2}\mathfrak{m}_c\}.
\end{aligned}
\end{equation}
Let $\pi^\pl_G=\pi_G(\cdot\mid \cE^\pl_G)$ and $\pi^\mi_G=\pi_G(\cdot\mid \cE^\mi_G)$ denote the distributions induced by the plus/minus  phases. 

At each step, the RC Glauber dynamics chooses an edge u.a.r. and resamples its state from $\pi_G$ conditional on the states of the other edges.
We let $(X_t^\pl)_{t\geq 0}$ denote the RC Glauber dynamics initialised from $X^\pl_0=E$ (the all-in edge set)  and $(X_t^\mi)_{t\geq 0}$ the RC Glauber dynamics initialised from $X^\mi_0=\emptyset$ (the all-out edge set). We use $\cG_{n,d}$ to denote the uniform distribution over $d$-regular graphs with $n$ vertices.

\begin{theorem}\label{thm:samplerRC1}
Fix integer $d\geq 3$ and real  $q>2$. There exists $\delta>0$ such that for any $K>0$, the following hold whp over $G\sim \cG_{n,d}$. Let  $\eps=1/n^K$ and $\beta_c=\beta_c(q,d)$.
\begin{enumerate}
    \item For any $\beta>\beta_c-\delta$, $(X_t^\pl)_{t\geq 0}$ on $G$ is within TV-distance $\eps$ from $\pi^\pl_G$ in $t=\tilde{O}(n^{3/2})$ steps.
    \item For any $\beta<\beta_c+\delta$, $(X_t^\mi)_{t\geq 0}$ on $G$ is within TV-distance $\eps$ from $\pi^\mi_G$ in $t=\tilde{O}(n^{3/2})$ steps.
\end{enumerate}
Moreover, there is an algorithm that, for any $\beta>0$, outputs in time $\tilde{O}(n^{2})$ an $\eps$-sample from the full distribution $\pi_G$.
\end{theorem}

\subsubsection{Results for the Ferromagnetic Ising model}

For real $\beta>0$, the Ising distribution on a graph $G=(V,E)$ is over all 2-spin configurations $\sigma\in \{0, 1 \}^V$ given by
\[
 \nu_G(\sigma)\propto \emm^{\beta m(\sigma)}
\]
where $m(\sigma)$ is the number of monochromatic edges under $\sigma$ (edges $\{u,v\}$  with $\sigma_u=\sigma_v$).

Analogously to the earlier picture explained for the bipartite hard-core model, it is well-known by now \cite{DM10,MMS12} that the Ising model exhibits a bimodality on random $d$-regular graphs iff $\beta>\beta_c(d)$ where $\beta_c(d)=\ln \frac{d}{d-2}$, causing an exponential slowdown on the mixing time of Glauber dynamics from worst case configurations. 
Recall $\cG_{n,d}$ is the uniform distribution over $d$-regular graphs with $n$ vertices. We define the phases
\[\Sigma^\pl=\{\sigma:|\sigma^{-1}(1)|\geq n/2\}\mbox{ and }
\Sigma^\mi=\{\sigma:|\sigma^{-1}(1)|<n/2\},\]
and write $\nu_G^\plm=\nu_G(\,\cdot\mid\Sigma^\plm)$
for the corresponding phase-conditioned distributions.

At each step, the Ising Glauber dynamics chooses a vertex u.a.r. and resamples its spin from $\nu_G$ conditional on the spins of its neighbours.
Analogously to the previous cases, we use  $(X_t^\pl)_{t\geq 0}$ to denote Glauber dynamics initialised at the all-one configuration, and   $(X_t^\mi)_{t\geq 0}$ for the chain initialised at the all-zero. As before, we use $\tilde{O}$ to denote asymptotics up to an $n^{o(1)}$ factor.
\begin{theorem}\label{thm:samplerIS}
Fix integer $d\geq 3$ and $\beta>\beta_c(d)$. For any $K>0$, the following holds whp over $G\sim \cG_{n,d}$. Let $\eps=1/n^K$. Then,
\begin{enumerate}
    \item   $(X_t^\pl)_{t\geq 0}$ on $G$ is within TV-distance $\eps$ from $\nu^\pl_G$ in $t=\tilde{O}(n^{3/2})$ steps. Similarly for $(X_t^\mi)_{t\geq 0}$.
    \item  There is an algorithm that outputs in time $\tilde{O}(n^{3/2})$ an $\eps$-sample from the full distribution $\nu_G$.
\end{enumerate}
\end{theorem}
Note here that the full-distribution algorithm is faster than in the previous two-cases due to the spin-flip symmetry of the Ising model.

\subsection{Previous Work}
There has been substantial progress on low-temperature sampling in recent years, below we summarise the state-of-the-art in the three applications we consider.  As mentioned earlier, low-temperature sampling is typically intractable on worst-case graphs, so the main benchmark in the literature for low-temperature sampling has been classes of random regular graphs, bipartite and non-bipartite, though some of the approaches have also been analysed on expander graphs and lattices. 

Overall, the main unresolved gap from all these approaches has been the absence of results that apply uniformly over the low-temperature regime.  Low-temperature results so far have required  large-degree or other large-parameter conditions to ensure some form of ``sufficiently strong'' correlations.  The main results of this paper address this gap. \vskip 0.2cm

\vskip 0.2cm \noindent \textbf{Bipartite Independent sets.} Sampling bipartite independent sets is a well-known problem which is conjectured to be hard, but whose approximation complexity is not yet established. On worst-case graphs of maximum degree $\leq d$, it is by now known that for $\lambda\leq \lambda_c(d)$ the problem admits an FPTAS/FPRAS even on non-bipartite graphs, see~\cite{Weitz06,PatelRegts17,PetersRegts19,LSS25} for deterministic approaches and the more recent  spectral independence work for Glauber dynamics~\cite{ALG20,CLV20,CLV21,CFYZ22,CCYZ25}. By contrast,  for $\lambda>\lambda_c(d)$ the problem becomes NP-hard for $d$-regular graphs \cite{Sly,SlySun,AndreasBIP} and \#BIS-equivalent for $d$-regular bipartite graphs~\cite{CaiGGGJSV16} (i.e., as hard as sampling independent sets on unbounded degree bipartite graphs).

For random bipartite graphs (and more generally on bipartite expanders), it was first demonstrated by Jenssen, Keevash, and Perkins~\cite{BIS1}  that one can sample for sufficiently large $\lambda$ by developing a new technique that is known as the algorithmic polymer method. Their approach has since been refined and improved in various ways, see \cite{LLLM19} for the unweighted case, \cite{BIS2,Chen} for Markov-chain approaches using polymer-based dynamics, and \cite{JPP23,JMP26} for graph-container methods on expanders. The best results in these works  require $\lambda= \Omega(\frac{\log d}{d})$.

A recent breakthrough by Kocurek, Oveis Gharan, and Tjowasi~\cite{KOGT26}
obtained polynomial-time sampling algorithms on random
regular bipartite graphs for $\lambda=O(d^{-1/2})$.
Combined with the polymer algorithms for larger fugacities,
their result gives the first all-$\lambda$ algorithms for
sufficiently large fixed $d$.
Narang and Perkins~\cite{NP26} subsequently obtained
all-fugacity algorithms under an efficiently checkable
spectral expansion condition, which holds with high
probability on random regular bipartite graphs of
sufficiently large degree.
See also~\cite{CCCYYZ25} for earlier rapid-mixing results
for Glauber dynamics on random non-bipartite regular
graphs in a comparable range of $\lambda$. The approaches of~\cite{KOGT26,NP26} overcome the
bottleneck between phases by sampling from auxiliary
distributions (fixed-occupation slices in~\cite{KOGT26}
and quadratically tilted measures in~\cite{NP26}) which incur additional
overhead to reconstruct the original distribution.
Our framework analyses convergence within the original
phases and gives substantially sharper running-time
guarantees for sampling, while removing the restriction
to sufficiently large degrees altogether.

\vskip 0.2cm\noindent \textbf{Random-Cluster/Ferromagnetic Potts model.}
Sampling from the ferromagnetic $q$-state Potts model (for $q>2$), or from the associated random-cluster model on graphs of  maximum degree $d$, is not as well understood as the hard-core model. For worst-case graphs, it is known that the problem is $\#\mathrm{BIS}$-hard for $\beta>\beta_c(q,d)$~\cite{BIShard} when $q$ is integer. A matching algorithmic result for $\beta<\beta_c(q,d)$ is not known in general (even for integer $q$), but see \cite{UG2,UG1} for positive results that approach this threshold when $q=d^{\Omega(1)}$.

For random regular graphs (and more generally expanders)  it was first demonstrated by  Helmuth, Jenssen, and Perkins~\cite{RCM-Helmuth2020} using the polymer method that one can obtain counting/sampling algorithms for all $\beta$, provided that $q$ is large relative to $d$ (roughly $q=\emm^{\Omega(d)}$). Since then, there have been various refinements but no approach managed to go all the way to $\beta_c$, without some assumption on $q,d$, see \cite{CDK24,UG2,arxivversion,RCrandom,BG25}. Of these, closer to our work is the initialisation/planting analysis approach of Galanis, Goldberg, and Smolarova~\cite{GGS26Plant} who considered the integer $q$ case (Potts) and proved fast mixing for random-cluster dynamics on random regular graphs starting from all-in or all-out state;  their result holds for all $\beta<\beta_c$ when  $q,d\geq 3$ are integers but requires $q\geq (5d)^5$ in the regime $\beta>\beta_c$. 

More recently, several works have addressed the mixing time of RC
Glauber dynamics from worst-case initial configurations outside the
metastability window. Blanca and Gheissari proved a mixing time of
$\Theta(n\log n)$ for $\beta<\beta_u$~\cite{BG} (see also the recent result of~\cite{BGPZ26Potts} for Potts Glauber dynamics) and rapid mixing for
sufficiently large $\beta$~\cite{BG25}. On the low-temperature side,
Blanca, Gheissari, Park, and Zhang~\cite{BGPZ26} proved
$n^{1+o(1)}$ mixing for every fixed $\beta>\beta_s>\beta_c$ when
$q\geq C\log d$, and Li~\cite{Li26} extended this bound to every
fixed real $q>2$. For integer $q\geq3$, the exponential lower bound of~\cite{coja2023} inside the metastability window
$(\beta_u,\beta_s)$ precludes any extension of these worst-case initialisation results and hence rapid convergence in this range
requires a suitable initialisation (and different ways to capture the evolution of the dynamics).

\vskip 0.2cm\noindent \textbf{Ferromagnetic Ising model.}
For the ferromagnetic Ising model, Jerrum and
Sinclair~\cite{JS93} established polynomial-time approximate
counting on arbitrary graphs at all temperatures via an auxiliary chain, see also~\cite{GuoJerrum} for a refinement using the RC Glauber dynamics (for $q=2$). 
More recently, Feng, Guo, and Zhang~\cite{FGZ26} obtained an
$\widetilde O(n^2)$-time sampler for the zero-field model
on arbitrary graphs,  using cut sparsification
and the RC dynamics. Near-linear sampling and mixing results are also known 
when the external fields consistently favour one spin
and are bounded away from zero, see ~\cite{FGW23,CZ23,FY26}.
These algorithms establish efficient sampling even in
regimes where single-spin Glauber dynamics has
exponential worst-case mixing time.

For random regular graphs, a more refined question that has been extensively studied is
whether single-spin Glauber dynamics converges rapidly
from natural initialisations.
Gheissari and Sinclair~\cite{SinclairsGheissari2022}
first introduced weak spatial mixing within a phase and proved
$O(n\log n)$ convergence at sufficiently low temperatures
from the all-plus and
all-minus configurations.
Gheissari, Sly, and Sohn~\cite{gheissari2025rapid}
extended rapid phase ordering to a broader class of
biased initial configurations, in the regime
$\beta\gtrsim(\log d)/d$.
Recent work of Paliy and Sohn~\cite{PS26} and Li and
Sly~\cite{LS26} further extends the temperature range
to inverse temperatures $\beta\geq C/d$ for some large  constant $ C>0$. 
These results leave open convergence throughout the
full low-temperature regime for every fixed degree.
Our result establishes $n^{3/2+o(1)}$ convergence from
all-plus or all-minus within the corresponding phase
for every fixed $d\geq3$ and $\beta>\beta_c(d)$.
By spin-flip symmetry, the same running time suffices
to sample from the full distribution.

\vskip 0.2cm\noindent \textbf{Weak Poincar\'e Inequalities} Our approach is motivated by the work of
Huang, Mohanty, Rajaraman, and Wu~\cite{HMRW25} who
developed a framework combining weak Poincar\'e
inequalities with simulated annealing to obtain sampling
guarantees beyond worst-case mixing; see also the recent
follow-up~\cite{DHMR26}.
Their route to these inequalities uses spectral
independence under typical pinnings, or more generally
covariance bounds along localization processes.
Establishing the required matrix bounds can be technically
demanding, particularly within a phase, where conditioning
can substantially change the correlation structure.

Our main  contribution is a different,
combinatorial route to weak Poincar\'e inequalities.
We require no analysis of typical pinnings: our structural
conditions concern only two independent samples from
the target phase distribution itself.
An elementary canonical-path argument converts the
geometry of their disagreements into the required
functional inequality, with exceptional pairs contributing
only to its additive error.
This makes the framework particularly suited to random
graphs, where equilibrium disagreements can be studied
directly through planted models, while controlling
correlations throughout a sequence of conditioned
distributions can be considerably harder.
The resulting approach turns a geometric understanding
of equilibrium configurations into guarantees for
Glauber dynamics from natural initialisations.

\vskip 0.4cm \noindent \textbf{Use of AI.} Various parts of this work have been produced with the help of  ChatGPT Astra~6.0. The AI contribution was  most substantial with  the canonical path analysis and the faster full-distribution sampler. The contribution for other arguments was less critical but the interaction has helped throughout to simplify and make more compact various arguments.  The authors  take full responsibility for the presentation and validity of the results.

\section{Proof Outline/Organization}\label{sec:framework}

Our starting point is that the variance of a function $f$ under
a distribution can be expressed through its difference on
two independent samples. This makes the geometry of
equilibrium disagreements (capturing the difference of the two samples) a natural tool for studying
convergence.

To make this a bit more concrete, for a phase of interest, write $\mu$ for the Gibbs
distribution conditioned on that phase and $P$ for Glauber
dynamics with moves leaving the phase rejected. We seek a
weak Poincar\'e inequality of the form
\[
 \Var_\mu(f)
 \leq A\,\cE_P(f,f)+b(f_{\max}-f_{\min})^2.
\]
Here, the Dirichlet form $\cE_P(f,f)$ measures the average
squared change of $f$ under one step of the dynamics.
Thus, the inequality controls global variation by local
variation, allowing a small additive error. In our
applications, we obtain $A=n^{1+o(1)}$ and can make $b$
an arbitrarily small inverse polynomial. This yields
rapid convergence from warm starts, with an error
controlled by $b$; see Lemma~\ref{lem:anneal}.

To establish this inequality, we study two independent
configurations drawn from the same phase. Although they
may disagree on a linear number of vertices, we show that
their disagreements, in an appropriate representation,
typically break into components of size $O(\log n)$
with bounded tree excess.
We first reduce the variance comparison to exchanges of
individual disagreement components, using the product
structure of their possible orientations. We then realise
each exchange by a sequence of Glauber moves.
The tree-like structure of a typical random graph allows us to sweep through the
component while keeping only a small boundary between
the processed and unprocessed vertices. This controls
the congestion of the resulting paths and gives the
coefficient $A$ above; exceptional pairs contribute to
the additive error $b$.
See Lemmas~\ref{lem:poincare} and~\ref{lem:poincareRC}
for the precise comparisons.
The required geometric properties are established using
planted models: conditioning on the vertex and edge
statistics makes the random graph amenable to exploration,
and transfer estimates carry these bounds back to
independent Gibbs samples on a typical graph.

To obtain convergence from the natural deterministic
initialisations, we combine this inequality with simulated
annealing. We begin at parameters where the desired phase
is strongly favoured and initialisation is easy, and then
gradually move to the target parameters. Bounds on the
variance of the observable driving the annealing allow
$n^{1/2+o(1)}$ stages, with successive phase distributions
close enough that each stage requires only
$n^{1+o(1)}$ Glauber updates.
Monotonicity then compares this auxiliary annealing
process with Glauber dynamics run at the fixed target
parameters from the corresponding extremal configuration.
This gives the $n^{3/2+o(1)}$ convergence bounds.

Finally, to sample from the full distribution, we  need to first select a phase with the right probability and then generate the corresponding phase sample; the tricky part is that the standard method to select a phase requires a high-accuracy estimate of the phase partition functions which is very time costly. 
The key ingredient to bypass this is that we only need  estimates with controlled upper tails, allowing accurate phase
selection without estimating the phase weights to the
final sampling accuracy; see Section~\ref{sec:finalcombo}.
\subsection{Preliminaries}
Let $P$ be a Markov chain on a state space $\Omega$ with stationary distribution $\pi$. For a function $f:\Omega\rightarrow \mathbb{R}$, the Dirichlet form $\cE_P(f,f)$ (local variance under $P$) and the global variance of $f$ are given by
\[
 \cE_P(f,f)=\frac{1}{2}\sum_{x,y\in \Omega}\pi(x)P(x,y)(f(x)-f(y))^2, \qquad \Var_\pi(f)=\frac{1}{2}\sum_{x,y\in \Omega}\pi(x)\pi(y)(f(x)-f(y))^2.
\]
We use $f_{\max}$ to denote $\max_{x\in \Omega} f(x)$, and similarly for $f_{\min}$. For $A,b>0$, we say that $P$ satisfies an $(A,b)$ weak-Poincar\'e inequality if 
\[
 \Var_\pi(f)\le A\cdot \cE_P(f,f)+b(f_{\max}-f_{\min})^2,
 \qquad \forall f:\Omega\rightarrow \mathbb{R}.
\]
We have the following Lemma which is proved in Appendix~\ref{sec:remainingproofs}.
\begin{lemma}\label{lem:anneal}
Let $A,R\geq 1$ and $b>0$. Let $P$ be a reversible chain with  state space $\Omega$, stationary distribution $\pi$ and nonnegative spectrum, which  satisfies an $(A,b)$ weak-Poincar\'e inequality.

Let $\sigma$ be a distribution on $\Omega$ with $\sigma(x)\le R\pi(x)$ for all $x\in \Omega$. Then, for every integer $t\ge0$,
\begin{equation}\label{eq:warm}
 \|\sigma P^t-\pi\|_{\TV}
 \le R(\emm^{-t/(2A)}
      +\sqrt{b}).
\end{equation}
Moreover, for  every probability distribution $\nu$ on $\Omega$,
\[
 \|\nu P^t-\pi\|_{\TV}
 \leq
 \sqrt{\chi^2(\nu\Vert\pi)}
 \sqrt{\tfrac14\emm^{-t/A}+b} \mbox{ where }
 \chi^2(\nu\Vert\pi)
 =\sum_x\pi(x)\left(\frac{\nu(x)}{\pi(x)}-1\right)^2.
\]
\end{lemma}
As an example of a weak Poincar\'e inequality, we state the following result on the hypercube (which will actually become handy later in our canonical-path arguments for Glauber).
\begin{lemma}\label{lem:cube-wpi}
Let $\nu$ be the uniform distribution on
$\Omega=\{-1,1\}^k$, and let $\mathcal B\subseteq\{-1,1\}^k $.
For $\sigma\in\Omega $, let $\sigma^{(i)}$ denote the
configuration obtained by flipping coordinate $i$.
Then every function $g:\Omega\to\mathbb R$ satisfies
\[
\operatorname{Var}_\nu(g)
\leq
\frac14\,
\mathbb E_{\sigma\sim\nu}\left[
 \mathbf 1_{\{\sigma\notin\mathcal B\}}
 \sum_{i=1}^k
 \bigl(g(\sigma)-g(\sigma^{(i)})\bigr)^2
\right]
+
\frac{k\,\nu(\mathcal B)}{4}(g_{\max}-g_{\min})^2.
\]
\end{lemma}
\begin{proof}
By tensorization of variance under the uniform product
measure $\nu$,
\[
\operatorname{Var}_\nu(g)
\leq
\sum_{i=1}^k
\mathbb E_\nu\left[
 \operatorname{Var}_\nu
 \bigl(g(\sigma)\mid(\sigma_j)_{j\neq i}\bigr)
\right]
=
\frac14\,\mathbb E_\nu\left[
 \sum_{i=1}^k
 \bigl(g(\sigma)-g(\sigma^{(i)})\bigr)^2
\right].
\]
Split the last expectation according to whether
$\sigma\in\mathcal B$. Since every squared difference
is at most $(g_{\max}-g_{\min})^2$, the contribution
from $\mathcal B$ is at most $\frac{k\,\nu(\mathcal B)}{4}(g_{\max}-g_{\min})^2$.
\end{proof}

\section{Proof for Bipartite Independent Sets}
In this section, we give the proof of Theorem~\ref{thm:samplerHC1} for the hard-core model. The proof of fast mixing for Glauber dynamics is via a comparison with a simulated annealing algorithm  where the activities are chosen to boost the $\pl $ side and suppress the $\mi$ side.

Let $G=(V^\pl\sqcup V^\mi,E)$ be a $d$-regular bipartite graph. For $s\geq 0$, we let $\mu^{\pl}_{G,s}$ be the distribution on $\cI^\pl_G$ where 
\[ \mu^{\pl}_{G,s}(I)\propto(\lambda \emm^{s})^{|I\cap V^\pl|}(\lambda \emm^{-s})^{|I \cap V^\mi |}=\lambda^{|I|} \emm^{s M(I)},\] 
where  $M(I)=|I \cap V^\pl|-|I\cap V^\mi|$ is the magnetisation of $I$.  Note that for $s=0$ we have our target plus-phase  distribution $\mu^{\pl}_G$.

We will obtain a sample from $\mu^\pl_G$ by first starting at some large $s=s_0$. For each value of $s$, we will be running Glauber dynamics for $O(n)$ steps, which will allow us to conclude (via a weak Poincar\'e inequality) that at the end of the round we have a good sample from $\mu^{\pl}_{G,s}$. The final bit will be to compare the evolution of the simulated annealing algorithm with the standard Glauber dynamics (with uniform activity $\lambda$) using the monotonicity of the model on bipartite graphs (and the annealing activities).

The analysis of the algorithm hinges on the following understanding of the distribution $\mu^\pl_{G,s}$. For two independent sets $I,J$, let $\cC(I,J)$ denote the connected components of the induced disagreement graph $G[I\oplus J]$. For a graph $C$, write
$\mathtt{exc}(C)=|E(C)|-|V(C)|+1$ for the tree-excess of $C$. For a bipartite graph $G$, let $\cB_{\ell,r}$ be the event that for $I,J\in \cI^\pl_G$ there is some component in $\cC(I,J)$ with size or excess greater than $\ell$ or $r$, respectively, i.e., we let 
\begin{equation}\label{eq:badpairs}
 \cB_{\ell,r}=\{(I,J):\exists C\in\mathcal C(I,J),\
 |V(C)|>\ell\ \text{or}\ \mathtt{exc}(C)>r\}.
\end{equation}
and set $b_{G,\ell,r}(s)=\Pr\big[(I,J)\in\cB_{\ell,r}\big]$ for independent samples $I,J\sim \mu^\pl_{G,s}$. For $\theta>0$, we let also $\cI^{\mathrm{bal}}_G(\theta)$ be the set of roughly balanced independent sets satisfying $M(I)\leq \theta n$ .
 \newcommand{\statelemrandomgraphHC}{Let $d\geq 3$, $\lambda>\lambda_c(d)$ and $s_0>0$ be fixed. There are constants $\theta,\rho,L,T >0$ such that, whp over $G\sim \cG^{\mathrm{bip}}_{n,d}$, for any  $\ell,r\geq 1$, the following hold uniformly over all real $s\in [0,s_0]$. Let $\mu^\pl=\mu^\pl_{G,s}$.
\begin{itemize}
    \item Small disagreement components: $ b_{G,\ell,r}(s)\leq n^T\big(\emm^{-\rho   \ell}+(\tfrac{L\ell^2}{ n})^r\big)$.
    \item Balanced independent sets are very unlikely: $\mu^\pl(\cI^{\mathrm{bal}}_G(\theta))\leq \emm^{-\rho n}$.
\end{itemize}
 }
\begin{theorem}\label{thm:randomgraphHC}
\statelemrandomgraphHC
\end{theorem}
The proof of Theorem~\ref{thm:randomgraphHC} is given in Section~\ref{sec:disHC}.

\subsection{A  weak-Poincare inequality}
To convert~\Cref{thm:randomgraphHC} to a sampling guarantee we will use the framework of weak-Poincare inequalities. Recall that 
\begin{equation*}
 \cB_{\ell,r}=\{(I,J):\exists C\in\mathcal C(I,J),\
 |V(C)|>\ell\ \text{or}\ \mathtt{exc}(C)>r\}.\tag{\ref{eq:badpairs}}
\end{equation*}
For $\theta>0$, define also $\cH_{\theta}=\{(I,J): I \mbox{ or  } J \mbox{ belongs to } \cI_G^{\mathrm{bal}}(\theta)\}$.

\begin{lemma}\label{lem:poincare}
Let $d\ge3$ and $\lambda>1$. The following holds for every $d$-regular
bipartite graph $G=(V^\pl\sqcup V^\mi,E)$ with
$|V^\plm|=n$ and vertex activities $0<\lambda_v\le\lambda$. Let $\mu$ be the hard-core distribution on $G$, 
and let $\mu^\pl=\mu(\cdot\mid\cI^\pl)$. 

Let $P^\pl$ be the 
single-site Glauber dynamics on $\cI^\pl_G$, with proposals
outside $\cI^\pl_G$ rejected. For integers $\ell,r\geq 2$ and real $\theta\in (0,1/2)$, let
$b=\Pr[(I,J)\in\cB_{\ell,r}\cup \cH_{\theta}]$, where $I,J$ are independent samples
from $\mu^\pl$. Then
\begin{equation}\label{eq:permanent-wpi-hc}
 \mathrm{Var}_{\mu^\pl}(f)
 \le A\,\mathcal E_{P^\pl}(f,f)+(b+\delta)(f_{\max}-f_{\min})^2
 \qquad (\forall f:\cI^\pl\to\mathbb R),
\end{equation}
where $A=8n (d\ell\lambda)^{k+1}$ with $k=r+4d\ln\ell$ and $\delta=4n\emm^{-\theta^2n/(4\ell)}$.
\end{lemma}
The rest of this section is devoted to the proof of~\Cref{lem:poincare}. We first begin with the following standard bound on the cutwidth of a tree-like graph whose proof is given for completeness in Appendix~\ref{sec:remainingproofs}. 

\begin{lemma}\label{lem:cutwidth}
Suppose $C$ is a connected graph on $ \ell$ vertices with maximum degree $\leq d$, and   $\mathrm{exc}(C)\leq r$. Then, there is an ordering $v_1,\hdots, v_\ell$ such that for any $i$ the number of edges in the cut induced by $v_1,\hdots,v_i$ is  $\leq d\lceil\log_2\ell\rceil+r$.
\end{lemma}

For a pair of independent sets $I,J$, observe that an edge of $G[I\triangle J]$ must join $I\setminus J$ to $J\setminus I$, since both sets are independent. Consequently, for a connected component $C$ of $G[I\oplus J]$  the sets $A=I\cap C$ and $B=J\cap C$ are exactly the two bipartition classes of $C$ (on a singleton, one is the vertex and the other is empty). Let $I^C=(I\backslash C)\cup (J\cap C)$.

For the following definition, fix an arbitrary vertex labelling.
\begin{definition}\label{def:sweeping}
Let $(I,J)\in \cI^\pl\times \cI^\pl$. For a connected component $C\in\cC(I,J)$ with $\ell$ vertices and tree excess $r$, the component switch $\gamma_{I,J,C}$ is a series  of Glauber moves $I_0\rightarrow I_1\rightarrow \cdots \rightarrow I_\ell$, where $I_0=I$ and $I_\ell=I^C$ (with $I^C=(I\backslash C)\cup (J\cap C)$). 

Namely, we sweep through the vertices in $C$ in the order given by~\Cref{lem:cutwidth}. When a vertex from $A=I\cap C$ is encountered, delete it if still occupied. When a  vertex $v$ from $B=J\cap C$ is encountered, first delete all of its still-occupied  $A$-neighbours, in vertex-label order, and then insert~$v$.
\end{definition}

\begin{lemma}\label{lem:mapping}
Consider a Glauber transition $e=(Y\rightarrow Y')$ with $Y\neq Y'$. Let $\mathcal{P}_{Y\rightarrow Y'}$ be the set of pairs
$(I,J)\in (\cI^\pl\times\cI^\pl)\setminus \cB_{\ell,r}$ with $M(I),M(J)\geq 2\ell$ for which the transition $Y\rightarrow Y'$ is traversed in a component switch $\gamma_{I,J,C}$
for some $C$ in $G[I\oplus J]$. Set $k:=d+d\lceil \log_2 \ell\rceil+r$. Then
\begin{equation}\label{eq:congestion}
\sum_{(I,J)\in \mathcal{P}_{Y\rightarrow Y'}}\mu^\pl(I)\mu^\pl(J)\leq 8n\lambda(d\ell\lambda )^{k}\mu^\pl(Y)P^\pl(Y,Y').
\end{equation}
\end{lemma}

\begin{proof}[Proof of Lemma~\ref{lem:poincare}]
 Let $R_f=(f_{\max}-f_{\min})^2$. We may assume $\ell\leq \theta n/4$: otherwise $ne^{-\theta^2 n/(4\ell)}\geq n\emm ^{-\theta }\geq 1/4$ and the conclusion follows from the trivial bound
$\mathrm{Var}_{\mu^\pl}(f)\leq \frac{1}{4}(f_{\max}-f_{\min})^2$. Similarly we may assume $\delta=4n\emm^{-\theta^2 n/(4\ell)}$ satisfies $\delta\leq 1/2$.

For $I,J\in \cI^\pl$, let $\cC(I,J)$ denote  the
components of $G[I\oplus J]$, and let $I^C=(I\setminus C)\cup(J\cap C)$ denote the independent set obtained by switching the component $C$ (cf. Definition~\ref{def:sweeping}). For convenience, let $\mathcal{Q}$ be the set of pairs $(I,J)$  in $(\cI^\pl\times\cI^\pl)\setminus \cB_{\ell,r}$ with $M(I),M(J)\geq 2\ell$, i.e., that $I,J$ have magnetisation at least $2\ell$ and that the disagreement components in $G[I\oplus J]$ have $\leq \ell$ vertices and tree-excess $\leq r$. We claim the one-component reduction inequality
\begin{align}
 \mathrm{Var}_{\mu^\pl}(f)
 &\le\frac12\mathbb E_{I,J\sim\mu^\pl}\bigg[
   \mathbf1_{\mathcal{Q}}
   \sum_{C\in\mathcal C(I,J)}\big(f(I^C)-f(I)\big)^2\bigg] +\left(b+
          \delta\right)R_f.
 \label{eq:sc-reduction}
\end{align}
Assuming this for the moment,  for $(I,J)\in \mathcal{Q}$ we have for $C\in \mathcal{C}(I,J)$
\[(f(I^C)-f(I))^2\leq \ell \sum_{e=(Y\rightarrow Y')\in \gamma_{I,J,C}}\big(f(Y')-f(Y)\big)^2\]
It follows that 
\[\mathbb{E}_{I,J\sim\mu^\pl}\bigg[
   \mathbf1_{\mathcal{Q}}
   \sum_{C\in\mathcal C(I,J)}\big(f(I^C)-f(I)\big)^2\bigg]\leq \ell\sum_{e=(Y\rightarrow Y')} \big(f(Y')-f(Y)\big)^2\sum_{(I,J)\in\mathcal{P}_{Y\rightarrow Y'}}\mu^\pl(I)\mu^\pl(J) \]
Using then the bound in~\Cref{lem:mapping} and plugging back into \eqref{eq:sc-reduction}, we obtain the desired weak Poincare inequality.

The main remaining step is therefore to prove \eqref{eq:sc-reduction}. For convenience, let $\mathcal{Q}'$ be the event $(I,J)\in (\cI^\pl\times\cI^\pl)\setminus(\cB_{\ell,r}\cup \cH_\theta)$, i.e., that $I,J$ are not $\theta$-balanced and that the disagreement components in $G[I\oplus J]$ have $\leq \ell$ vertices and tree-excess $\leq r$.  For $I,J\sim \mu^\pl$, we have  $\Pr[(I,J)\notin \mathcal{Q}']=\Pr[(I,J)\in\cB_{\ell,r}\cup \cH_\theta]\leq b$, so 
the variance identity gives
\begin{align}
\operatorname{Var}_{\mu^\pl}(f)=\frac12\mathbb E\left[
 \big(f(I)-f(J)\big)^2
\right]
&\leq
\frac12\,
\mathbb E\left[
 \mathbf 1_{\mathcal{Q}'}
 \big(f(I)-f(J)\big)^2
\right]
+
b\, R_f.
\label{eq:initial-pair-truncation}
\end{align}
We now group the remaining pairs according to the intersection
and symmetric difference of the pair. For a pair $(I,J)\in \mathcal{Q}'$, call $O=(I\cap J, I\oplus J)$ its skeleton.  Let $C_1,\ldots,C_q$ be the connected components
of $G[I\oplus J]$. Each component has two possible orientations,
according to which of its bipartition classes belongs to
the first independent set. Thus the ordered pairs with
skeleton $O$ are indexed by $\sigma\in\{-1,1\}^q$; write
them as $(I_\sigma,J_\sigma)$, with the convention that
$\sigma_i=1$ assigns $C_i\cap V^\pl$ to $I_\sigma$.
In particular, note that $
J_\sigma=I_{-\sigma},
I_{\sigma^{(i)}}=I_\sigma^{C_i}$, 
where $\sigma^{(i)}$ is obtained by reversing coordinate $i$.

Let $\mathcal{O}$ be the collection of skeletons represented
by at least one pair in $\mathcal{Q}'$.
For $O=(O_{\mathrm{agr}}, O_{\mathrm{dis}})\in\mathcal{O}$, define $\mathcal{F}_O=\{\sigma:I_\sigma,J_\sigma\in\cI^\pl\}$ and  
\begin{align*}
\cH_O
=\{\sigma:\min\{M(I_\sigma),M(J_\sigma)\}\geq\theta n\},\quad
\cH_O'
=\{\sigma:\min\{M(I_\sigma),M(J_\sigma)\}\geq2\ell\}.
\end{align*}
Since $2\ell\leq\theta n$, we have $\cH_O\subseteq\cH'_O\subseteq\mathcal F_O$. 
% Since $2\ell\leq\theta n$, we have $\cH'_O\subseteq\cH_O\subseteq\mathcal F_O$.         CHANGED
The event $\cB_{\ell,r}$ depends only on the component
structure of $G[I\oplus J]$, so all orientations of a skeleton
in $\mathcal{O}$ satisfy the component bound $\leq \ell$.

For every $\sigma\in\mathcal{F}_O$, we have
$\mu^\pl(I_\sigma)\mu^\pl(J_\sigma)=a_O$ where  $a_O:=\frac{\prod_{v\in O_{\mathrm{agr}}} \lambda_v^2\ \prod_{v\in O_{\mathrm{dis}}}\lambda_v}{Z_\pl^2}$. Consequently, if $p_O$ denotes the probability that
$(I,J)$ has skeleton $O$, then
\begin{equation}
p_O=a_O|\mathcal{F}_O|,
\qquad
\sum_{O\in\mathcal{O}}p_O\leq1.
\label{eq:skeleton-weights}
\end{equation}
In particular, for the expectation in
\eqref{eq:initial-pair-truncation} we have 
\begin{equation}
\frac{1}{2}\,
\mathbb E\left[
 \mathbf 1_{\mathcal{Q}'}
 \bigl(f(I)-f(J)\bigr)^2
\right]\leq \frac12
\sum_{O\in\mathcal{O}}a_O
\sum_{\sigma\in\mathcal{H}_O}
 \bigl(f(I_\sigma)-f(J_\sigma)\bigr)^2=:E.
\label{eq:grouped-original-pairs}
\end{equation}

We next show that almost every orientation of each
$O\in\mathcal{O}$ belongs to $\mathcal{H}_O'$. We have $M(I_\sigma)-M(J_\sigma)
=\sum_{i=1}^q\sigma_i|C_i|$, so $
\min\{M(I_\sigma),M(J_\sigma)\}
=
\frac12\left(M(I_\sigma)+M(J_\sigma)-\left|\sum_{i=1}^q\sigma_i|C_i|\right|\right)$. Using $\ell\leq\theta n/4$ and
$\sum_i|C_i|^2\leq\ell\sum_i|C_i|\leq2n\ell$,
Azuma's inequality gives  that for  a uniformly random $\sigma\in\{-1,1\}^q$ it holds that $\Pr_\sigma[\sigma\notin\cH_O']\leq \delta/(2n)$ where  recall that $\delta= 4n \emm^{-\theta^2 n/(4\ell)}\leq 1/2$.
It follows that
\begin{equation}
|\mathcal F_O|\geq|\cH_O'|
\geq(1-\delta/(2n))2^q.
\label{eq:valid-orientation-fraction}
\end{equation}

Extend $f$ to $\cI$ by setting $\hat f(I)=f(I)$ if $I\in \cI^\pl$ and $\hat f(I)=f_{\min}$ otherwise. Note that for $\sigma\in \cH_O'$ we have $I_\sigma^{C_i}\in \cI^\pl$. Therefore, applying the weak Poincare inequality on the hypercube to the function $\hat f$ and the event $\cH_O'$ gives
\begin{equation}
\sum_{\sigma\in\cH_O}
 \bigl(f(I_\sigma)- f(J_\sigma)\bigr)^2\leq \sum_{\sigma\in\{-1,1\}^q}
 \bigl(\hat f(I_\sigma)-\hat f(J_\sigma)\bigr)^2
\leq
\sum_{\sigma\in\cH_O'}\sum_{i=1}^q
 \bigl(f(I_\sigma)-f(I_\sigma^{C_i})\bigr)^2
+
\frac{q \delta 2^q}{2n}  R_f.
\label{eq:antipodal-cube-comparison}
\end{equation}
Using that $q\leq 2n$ and $|\mathcal{F}_O|\geq (1-\delta/(2n))2^q$ from \eqref{eq:valid-orientation-fraction}, we can therefore bound the expression $E$ as 
\[E\leq \frac{1}{2}\sum_{O\in\mathcal{O}}a_O\sum_{\sigma\in\cH_O'}\sum_{i=1}^q
 \bigl(f(I_\sigma)-f(I_\sigma^{C_i})\bigr)^2
+
 \tfrac{\delta}{2(1-\delta/(2n))} R_f\sum_{O\in\mathcal{O}}a_O |\mathcal{F}_O|.\]
Using that $\sum_{O\in\mathcal{O}}a_O |\mathcal{F}_O|\leq 1$ by
\eqref{eq:skeleton-weights} and $\delta/[2(1-\delta/(2n))]\leq\delta$, we therefore obtain \eqref{eq:sc-reduction}, as claimed. 
\end{proof}

\begin{proof}[Proof of~\Cref{lem:mapping}]
Suppose wlog that $Y'=Y\backslash u$; the case $Y'=Y\cup \{u\}$ is analogous and is discussed in the end. The key step is to define a function $f:=f_{Y,Y'}$ mapping $(I,J)\in \mathcal{P}_{Y\rightarrow Y'}$ to $(X,D)$ where $X\in \cI^\pl$ and $D$ is a vertex set with $|D|\leq k$ so that 
\begin{enumerate}
\item \label{it:xd} $I\cap J= X\cap Y$ and $I\oplus J=(X\oplus Y) \sqcup D$.
\item \label{it:xda}for any $\hat X\in \cI^\pl$, at most $2(d\ell)^k$ pairs $(I,J)$ have images whose first coordinate equals~$\hat X$.
\end{enumerate}
To see why this is enough to conclude the lemma, first observe that from~\Cref{it:xd} and $|D|\leq k$, we obtain that $\mu^\pl(I)\mu^\pl(J)=\mu^\pl(X)\mu^\pl(Y)\prod_{v\in D}\lambda_v\leq \lambda^{k}\mu^\pl(X)\mu^\pl(Y)$ (using that $\lambda_v\leq \lambda$ and $\lambda>1$). From~\Cref{it:xda} and $P(Y\rightarrow Y')=\frac{1}{2n(1+\lambda_u)}\geq \frac{1}{4n\lambda}$, we can therefore bound 
\[
\sum_{(I,J)\in \mathcal{P}_{Y\rightarrow Y'}}
\mu^\pl(I)\mu^\pl(J)
\leq 2(d\ell\lambda)^{k}\mu^\pl(Y)
\sum_{X\in \cI^\pl}\mu^\pl(X)
\leq 8n\lambda(d\ell\lambda)^{k}
\mu^\pl(Y)P^\pl(Y\rightarrow Y').
\]

Consider $(I,J)\in \mathcal{P}_{Y\rightarrow Y'}$ and let $C$ in $G[I\oplus J]$ be the component whose switch traverses the transition $Y\rightarrow Y'$. Let $A=I\cap C$, $B=J\cap C$ and recall that the component switch from $I$ to $I^C$ ultimately deletes all vertices from $A$ and replaces them with the vertices in $B$. 

Consider the vertex set $\widetilde X=(I\cap J)\cup\bigl((I\oplus J)\setminus Y\bigr)$.
Now $\widetilde X$ is not quite an independent set but can be turned into one with few vertex deletions. It agrees on $I\cap J$ with both $I$ and $J$ (trivially), and with $J$ on all other components in $G[I\oplus J]$ other than $C$ (note that $Y$ only differs with $I$ on some vertices in $C$). Inside $C$, we have $\widetilde X\cap C=(A\backslash Y)\cup (B\backslash Y)$, so the possible edge-conflicts are precisely between deleted vertices from $A$ and vertices from $B$ not yet inserted. Let $D$ be the vertices in $A\backslash Y$ that have some neighbour in $B\backslash Y$; by removing $D$ from $\widetilde X$ we therefore obtain an independent set $X=\widetilde X\backslash D$ satisfying~\Cref{it:xd}. 

To bound $|D|$, let $P$ be the prefix of the original
cutwidth ordering preceding the vertex $w$ currently
being processed. Let $D_w\subseteq A\setminus P$
consist of the vertices deleted while processing $w$
up to the current state. We have $|D_w|\leq d$:
processing an $A$-vertex deletes at most that vertex,
and processing a $B$-vertex deletes at most its
$d$ neighbours. For each $a\in D\cap P$, choose a neighbour in
$B\setminus Y$. This neighbour lies outside $P$,
since every $B$-vertex in $P$ has already been inserted.
For each $a\in D\setminus(P\cup D_w)$, its deletion
occurred before processing $w$, although its own
position in the ordering has not yet been reached.
Hence it was deleted while processing a neighbour
in $B\cap P$. Thus every vertex in $D\setminus D_w$ can be assigned
an edge of the cut $(P,C\setminus P)$ incident to it.
These assigned edges are distinct, since their
endpoints in $A$ are distinct. Then, with $E_C(P,C\setminus P)$ denoting the edges of $C$
crossing the cut $(P,C\setminus P)$, we obtain by Lemma~\ref{lem:cutwidth} that
\[
 |D|
 \leq |E_C(P,C\setminus P)|+|D_w|
 \leq d\lceil\log_2\ell\rceil+r+d=k.
\]

It remains to bound the number of pairs assigned a fixed
companion $\widehat X$. Let $v$ be the vertex that was updated in the $Y\rightarrow Y'$ transition (note that $\{v\}=Y\oplus Y'$). Let also $S=\widehat X\oplus Y$.
For every pair assigned $\widehat X$, we have $v\in S$:
if the transition deletes $v$, then $v\in Y\setminus\widehat X$;
if it inserts $v$, then $v\in\widehat X\setminus Y$. For fixed $\widehat X,D$, the identity
\[
 I\oplus J=S\sqcup D
\]
recovers the disagreement graph, and its component containing
$v$ is the switched component.
The direction of $Y\to Y'$ determines its orientation:
a deleted vertex belongs to $I$, whereas an inserted vertex
belongs to $J$.
Hence its source class $A$ and target class $B$ are determined.
Outside this component, $I$ agrees with $Y$ and $J$ with
$\widehat X$, so
\[
 I=(Y\setminus C)\cup A,
 \qquad
 J=(\widehat X\setminus C)\cup B.
\]
Thus each fixed $\widehat X,D$ determines at most one pair.

We now count the possible sets $D$ of size $j$.
Start with the component $R$ of $v$ in $G[S]$.
At each step, choose a vertex outside $R$ adjacent to $R$,
add it, and enlarge $R$ to include all newly connected
vertices of $S$.
Discard any exploration for which $|R|>\ell$.

Every admissible $D$ can be recovered by $j$ such additions.
Indeed, throughout its reconstruction $R\subseteq C$.
If some vertex of $D$ remains unrecovered, connectivity
of $C$ supplies an unrecovered vertex of $D$ adjacent to $R$,
because every visible vertex adjacent to $R$ has already
been absorbed.

At each retained step, the number of choices is at most $|N_G(R)\setminus R|\le d|R|\le d\ell$. 
Therefore there are at most $(d\ell)^j$ exploration
sequences of length $j$. Each sequence determines its set
of added vertices, and every admissible $D$ occurs along
at least one sequence. Hence there are at most $(d\ell)^j$
possible sets $D$ of size $j$. Combining this with reconstruction, the number of pairs
assigned $\widehat X$ is at most $
 \sum_{j=0}^{k}(d\ell)^j
 \le 2(d\ell)^k$,
as required.

The proof for an insertion transition $Y'=Y\cup\{u\}$ is obtained by applying the
same encoding with $Y'$ in place of $Y$, keeping
$A=I\cap C$ and $B=J\cap C$. The updated vertex satisfies
$u\in Y'\setminus X$, so it again gives the starting
vertex for reconstruction, and the multiplicity bound
is unchanged. Moreover, in both cases $X$ agrees with
$J$ outside $C$, so $M(X)\geq M(J)-|C|\geq\ell>0$ and  hence  $X\in\cI^\pl$. The insertion bound now follows from the deletion bound at the occupied endpoint and detailed balance, i.e., $\mu^\pl(Y')P^\pl(Y',Y)
 =\mu^\pl(Y)P^\pl(Y,Y')$.
\end{proof}

\subsection{Proof of Rapid Mixing for Glauber Dynamics}\label{sec:GlauberHC}

For large enough initial bias $s_0$ the chain, started from $V^\pl$, can be easily analyzed using coupling. We have the following mixing result which is shown in Appendix~\ref{sec:initialization}.
\newcommand{\stateleminit}{Let $d\geq 3$ and $\lambda>0$. Let $s_0>0$ be such  that $\lambda \emm^{-s_0}\le\frac{1}{4d^2}$ and $\lambda \emm^{s_0}\geq 10(\lambda+1)^2$. 

Let $G=(V^\pl\sqcup V^\mi,E)$  be a $d$-regular bipartite graph 
with activities $\lambda \emm^{s_0}$ on $V^\pl$ and $\lambda \emm^{-s_0}$ on $V^\mi$. Then, for  any $\varepsilon\geq \emm^{-n/2}$,  Glauber dynamics initialised from $X_0=V^\pl$, with proposals outside of  $\cI^\pl_G$ rejected, outputs in $\leq 4n \log (4n/\varepsilon)$ steps an $\varepsilon$-sample 
from $\mu^\pl_{G,s_0}$.}
\begin{lemma}\label{lem:init}
\stateleminit
\end{lemma}
The proof of~\Cref{lem:init} is given in Section~\ref{sec:initialization}. We now have all the pieces to complete the proof of fast mixing for  Theorem~\ref{thm:samplerHC1}. We state it here as a separate lemma, so that we can refer to it when we complete the proof of the theorem later. Recall that $(X_t^\pl)$ is the standard Glauber dynamics (no rejection) initialised at $X_0=V^\pl$.
\begin{lemma}\label{lem:samplerbHC1}
Let $d\geq 3$ be an integer and $\lambda>\lambda_c(d)$. For any $K>0$, whp over $G\sim \cG^{\mathrm{bip}}_{n,d}$,  the chain $(X_t^\pl)_{t\geq 0}$ on $G$ is within TV-distance $\eps=1/n^K$ from $\mu^\pl_G$ in $t=\tilde{O}(n^{3/2})$ steps. Similarly for $(X_t^\mi)_{t\geq 0}$.
\end{lemma}

\begin{proof}
We prove the plus-phase assertion; the minus-phase assertion is completely analogous. Fix  $\delta, K>0$, and $\lambda>\lambda_c(d)+\delta$; set
$\eps=n^{-K}$. Choose a constant $s_0\geq1$ satisfying the assumptions
of Lemma~\ref{lem:init} and let $\Lambda=\max\{2,\lambda\emm^{s_0}\}$. Let $\theta,\rho,L,T>0$ be the constants from
Theorem~\ref{thm:randomgraphHC}, applied on $[0,2s_0]$.
Set $\ell=\left\lceil\frac{2K+T+40}{\rho}\log n\right\rceil$
 and $r=\lceil2K+T+20\rceil$ .

 Apply
Theorem~\ref{thm:randomgraphHC} with the interval
$[0,2s_0]$, and fix a graph $G$ satisfying its conclusions.
Throughout the proof, let $P_s$ denote the unrestricted
Glauber dynamics with activities $\lambda\emm^s$ on $V^\pl$
and $\lambda\emm^{-s}$ on $V^\mi$, and let $P_s^\pl$
denote its restriction to $\cI_G^\pl$, with moves leaving
$\cI_G^\pl$ rejected. Theorem~\ref{thm:randomgraphHC} and
Lemma~\ref{lem:poincare} imply, uniformly for $s\in[0,s_0]$,
\begin{equation}\label{eq:HC-annealing-wpi}
 \Var_{\mu_{G,s}^\pl}(f)
 \leq A\, \cE_{P_s^\pl}(f,f)
       +b(f_{\max}-f_{\min})^2,
\end{equation}
where $A=8n(d\ell\Lambda)^{r+4d\ln\ell+1}
     =n^{1+o(1)}$ and $b= n^{-(2K+10)}$. For the variance bound, apply
\eqref{eq:sc-reduction} with $f=M$. On $\mathcal Q$ we have
$|M(I^C)-M(I)|\leq |C|$ and $\sum_C |C|^2\leq 2n\ell$.
Since $0\leq M\leq n$ on $\cI_G^\pl$, this gives, for all $s\in[0,2s_0]$,
\[
 \Var_{\mu_{G,s}^\pl}(M)\leq n\ell+bn^2\leq 2n\ell=:v.
\]

We now choose the annealing schedule to cool $s_0$ down to $s=0$, and define a chain $(Y_t)$ that tracks the evolution of the algorithm. Define in particular
\[m=\lceil s_0\sqrt{v}\rceil,\qquad
 h=\frac{s_0}{m},\qquad
 s_j=s_0-jh\quad(0\leq j\leq m),
\]
and let  $\eta=\eps/(20n m)$, $t_{\mathrm{round}}=2A\log \frac{4}{\eta}$. Consider the following chain $(Y_t)$.
Start from $Y_0=V^\pl$ and first run $P_{s_0}^\pl$ for
time $t_{0}= \lceil4n\log\frac{4n}{\eta}\rceil$. Next, for $j=1,\ldots,m$, run $P_{s_j}^\pl$ for
$t_{\mathrm{round}}$ steps starting from the configuration $Y_{T_{j}}$ where $T_j=t_0+t_{\mathrm{round}}(j-1)$.

 We show by induction that for $j=0,\hdots, m$ it holds that
\begin{equation}\label{eq:HC-one-round}
 \big\|
 Y_{T_{j+1}}
 -\mu_{G,s_j}^\pl
 \big\|_{\TV}
 \leq (j+1)\eta.
\end{equation}
For $j=0$, this follows by the initialisation   
Lemma~\ref{lem:init}, so consider $j\geq 1$. First observe that the $\chi^2$ distance between $\mu^\pl_{G,s_{j-1}}$ and $\mu^\pl_{G,s_{j}}$ is less than $\emm-1$. Indeed, with $F(s)=\log Z^\pl_{G,s}$, we have $ F''(s)=\Var_{\mu_{G,s}^\pl}(M)\leq v$ for all $s\in[0,2s_0]$, so 
\begin{align}
 1+\chi^2\bigl(
       \mu_{G,s_{j-1}}^\pl\Vert\mu_{G,s_j}^\pl
       \bigr)
=
 \frac{Z^\pl_{G,s_j+2h}Z^\pl_{G,s_j}}
      {(Z^\pl_{G,s_j+h})^2}\leq 
 \emm^{F(s_j+2h)+F(s_j)-2F(s_j+h)}\leq \emm^{h^2 v}\leq\emm.
\end{align}
Invoking the weak Poincar\'e inequality\footnote{\label{foo:spectrum}Note that $P_s^\pl$ has nonnegative spectrum, since it is an average of single-site projections, i.e., we can decompose $P_s^\pl=\frac{1}{|V|}\sum_{v\in V} P_{v,s}^\pl$ where $(P_{v,s}^\pl)^2=P_{v,s}^\pl$ for every $v\in V$.} and the $\chi^2$-bound in Lemma~\ref{lem:anneal}, we therefore obtain that
 \[\big\|
 Y_{T_{j+1}}
 -\mu_{G,s_j}^\pl
 \big\|_{\TV}\leq \big\|
 Y_{T_{j}}
 -\mu_{G,s_{j-1}}^\pl
 \big\|_{\TV}+\sqrt{\chi^2\bigl(
       \mu_{G,s_{j-1}}^\pl\Vert\mu_{G,s_j}^\pl
       \bigr)}
 \sqrt{\tfrac14\emm^{-t_{\mathrm{round}}/A}+b}\]
Since $b=n^{-(2K+10)}$, the final product is less than $\eta$, completing the inductive proof of \eqref{eq:HC-one-round}.

Since $s_m=0$ and $\mu_{G,s_m}^\pl=\mu_{G}^\pl$, it follows that at time $T=t_{0}+m t _{\mathrm{round}}=\tilde{O}(n^{3/2})$, 
we have
\begin{equation}\label{eq:HC-annealing}
 \big\|Y_T-\mu_G^\pl\big\|_{\TV}
 \leq
 (m+1)\eta\leq 2m\eta.
\end{equation}

It remains to transfer this conclusion to the unrestricted
Glauber dynamics at the fixed activity $\lambda$.
Let $(X_t^\pl)$ be this chain, started from $V^\pl$. Consider also another chain $(Z_t)$ which is obtained by picking $Z_0\sim \mu^\pl_{G}$ and running the restricted chain $P^\pl$ for $T$ steps, so that $Z_t\sim \mu^\pl_{G}$ for all $t$.

 For $I,J\in \cI_G$ we write $I\preceq J$ if  $I\cap V^\pl\subseteq J\cap V^\pl$ and 
 $I\cap V^\mi\supseteq J\cap V^\mi$. Let $\cE_t$ be the event that no rejection has occurred during the chain $Z_t$. We can couple $Z_t, X_t^\pl,Y_t$ using the same vertex updates so that 
\[\mbox{ whenever $\cE_t$ is true, }  Z_t\preceq X_t^\pl\preceq Y_t\]
The unrestricted Glauber updates preserve the order
$\preceq$ and are increasing in the bias $s$, i.e., for $s\geq 0$ we have $\frac{\lambda }{1+\lambda }\leq \frac{\lambda \emm^s}{1+\lambda \emm^s}$ and $\frac{\lambda \emm^{-s}}{1+\lambda \emm^{-s}}\leq \frac{\lambda }{1+\lambda }$.
Moreover,  a proposal that is rejected  is a downward move in the partial order. So, using common update vertices and the same uniform random
variables to determine the proposals, we can couple the chains so that
$X_t^\pl\preceq Y_t$ always, and
$Z_t\preceq X_t^\pl$ whenever $\cE_t$ holds.

A Glauber move can leave $\cI_G^\pl$ only from a configuration
with $M(I)=0$. For sufficiently large $n$, such
configurations belong to $\cI_G^{\mathrm{bal}}(\theta)$. Since $Z_t$ is distributed as $\mu^\pl_G$, we can bound
\[\Pr\big(\overline{\cE_T}\big)\leq T \mu_{G}^\pl\,\big(\cI_G^{\mathrm{bal}}(\theta)\big)\leq T\emm^{-\rho n}.\]
On the event $\cE_T$, the coupling satisfies
$Z_T\preceq X_T^\pl\preceq Y_T$, so 
 \[\mathbf1_{\{Z_T\neq X_T^\pl\}}
 \leq
 \mathbf{1}_{\overline{\cE_T}}
 +\bigl(M(Y_T)-M(Z_T)\bigr)\mathbf1_{\cE_T}=\mathbf{1}_{\overline{\cE_T}}
 +\big(M(Y_T)-M(Z_T)\big)(1-\mathbf{1}_{\overline{\cE_T}}).\]
Taking expectations and using $0\leq M(I)\leq n$ for $I\in \cI^\pl$, we obtain using  \eqref{eq:HC-annealing} that
\begin{align*}
 \| X_T^\pl-\mu_G^\pl\|_{\TV}
 \leq \Pr[X_T^\pl\neq Z_T]\leq
 (n+1)\Pr\big(\overline{\cE_T}\big)
 +n \big\|Y_T-\mu_G^\pl\big\|_{\TV}\leq  2nT\emm^{-\rho n}+2nm\eta\leq \eps,
\end{align*}
as wanted. The minus-phase follows by an analogous argument.
\end{proof}

\subsection{From Phase Samplers to Fast Global Samplers}

We state the results in this section in a more generic setup.  We have a finite space $\Omega$, a weight function $w:\Omega\to\mathbb R_{>0}$, and a Hamiltonian
$H:\Omega\to[-H_{\max},H_{\max}]$; we use $R_H$ to denote the range of $H$.  For $t\in\mathbb R$, let
\[
 Z_t=\sum_{x\in\Omega}w(x)\emm^{tH(x)},\qquad
 \mu_t(x)=\frac{w(x)\emm^{tH(x)}}{Z_t}
\] be the corresponding partition function $Z_t$ and distribution $\mu_t$. 

To accommodate for our sampling-from-phases context, let $\Omega=\Omega^\pl\,\sqcup\,\Omega^\mi$ be a partition of $\Omega$. Let $Z^\pl_t,Z^\mi_t$ be the restricted partition functions on $\Omega^\pl,\Omega^\mi$ and let $\mu^\pl_t,\mu^\mi_t$ denote the restricted distributions on the two phases.

The key component in combining samples from $\mu^\pl_t$ and $\mu^\mi_t$ will be the following concept of an upper-tail estimator. The interesting feature is that it relaxes the standard need to have small relative error but only needs a constant-factor  estimate with small tail probability. We will show in the upcoming section how to obtain such estimators using simulated annealing.
\begin{definition}
For $\delta\in (0,1/4)$, we say that a non-negative random variable $W$ is an upper-$\delta$-tailed estimator for a quantity $Z$  if $\E[W]=Z$  and $W$ satisfies the concentration bounds
\[\Pr\big[\tfrac{1}{2}Z\leq W\leq 2Z\big]\geq 1-\delta \mbox{ and } \mathbb E[W\, \mathbf{1}_{\{W>4Z\}}]\leq \delta Z.\]
\end{definition}

The next lemma shows how to combine upper-tail estimators for the phase partition functions into a high-accuracy ``selector''  for the two phases; then one can produce a sample from the mixture by just sampling from the selected phase. 
\begin{lemma}
\label{lem:anneal-rejection}
Suppose we have upper-$\delta$-tailed estimators $W^\pl,W^\mi$ for $Z^\pl,Z^\mi$ respectively.

Then there is an algorithm using at most $2L+2$ independent samples from the estimators where $L=\left\lceil64\log(1/\delta)\right\rceil$, 
and whose output $\sigma\in\{\pl,\mi\}$ is within $4\delta$ in TV-distance of the distribution
\[
 \Pr(\sigma=\pl)=\frac{Z^\pl}{Z^\pl+Z^\mi},\qquad
 \Pr(\sigma=\mi)=\frac{Z^\mi}{Z^\pl+Z^\mi}.
\]
\end{lemma}

\begin{proof}
 For $i=0,1,\hdots,L$, let $W^\pl_i,W^\mi_i$ be samples from $W^\pl$ and $W^\mi$
and set $B=8\max\{W^\pl_0,W^\mi_0\}$.  If $B=0$, return arbitrary $\sigma\in\{\pl,\mi\}$. Otherwise, make at most $L$ attempts of the following process. At attempt $i$, choose $\sigma\in\{\pl,\mi\}$ uniformly at random and with probability $\min\{W^\sigma_i/B,1\}$ accept and return $\sigma$.
If all  attempts fail we return $\pl/\mi$ arbitrarily. 

We now analyze the algorithm. With probability at least $1-2\delta$, both $W^\pl_0$ and $W^\mi_0$ are within a factor $2$ of $Z^\pl$ and $Z^\mi$ (respectively) and we have
\begin{equation}\label{eq:es1}
 4\max\{Z^\pl,Z^\mi\}\leq B\leq16\max\{Z^\pl,Z^\mi\}.
\end{equation}

Condition on the value of $B$. Let $a^\plm_i:=\mathbb E\min\{W^\plm_i,B\}$. Whenever we have \eqref{eq:es1}, it holds that $B\geq 4 Z^\pl$, so the upper-tail bound for $W^\pl$ implies $\E[W^\pl_i\, \mathbf{1}_{W^\pl_i\leq 4Z^\pl}]\leq a^\pl_i\leq \E[W^\pl_i]$. We thus obtain
\[
 (1-\delta)Z^\pl\leq a^\pl_i\leq Z^\pl, \text{ and similarly } (1-\delta)Z^\mi\leq a^\mi_i\leq Z^\mi.
\]
Note that the probability that the filter accepts sign $\sigma\in \{\pl,\mi\}$ is $
\frac{1}{2}\mathbb E\min\{W^\sigma/B,1\} = a^\sigma/(2B)$
and hence conditioned on the algorithm accepting a sign, the sign $\sigma$ has distribution $a^\sigma/(a^\pl+a^\mi)$;
this is TV distance $\leq\delta$ from $Z^\sigma/(Z^\pl+Z^\mi)$ using that
\[\big|\frac{a^\pl}{a^\pl+a^\mi} -\frac{Z^\pl}{Z^\pl+Z^\mi}\big|=
\frac{|a^\pl Z^\mi-a^\mi Z^\pl|}
     {(a^\pl+a^\mi)(Z^\pl+Z^\mi)}
\leq
\frac{\delta Z^\pl Z^\mi}
     {(1-\delta)(Z^\pl+Z^\mi)^2}
\leq \frac{\delta}{4(1-\delta)}
\leq\delta.\]
The probability of acceptance in each attempt is
$\frac{a^\pl+a^\mi}{2B}
 \geq\frac{(1-\delta)(Z^\pl+Z^\mi)}{32\max\{Z^\pl,Z^\mi\}}
 \geq\frac1{64}$,
and hence the probability of having no acceptance in $L$ attempts is at most $\emm^{-L/64}\leq\delta$. The total error is now 
bounded by $4\delta$ (where $2\delta$ is for the failure of~\eqref{eq:es1}, $\delta$ is for the no accept event in the $L$ attempts, and $\delta$ is the TV-distance between the two distributions of the sign $\sigma$).
\end{proof}

\subsubsection{Fast upper-tail estimators from annealing and weak Poincar\'e inequalities}

\begin{lemma}
\label{lem:anneal-product}
Let $\delta\in (0,1/4)$ and $h$ satisfy $1/h\geq 10H_{\max} \log(4/\delta)$. For $t_0\in\mathbb R$ and an integer $m\geq1$ let
$t_i=t_0+ih$. 

Consider independent $X_i\sim\mu_{t_i}$ for $i=0,1,\hdots, m-1$ and let  $W=Z_{t_0}\prod_{i=0}^{m-1}\emm^{hH(X_i)}$. Then,  $W$ is an upper-$\delta$-tailed estimator for $Z_{t_m}$.
\end{lemma}

\begin{proof}
From independence of the $X_i$'s, we have that
\[
\mathbb{E} W=Z_{t_0}\prod^{m-1}_{i=0}\mathbb E_{\mu_{t_i}}\emm^{hH(X_i)}=Z_{t_0}\prod^{m-1}_{i=0}\frac{Z_{t_{i+1}}}{Z_{{t_i}}}=Z_{t_m}.\]
Let $U=W/Z_{t_m}$, so that our remaining goal is to show that
\begin{equation}\label{eq:anneal-product-tails}
 \Pr\big[U\notin[1/2,2]\big]\leq\delta,
 \qquad
 \mathbb E[U\mathbf1_{\{U>4\}}]\leq\delta.
\end{equation}
 Since $ H\in [-H_{\max},H_{\max}]$ we have for every $t\in{\mathbb R}$ and $j\geq 0$ the crude bounds
\begin{equation}\label{epp}
\emm^{-jH_{\max}}\leq \frac{Z_{t+j}}{Z_t}=\E_{\mu_t}[\emm^{ j H}]\leq \emm^{ j H_{\max}}.
\end{equation}
Applying~\eqref{epp}  we obtain for integers $p\geq 1$, with $Z_i=Z_{t_i}$, 
\begin{equation}\label{eq:anneal-product-moments1}
 \E\big[U^p\big]
 =\left(\frac{Z_0}{Z_m}\right)^p
   \prod_{i=0}^{m-1}\frac{Z_{i+p}}{Z_i}
 =\prod_{k=0}^{p-1}
   \frac{Z_{m+k}Z_0}{Z_kZ_m}
   \leq\prod_{k=0}^{p-1}\emm^{2H_{\max} kh}
 =\emm^{H_{\max} hp(p-1)},
\end{equation}
and similarly  $\E U^{-p}\leq \emm^{H_{\max} hp(p+1)}$. 

Then, for $L=\log(4/\delta)>1$ and $p=\lfloor 3 L\rfloor$, we have $p(p+1)\leq 10L^2$, so we obtain the bounds on  $\E\big[U^{\pm p}\big]\leq \emm^{L}$. Markov's inequality then gives
\[
 \Pr(U\geq2)+\Pr(U\leq1/2)\leq 2^{-p}\E U^p + 2^{-p}\E U^{-p}
 \leq 4\emm^{-(3\log2-1)L}
 \leq\delta.
\]
Moreover, writing $U=U^{1-p}U^{p}$ and using $p\geq 1$, we have $\mathbb E[U\mathbf1_{\{U>4\}}]
 \leq4^{1-p}\mathbb E[U^p]
 \leq\delta$. 
\end{proof}

The next lemma shows that the values of a finite-valued statistic recorded
along a warm-start annealing run can be coupled to independent values of the
statistic. Note that since Lemma~\ref{lem:anneal-product} only uses the value 
of the hamiltonian in the estimator we can use it for samples generated
by warm-start annealing run (with the only ``cost'' being the coupling bound).

Recall that $R_H$ denotes the range of the Hamiltonian $H:\Omega\rightarrow [-H_{\max},H_{\max}]$.
\begin{lemma}\label{lem:anneal-statistic}
Let $\delta\in (0,1/4)$ and $h$ satisfy $1/h\geq 10H_{\max} \log(4/\delta)$. For $t_0\in\mathbb R$ and an integer $m\geq1$ let
$t_i=t_0+ih$. Let $P_i$ be a reversible Markov chain with stationary
distribution $\mu_{t_i}$ and nonnegative spectrum, satisfying an $(A,b)$ weak-Poincar\'e inequality for some fixed $A,b>0$ independent of $i$.

Suppose that  $X_0\sim\mu_{t_0}$, and for
$i=1,\hdots,m$ obtain $X_i$ by running $P_i$ for $T$ steps from $X_{i-1}$. Then, we can couple
$H(X_0),\ldots,H(X_m)$ with random variables $H(Y_0),\ldots,H(Y_m)$ such that $Y_0,\ldots,Y_m$ are independent, $Y_i\sim\mu_{t_i}$, and
\[
 \Pr\!\big[ H(X_i)\ne H(Y_i) \mbox{ for some } i=0,1,\hdots,m\big]
 \le
 m^2 |R_H|(\emm^{-T/(2A)}+\sqrt b).
\]
\end{lemma}
\begin{proof}
Write for convenience $\mu_i=\mu_{t_i}$ and $P_i=P_{t_i}$, and $u=\emm^{-T/(2A)}+\sqrt b$. Since $t_{i}-t_{i-1}\leq h$ for every
$i=1,\ldots,m$, we obtain that for $x\in\Omega$
\begin{equation}\label{eq:ratiobrf}
 \frac{\mu_{i-1}(x)}{\mu_{i}(x)}
 =\emm^{-hH(x)}\frac{Z_{i}}{Z_{i-1}}
 \leq\emm^{2hH_{\max}}
 \leq2.
\end{equation}
Let $\hat \mu_i$ denote the distribution of $X_i$. By Lemma~\ref{lem:anneal}, we have
 $\big\|\mu_{i-1}P_i^T-\mu_{i}\big\|_{\TV}
 \leq2u$, so  by induction
\begin{equation}\label{eq:anneal-statistic-marginals}
 \big\|\hat \mu_i-\mu_{i}\big\|_{\TV}
 \leq 2iu.
\end{equation}

For $a\in R_H$,  consider the function
$f_a(x)=\mathbf1_{\{H(x)=a\}}$ for $x\in \Omega$  and write $p_a=\E_{\mu_i}[f_a]$ for the probability that $H(Y_i)=a$ (where $Y_i\sim \mu_i$). For $x\in \Omega$, let $\delta_x$ denote the probability distribution on $\Omega$ with all of its mass on $x$. Then we claim that
\begin{equation}\label{ezooo}
 \E_{x\sim \hat\mu_{i-1}}|\delta_x P_i^T f_a-p_a|
 \leq 2iu.
\end{equation}
Using the TV bound in \eqref{eq:anneal-statistic-marginals} and the pointwise bound \eqref{eq:ratiobrf}, we only need to show $\E_{x\sim \mu_{i}}|\delta_x P_i^T f_a-p_a|
 \leq u$. Consider $a\in R_H$ with $p_a\neq 0$ (otherwise the inequality is trivial). Let 
$\nu_a=\mu_i(\,\cdot\mid H(x)=a)$. From reversibility, $\mu_i(x)\delta_xP_i^Tf_a=\sum_y\mu_i(x)P_i^T(x,y)f_a(y)=\sum_y\mu_i(y)P_i^T(y,x)f_a(y)=p_a (\nu_aP_i^T)(x)$, so
\[
\E_{x\sim \mu_i}|\delta_x P^T_i f_a-p_a|=
 p_a\sum_x
   \big|(\nu_aP_i^T)(x)-\mu_i(x)\big|=2p_a\|\nu_aP_i^T-\mu_i\|_{\mathrm{TV}}.
\]
We have $\chi^2(\nu_a\Vert\mu_i)=\sum_{x} \mu_i(x)(\frac{\nu_a(x)}{\mu_i(x)}-1)^2=(1-p_a)+(\frac{1}{p_a}-1)^2p_a=(1-p_a)/p_a$,
so the $\chi^2$ bound in \Cref{lem:anneal} gives
\[
 \mathbb E_{\mu_i}\big|P_i^T f_a-p_a\big|
 \leq
 2u\sqrt{p_a(1-p_a)}
 \leq u
\]
This finishes the proof of \eqref{ezooo}.

We next construct the desired coupling.
We will use the following auxiliary independent couplings. For step $i>0$, consider an independent copy $(Z_t)$ of $P_i$ and let $\hat p_a(x)=\Pr[H(Z_T)=a\mid Z_0=x]=\delta_x P^T_i f_a$. For each possible starting state $x$,
maximally couple the distributions $\hat{\mathbf{p}}(x)=(\hat p_a(x))_a$ and $\mathbf{p}=(p_a)_a$.

Now, generate $Y_0,Y_1,\hdots, Y_m$ independently so that $Y_i\sim \mu_i$. We then couple $X_0,X_1,\hdots, X_m$ to them. Since $X_0\sim\mu_0$, we can couple $H(X_0)$ and $H(Y_0)$ perfectly (by coupling for example $X_0=Y_0$).  For $i\geq 1$, conditioned on $X_{i-1}=x$,  maximally couple $H(X_i)$ with $H(Y_i)$ using the auxiliary coupling of $\hat{\mathbf{p}}(x)$ with $\mathbf{p}$, conditioned on the value of the second coordinate.  The second marginal is $\mathbf{p}=(p_a)_a$, so we can then finish sampling $X_i$ from the distribution 
$P_i^T(x,\cdot)$ conditioned on its sampled statistic. Hence, both $X_0,\hdots, X_m$ and $Y_0,\hdots, Y_m$  have the desired distributions. 

Using the coupling and the bound \eqref{ezooo}, we conclude that \[\Pr[H(X_i)\neq H(Y_i)]=\E_{x\sim \hat \mu_{i-1}} \tfrac{1}{2}\sum_a |\delta_x P_i^T f_a-p_a|\leq i|R_H|u.\]
A union bound over $i$  gives the desired bound 
$
 \Pr\Big[\bigcup^m_{i=0}\{H(X_i)\neq H(Y_i)\}\Big]\leq m^2 |R_H|u$.
\end{proof}

\subsubsection{From phase estimators and samplers to full samplers}\label{sec:finalcombo}

We are now able to combine all the pieces and complete the proof of Theorem~\ref{thm:samplerHC1}.

\begin{proof}[Proof of Theorem~\ref{thm:samplerHC1}]
We just do the second item of the Theorem here, we have covered the first part in Lemma~\ref{lem:samplerbHC1}.
Fix $\lambda>\lambda_c(d)$ and $K>0$, and set
$\eps=n^{-K}$. 

Let $\eta=\frac{\eps}{100}$ and $
 L=\left\lceil64\log(1/\eta)\right\rceil$.
Choose a constant $s_0\geq1$ satisfying the assumptions
of Lemma~\ref{lem:init}, and set the length $m_0$ of the cooling schedule to be $ m_0=\left\lceil10ns_0\log(4/\eta)\right\rceil$ and the increments $h=s_0/m_0$. Choose an integer $k=O(\log n)$ sufficiently large that
$n(\lambda+\lambda^{-1})\emm^{-ks_0}\leq\eta$,
and define
\[
 \hat s_0=ks_0,\qquad m=km_0,\qquad
 \hat s_i=\hat s_0-ih\quad(0\leq i\leq m).
\]
Thus $\hat s_m=0$, the schedule includes $s_0$, and
$m=O(n\log^2 n)$. Let for convenience $\mu^\pl_i=\mu^\pl_{G,\hat s_i}$, $Z^\pl_i=Z^\pl_{G,\hat s_i}$ and $Z^\pl=Z^\pl_G$.

We first describe an ideal estimator for the plus phase. 
For $i=0,\ldots,m-1$, let  $I_i\sim\mu^\pl_{i}$ be independent samples and set
\[
 W^\pl
 =(\lambda\emm^{\hat s_0})^n
   \prod_{i=0}^{m-1}\emm^{-hM(I_i)}.
\]
Apply Lemma~\ref{lem:anneal-product} on $\cI^\pl_G$
with Hamiltonian $H(I)=-M(I)$, parameter $t_i=-\hat s_i$, and
$H_{\max}=n$.
By scaling the estimator in that lemma, $W^\pl$ is
an upper-$\eta$-tailed estimator for
$\widetilde Z^\pl
 =\frac{(\lambda\emm^{\hat s_0})^n}{Z^\pl_{0}}Z^\pl$.
Moreover, the contribution of $V^\pl$ gives the lower bound $Z^\pl_{0}\geq (\lambda\emm^{\hat s_0})^n
$, whereas the sum
over all vertex subsets gives the upper bound
$Z^\pl_{0}
 \leq(1+\lambda\emm^{\hat s_0})^n(1+\lambda\emm^{-\hat s_0})^n
 \leq(\lambda\emm^{\hat s_0})^n
       \emm^{n(\lambda^{-1}+\lambda)\emm^{-\hat s_0}}
 \leq(\lambda\emm^{\hat s_0})^n\emm^\eta$.
Consequently,
\begin{equation}\label{eq:HC-scaled-partitions}
 \emm^{-\eta}Z^\pl_G
 \leq\widetilde Z^\pl_G
 \leq Z^\pl_G.
\end{equation}
Similarly, we obtain an upper-$\eta$-tail estimator $W^\mi$
for the analogous $\widetilde Z^\mi_G$ satisfying the same bounds.

We next implement these estimators using Glauber dynamics.
As in the proof of Lemma~\ref{lem:samplerbHC1}, choose
the parameters in Theorem~\ref{thm:randomgraphHC} and
Lemma~\ref{lem:poincare} so that, whp over $G$, the
restricted chains in both phases satisfy, uniformly
for $s\in[0,s_0]$, a weak Poincar\'e inequality with $A=n^{1+o(1)}$ and $b= 1/n^{2K+20}$. Fix a graph satisfying these conclusions, and let $ T=\left\lceil2(K+10)A\log n\right\rceil$.

For each temperature $\hat s_i>s_0$, generate an independent
$1/n^{K+10}$-sample from $\mu^\pl_{G,\hat s_i}$ using
Lemma~\ref{lem:init}, starting from $V^\pl$.
At temperature $s_0$, initialise another chain in the
same way. For the remaining temperatures, run the
restricted Glauber dynamics for $T$ steps at each
temperature, starting from the preceding configuration.

The samples generated by Lemma~\ref{lem:init} can be
coupled to independent exact samples with total
failure probability at most $(m+1)n^{-K-10}$.
For the final block of $m_0$ temperatures, apply
Lemma~\ref{lem:anneal-statistic} (with Hamiltonian $H(I)=-M(I)$).
Here the Hamiltonian is integer-valued and the
restricted Glauber chains have nonnegative spectrum.
Thus the entire sequence of recorded magnetisations
can be coupled to those of the ideal independent
samples with failure probability at most
\[
 (m+1)/n^{K+10}
 +m_0^2(n+1)
   \bigl(\emm^{-T/(2A)}+\sqrt b\bigr)
 \leq \frac{\eta}{2L+2},
\]
for all sufficiently large $n$.
Using the recorded magnetisations in the product
defining $W^\pl$ therefore gives an implementation
that agrees with the ideal estimator under this
coupling except with probability $\eta/(2L+2)$.
The minus-phase estimator is implemented analogously.

Now apply Lemma~\ref{lem:anneal-rejection}, using
independent runs for its at most $2L+2$ estimator
samples. A union bound shows that all implemented
estimators agree with their ideal counterparts
except with probability at most $\eta$.
The resulting sign is therefore within TV-distance
$5\eta$ of a Bernoulli  with probability $\tilde p=\widetilde Z^\pl/(\widetilde Z^\pl+\widetilde Z^\mi)$. 
By \eqref{eq:HC-scaled-partitions} and its minus-phase
analogue, this distribution is within TV-distance
$\emm^\eta-1\leq2\eta$ of the desired phase probabilities $p=Z^\pl/(Z^\pl_G+Z^\mi_G)$. 
Finally, conditional on the selected sign, use
Lemma~\ref{lem:samplerbHC1} (with fresh randomness) to
generate an $\eta$-sample from the corresponding
phase. Since $\mu_G$ is the mixture of the two phase
distributions with these probabilities, the output
is within TV-distance $8\eta\leq\eps$ of $\mu_G$.

Each estimator requires at most $m+1$ applications
of Lemma~\ref{lem:init}, costing $O(n\log n)$ steps
each, and $m_0T$ steps for the final annealing block.
Its total running time is therefore $O(mn\log n+m_0T)=\tilde O(n^2)$. There are $O(\log n)$ estimator samples, and the
final phase sample costs $\tilde O(n^{3/2})$.
The total running time remains $\tilde O(n^2)$,
as required.
\end{proof}

\section{Proof for the Ferromagnetic Ising Model}
\label{sec:GlauberIS}

In this section, we give the proof of Theorem~\ref{thm:samplerIS}. The details resemble largely those for the hard-core model so we highlight the key differences mainly and do the proofs a bit faster. 

As in the hard-core model, we compare Glauber dynamics with a simulated
annealing algorithm that initially favours the plus phase. Let $G=(V,E)$
be a $d$-regular graph with $n$ vertices. For $s\geq0$, we consider the magnetization-weighted distribution
\[
 \nu_{G,s}(\sigma)\propto\emm^{\beta m(\sigma)+sM(\sigma)} \mbox{ for } \sigma\in \{0,1\}^V,
\]
where $M(\sigma)=\sum_{v\in V}\sigma_v-n/2$ is the magnetization of the configuration. 

We can write the phases $\Sigma^\pl,\Sigma^\mi$ as $\Sigma^\pl=\{\sigma\in \{0,1\}^V\mid M(\sigma)\geq0\}$ and $\Sigma^\mi=\{\sigma\in \{0,1\}^V\mid M(\sigma)<0\}$.  We then consider the phase-conditioned distributions $\nu_{G,s}^\plm=\nu_{G,s}(\,\cdot\mid\Sigma^\plm)$. At $s=0$ we recover the target distribution $\nu_G^\plm$. Analogously to the hard-core model, we refer to $s$ as the field.

We start at a sufficiently large constant field $s_0$ and gradually
decrease $s$ to zero, running Glauber dynamics within the phase at each
stage. A weak Poincar\'e inequality controls the error along this
schedule. Monotonicity in the external field then lets us compare the
annealed chain with the usual Glauber dynamics at the fixed target
temperature. Write $P_s$ for the single-site heat-bath chain and
$P_s^\pl$ for its restriction to $\Sigma^\pl$, rejecting moves that
leave the phase.

We first describe the disagreement components used in the comparison.
For $\sigma,\sigma'\in\Sigma^\pl$, let
$\sigma\oplus \sigma'=\{v:\sigma_v\neq\sigma'_v\}$. Note that the edges in the induced graph $G[\sigma\oplus \sigma']$ that are monochromatic under $\sigma$ are also monochromatic under $\sigma'$. 
Let $F_{\sigma,\sigma'}$ be the random edge set obtained by independently
keeping each edge of  $G[\sigma\oplus \sigma']$ that is monochromatic
under $\sigma$ with probability $1-\emm^{-2\beta}$.
Conditioned on $F_{\sigma,\sigma'}=F$, let $\mathcal C(\sigma,\sigma',F)$ be the
components of $(\sigma\oplus \sigma',F)$, including isolated vertices.

The main observation is that each component in $\mathcal C(\sigma,\sigma',F)$ is monochromatic in both $\sigma,\sigma'$, with opposite
spins in the two. The motivation behind  the percolation step in determining the edges $F_{\sigma,\sigma'}$ is that these
components can be exchanged between the configurations without changing
the joint weight of the initial and the final triples (under the joint distribution of $\sigma,\sigma',F_{\sigma,\sigma'}$), provided both configurations remain in the phase. The percolation step further ensures that the disagreement component steps have size $O(\log n)$ and $O(1)$ tree-excess with high probability over the choice of independent $\sigma,\sigma'\sim\nu^\pl_{G,s}$.

Let $\cB_{\ell,r}$ consist of triples $(\sigma,\sigma',F)$ for which
some $C\in\mathcal C(\sigma,\sigma',F)$ has $|C|>\ell$ or
$\mathtt{exc}(C)>r$. For independent $\sigma,\sigma'\sim\nu_{G,s}^\pl$ and the percolated edge-subset $F_{\sigma,\sigma'}$,
set $b_{G,\ell,r}(s)=\Pr[(\sigma,\sigma',F_{\sigma,\sigma'})\in\cB_{\ell,r}]$.
All probabilities below include this auxiliary sampling when relevant.

\begin{lemma}\label{lem:randomgraphISb}
Fix integer $d\geq3$, and real $\beta>\beta_c(d)$ and $s_0>0$. For every $D>0$, there are constants $C,\theta,\rho>0$ and an
integer $r\geq1$ such that the following holds whp over $G\sim\cG_{n,d}$.

For all $s\in[0,s_0]$, with $\ell=\lceil C\log n\rceil$, we have $b_{G,\ell,r}(s)\leq n^{-D}$ and $\nu_{G,s}^\pl(M(\sigma)\leq\theta n)\leq\emm^{-\rho n}$.
\end{lemma}
The proof of Lemma~\ref{lem:randomgraphISb} is given in Section~\ref{sec:randomgraphIS}.

\subsection{A weak Poincar\'e inequality}

We convert these estimates into a weak Poincar\'e inequality by first
reducing the comparison to switches of individual components and then
implementing each switch by single-spin updates, as in the hard-core proof.

\begin{lemma}\label{lem:poincareIS}
Let $G$ have maximum degree at most $d$, let $0\leq s\leq s_0$,
and let $\ell,r\geq1$ be integers. Suppose $\theta>0$ and
$\ell\leq\theta n/8$.

For independent $\sigma,\sigma'\sim\nu_{G,s}^\pl$ and the percolated edge-subset $F_{\sigma,\sigma'}$, 
let $b$ be the probability of
the event $\{(\sigma,\sigma',F_{\sigma,\sigma'})\in\cB_{\ell,r}\}
\cup\{\min(M(\sigma),M(\sigma'))\leq\theta n\}$.
Then, 
\[
 \Var_{\nu_{G,s}^\pl}(f)
 \leq A\cE_{P_s^\pl}(f,f)
 +(b+\delta) (f_{\max}-f_{\min})^2,\qquad (\forall f:\Sigma^\pl\to\mathbb R)
\]
where $k=d\lceil\log_2\ell\rceil+r$,
$A=4n\ell\emm^{\beta d+2s_0}(d\ell\emm^{2\beta})^k$ and
$\delta=4n\emm^{-\theta^2n/(8\ell)}$.
\end{lemma}

For $U\subseteq V$, let $\sigma^U$ denote the configuration obtained
by flipping the spins of $\sigma$ on $U$. In particular, for a component
$C\in\mathcal C(\sigma,\sigma',F)$, the configuration $\sigma^C$
replaces the spins on $C$ with those of $\sigma'$.
Fix vertex and edge labellings. The component switch
$\gamma_{\sigma,\sigma',F,C}$ flips the vertices of $C$, one by one,
in the cutwidth order from Lemma~\ref{lem:cutwidth}, applied to the
retained graph $(C,F[C])$ where $F[C]$ denotes the edges in $F$ with both endpoints in $C$. If $|C|\leq\ell$ and $M(\sigma)\geq4\ell$,
all configurations on this path belong to $\Sigma^\pl$.

\begin{proof}[Proof of Lemma~\ref{lem:poincareIS}]
Write $R_f=(f_{\max}-f_{\min})^2$. Let $\mathcal Q$ be the event
that the triple $(\sigma,\sigma', F)$ is outside $\cB_{\ell,r}$ and both magnetisations $M(\sigma),M(\sigma')$
exceed $\theta n$. Let also $\mathcal Q'$ be the same event with
the magnetisation requirement replaced by $\min(M(\sigma),M(\sigma'))\geq4\ell$.
The variance identity bounds the contribution outside $\mathcal Q$
by $bR_f/2$. We claim the following one-component reduction:
\begin{equation}\label{eq:IS-component-reduction}
 \Var_{\nu_{G,s}^\pl}(f)
 \leq\frac12\E\bigg[\mathbf1_{\mathcal Q'}
  \sum_{C\in\mathcal C(\sigma,\sigma',F_{\sigma,\sigma'})}
        (f(\sigma^C)-f(\sigma))^2\bigg]+(b+\delta)R_f.
\end{equation}
The larger event $\mathcal Q'$ allows component exchanges while
leaving enough room for the intermediate Glauber moves.

To prove \eqref{eq:IS-component-reduction}, group triples $(\sigma,\sigma',F)$ in $\mathcal{Q}$
by their disagreement set $D=\sigma\oplus \sigma'$, the common spins on $V\setminus D$, and the edge-set $F$. We refer to $O=(D,\sigma_{V\setminus D},F)$ as the skeleton of the triple $(\sigma,\sigma', F)$. If its components are
$C_1,\ldots,C_j$, each has two orientations, according to its spin
in the first configuration; its spin in the second is opposite.
The joint probability mass of a triple is
\[
 \varrho_s(\sigma,\sigma',F)
 =\nu_{G,s}^\pl(\sigma)\nu_{G,s}^\pl(\sigma')
       \Pr[F_{\sigma,\sigma'}=F].
\]
This mass is constant over orientations for which both configurations
are in the phase. Indeed, the fields cancel on $D$, an edge from $D$
to its complement contributes $\emm^\beta$, and an edge within $D$
contributes $\emm^{2\beta}-1$ if retained and $1$ otherwise.
For the last assertion, an unretained monochromatic edge contributes
$\emm^{2\beta}\emm^{-2\beta}=1$, just as a bichromatic edge.

For a skeleton represented in $\mathcal Q$, write
$a=\sum_{v\notin D}(\sigma_v-1/2)>\theta n$. Under independent uniform
orientations $\varepsilon_i\in\{-1,1\}$, the two magnetisations are
\[
 a+\frac{1}{2}\sum_i\varepsilon_i|C_i|\quad\text{and}\quad
 a-\frac{1}{2}\sum_i\varepsilon_i|C_i|.
\]
Since $\sum_i|C_i|^2\leq n\ell$ and $4\ell\leq\theta n/2$,
Hoeffding's inequality bounds the fraction of orientations outside
$\mathcal Q'$ by $2\emm^{-\theta^2n/(8\ell)}$.
We can therefore apply the same hypercube argument as in the proof of
Lemma~\ref{lem:poincare}: reversing every coordinate exchanges the
two configurations, whereas reversing one coordinate switches one
component. Extend $f$ outside the phase by $f_{\min}$, apply the
antipodal hypercube inequality, and discard the unsafe orientations.
There are at most $n$ coordinates, and the feasible orientations
occupy at least the safe fraction of the cube, so the discarded
contribution is at most $\delta\, R_f$. Enlarging the resulting
component sum to all triples in $\mathcal Q'$ proves
\eqref{eq:IS-component-reduction}. If the safe fraction is less than
$1/2$, the claimed inequality already follows from $\Var(f)\leq R_f/4$.

It remains to bound congestion along the component switches.
For a transition $Y\to Y^v$, let $\mathcal P_{Y\to Y^v}$ be the
triples in $\mathcal Q'$ whose component switch traverses it.
The analogue of Lemma~\ref{lem:mapping} is
\begin{equation}\label{eq:IS-congestion}
 \sum_{(\sigma,\sigma',F)\in\mathcal P_{Y\to Y^v}}
       \varrho_s(\sigma,\sigma',F)
 \leq\frac{A}{\ell}\nu_{G,s}^\pl(Y)P_s^\pl(Y,Y^v).
\end{equation}
Here the updated vertex determines the switched component.
If $U\subseteq C$ is its processed prefix, encode the triple by
$X=(\sigma')^U$, $F_0=F\setminus L$ and $L=F[U,C\setminus U]$.
Thus $Y=\sigma^U$, both $Y,X$ remain in the phase, and $|L|\leq k$.
Deleting $L$ leaves every retained edge monochromatic in the encoded
pair. The weight calculation above gives the exact identity
\[
 \varrho_s(\sigma,\sigma',F)
 =(\emm^{2\beta}-1)^{|L|}\varrho_s(Y,X,F_0)
 \leq\emm^{2\beta k}\varrho_s(Y,X,F_0).
\]
For fixed $Y,X,F_0,v$, record the query positions of the missing edges
in a deterministic search from $v$. Traverse edges of $F_0$ and edges
whose query indices are recorded, inspecting each incident edge once,
including edges between discovered vertices. This reconstructs $C$.
There are at most $d\ell$ queries and $k$ missing edges,
hence at most $2(d\ell)^k$ records, as in Lemma~\ref{lem:mapping}.
Restoring $L$ recovers $C$; the original spin throughout $C$ is
$Y_v$, since $v$ is not yet processed. This recovers the original
pair. Summing over $X,F_0$ gives $\nu_{G,s}^\pl(Y)$, and
$P_s^\pl(Y,Y^v)\geq\emm^{-\beta d-2s_0}/(2n)$ proves
\eqref{eq:IS-congestion}. Cauchy--Schwarz along paths of length
at most $\ell$, followed by \eqref{eq:IS-component-reduction},
now gives the desired weak Poincar\'e inequality (the full details are analogous to those used in the proof of Lemma~\ref{lem:mapping} for the hard-core model).
\end{proof}

\subsection{Proof of rapid mixing for Glauber}

We are now able to complete the proof of Theorem~\ref{thm:samplerIS}. We will use the following initialisation lemma at a sufficiently large constant field, which has an  elementary coupling proof that is given for completeness in Appendix~\ref{sec:initialization}.
\newcommand{\stateleminitIS}{Let $G$ have $n$ vertices and maximum degree at most $d$, and choose
$s_0\geq\max\{\beta d+\log(2d),\log4+6\}$.
For $\emm^{-n/2}\leq\eta\leq1$, the chain $P_{s_0}^\pl$ from all-plus
is within distance $\eta$ of $\nu_{G,s_0}^\pl$ after
$\lceil2n\log(4n/\eta)\rceil$ steps, for all sufficiently large $n$.
}
\begin{lemma}\label{lem:initIS}
\stateleminitIS
\end{lemma}

\begin{proof}[Proof of Theorem~\ref{thm:samplerIS}]
We prove the plus-phase assertion. Fix $\beta>\beta_c(d)$
and $\eps=1/n^{K}$, and choose $s_0$ as in Lemma~\ref{lem:initIS}.
Apply Lemma~\ref{lem:randomgraphISb} on $[0,2s_0]$ with sufficiently
large $D$, and fix a graph satisfying its conclusions.
Lemma~\ref{lem:poincareIS} gives a weak Poincar\'e inequality with
$A=n^{1+o(1)}$ and additive error at most $n^{-(2K+10)}$. Applying \eqref{eq:IS-component-reduction} with $f=M$ and using
$|M(\sigma^C)-M(\sigma)|=|C|$ and $\sum_C |C|^2\leq n\ell$ on
$\mathcal Q'$, we obtain, for all $s\in[0,2s_0]$,
\[
 \Var_{\nu_{G,s}^\pl}(M)\leq \tfrac12 n\ell+1\leq 4Cn\log n=:v.
\]
As in Section~\ref{sec:GlauberHC}, take $
 m=\lceil s_0\sqrt v\rceil$,  $h=s_0/m$, $
 s_j=s_0-jh$ and $\eta=\eps/(20nm)$.

Initialise at $s=s_0$ using Lemma~\ref{lem:initIS}, and then run
$P_{s_j}^\pl$ for $\lceil2A\log(4/\eta)\rceil$ steps at each stage.
For $F(s)=\log Z_{G,s}^\pl$, the bound $F''(s)\leq v$ gives
$1+\chi^2(\nu_{G,s_j+h}^\pl\Vert\nu_{G,s_j}^\pl)\leq\emm^{h^2v}\leq\emm$.
Lemma~\ref{lem:anneal} therefore bounds the final error of the annealed
chain by $\|Y_T-\nu_G^\pl\|_{\TV}\leq2m\eta$, where $T=n^{3/2+o(1)}$.
Its nonnegative-spectrum hypothesis holds because restricted heat-bath
kernels are averages of single-site conditional-expectation projections.

To compare with the fixed-temperature chain, couple $Y_t$, the
zero-field chain $X_t^\pl$ from all-plus, and a stationary restricted
zero-field chain $Z_t$ by common updates. The update probability of
a plus spin is increasing in the field and in the neighbouring spins;
rejecting a move of $Y_t$ out of the phase only increases its state.
Consequently $Z_t\leq X_t^\pl\leq Y_t$ until the first rejected
proposal of $Z_t$. Such a rejection requires $0\leq M(Z_t)\leq2$,
so its probability by time $T$ is at most $T\emm^{-\rho n}$.
The magnetisation comparison from Section~\ref{sec:GlauberHC}, using
the number of plus spins, gives
\[
 \|X_T^\pl-\nu_G^\pl\|_{\TV}
 \leq(n+1)T\emm^{-\rho n}+n\|Y_T-\nu_G^\pl\|_{\TV}\leq\eps.
\]
Spin reversal proves the minus-phase assertion, with only an
exponentially small correction for the balanced configurations when
$n$ is even. Finally, a plus-phase sample to accuracy $\eps/2$,
followed by a fair global spin flip, is an $\eps$-sample from $\nu_G$.
This proves the full-sampling assertion with the same running time.
\end{proof}

\section{Proofs for the random-cluster model}
\label{sec:RC-wpi}

\subsection{Lifting edge subsets to edge-vertex configurations}
Recall that for real parameters $\beta,q>0$ the random-cluster (RC) distribution on a graph $G=(V,E)$ is given on edge subsets $F\subseteq E$ by
$\pi(F) = \frac{(\emm^\beta-1)^{|F|}q^{\kappa(F)}}{Z}$, 
where $\kappa(F)$ is the number of connected components
of $(V,F)$, including isolated vertices.

 We lift an edge configuration $F$ to a pair $(F,S)$ where 
$S\subseteq V$ is a random collection of  connected components
of $(V,F)$. It will be convenient to use $F[S]$ to denote the edges in $F$ with both endpoints in $S$. In the case of integer $q$, the vertex lifting is analogous to considering a single  color class of the Potts model.

More formally, each of the components will get independently label 1  with probability $1/q$. We will refer to components with label 1 as active, and the rest as inactive. Then, $S$ is the set of \textit{active} vertices. We let $\Omega\subseteq 2^{E}\times 2^V$ denote the pairs $(F,S)$ that are compatible; we will also use $\omega$ to denote an element $(F,S)$ of $\Omega$, and use $\omega_{\edge}$ and $\omega_{\ver}$ to denote the sets $F,S$ respectively. Finally we use $\omega_{\mathrm{inact}}$ to denote the edges whose both endpoints are in $V\backslash S$  and  $\kappa_0(\omega)$ to denote the number of inactive components (i.e., components of $(V\backslash S,\omega_{\mathrm{inact}})$). The resulting distribution on $\Omega$ is given by
\[\hat\pi(\omega)\propto (\emm^{\beta}-1)^{|\omega_\edge|}(q-1)^{\kappa_0(\omega)}
  \mbox{ for all } \omega\in \Omega
 \]
Note that the edge marginal of $\hat\pi$ is exactly the RC distribution $\pi$. 

To obtain a closer connection to the disordered/ordered phases we will need to consider $\Omega^\plm\subseteq \Omega$ with $\Omega^\plm=\{(F,S)\in \Omega\mid F\in \cE^\plm_G\}$. We thus obtain the ``disordered'' distribution
\[\hat\pi^\mi(\omega)= \hat\pi(\omega\mid \Omega^\mi) 
 \]
In the case of the ordered phase, we will use a slight variant of this lifting scheme, where the largest component of $(V,F)$ gets deterministically the label 1 (i.e., it is active). We denote the resulting distribution by $\hat\pi^\pl(\omega)$; in particular, using $C_1(\omega)$ to denote the largest component of $\omega$ (breaking ties by a deterministic rule), we have that
\[\hat\pi^\pl(\omega)=q\, \hat\pi(\omega \mid \Omega^\pl)\,\mathbf{1}\{C_1(\omega) \text{ is active}\}\]
Note that, analogously to $\hat \pi$ and $\pi$, the edge marginals of $\hat \pi^\mi$ and $\hat \pi^\pl$ are precisely $\pi^\mi$ and $\pi^\pl$.

For a pair of lifted configurations, we will be interested in the disagreements between a pair of lifted configurations. There are actually two types of disagreements:
\begin{enumerate}
    \item \label{it:simpleedge} A simple-edge disagreement is an edge $e\in F\oplus F'$ whose endpoints $u,v$ are either both active (i.e., $u,v\in S\cap S'$) or both inactive (i.e., $u,v\in (V\setminus S)\cap(V\setminus S')$).
    \item A component disagreement is a component $C$ of the graph
$H(\omega,\omega')$, defined as the subgraph of $(V,F\cup F')$
induced by the disagreeing vertices $S\oplus S'$.
\end{enumerate}
The following fact is the key observation behind the canonical path argument, allowing us to handle the more intricate component disagreements and move between configurations $\omega,\omega'\in \Omega$ using relatively simple component switches (analogous to those in the hard-core model).
\begin{observation}\label{obs:actinact}
Every component $C$ of $H(\omega,\omega')$ is active in exactly one of $\omega,\omega'$, i.e., it satisfies 
either $C\subseteq S$ and $C\cap S'=\emptyset$, or the other way around. Indeed, if $0<|C\cap S|<|C|$ there must be an edge connecting a vertex from $S\backslash S'$ to a vertex in $S'\backslash S$ using an edge either in $F$ or $F'$. But this would contradict the compatibility of $(F,S)$ or  $(F',S')$ (respectively). 
\end{observation}

\subsection{The random graph estimate}
Having defined the lifted configuration space $\Omega$, we are now ready to state the random graph properties we will use. For integers $\ell, R,r\geq 0$,  let 
\begin{equation}\label{eq:RC-badpairs}
\begin{aligned}
\cA_{G,R}&=\{(\omega,\omega')\in \Omega^2 \mid\mbox{ one of the inactive graphs $(V,\omega_{\inact}), (V,\omega_{\inact}')$ has tree-excess $>R$}\},\\
 \cB_{G,\ell,r}
 &=
 \{(\omega,\omega')\in \Omega^2\mid  \mbox{ $H(\omega,\omega')$ contains a component $C$ with 
       } |V(C)|>\ell\text{ or }\mathtt{exc}(C)>r\}.
\end{aligned}
\end{equation}
The event $\cA_{G,R}$ occurs if at least one of the two inactive
graphs has total tree-excess greater than $R$, summed over all its
components. We will show that this event has small probability for
$R=C\log n/\log\log n$ with $C$ sufficiently large. This phenomenon is familiar from percolation: non-giant components are mostly tree-like and rarely contain cycles. The event $\cB_{G,\ell,r}$ is the same phenomenon in the graph induced by disagreements of a pair of samples.

For $t>0$, define also 
\begin{align*}
\cH^\pl_t&=\{(\omega,\omega')\in \Omega^\pl\times \Omega^\pl: |\omega_{\edge}|\leq \tfrac{dn}{2}\mathfrak{m}_c+d t \mbox{ or  } |\omega'_{\edge}|  \leq \tfrac{dn}{2}\mathfrak{m}_c+d t\},\\
\cH^\mi_t&=\{(\omega,\omega')\in \Omega^\mi\times \Omega^\mi: |\omega_{\edge}|\geq \tfrac{dn}{2}\mathfrak{m}_c-d t \mbox{ or  } |\omega'_{\edge}|  \geq\tfrac{dn}{2}\mathfrak{m}_c-d t\}.
\end{align*}

 \newcommand{\statelemrandomgraphRC}{Let $d\geq 3$ be an integer, $q>2$ and $\beta_0>0$ be reals. Then, for any sufficiently small $\delta>0$ and any $D>0$, there are constants $\theta,C,C'>0$ and an integer $r\geq 1$ such that the following hold whp over $G\sim \cG_{n,d}$ with $R= C\frac{\log n}{\log\log n}$ and $\ell=\lceil C'\log n\rceil$.
 
For  all $\beta\in [\beta_c-\delta,\beta_0]$,  we have $\Pr[(\omega,\omega')\in \cA_{G,R}\cup \mathcal{B}_{G,\ell,r}\cup \cH^{\pl}_{\theta n}]\leq 1/n^D$ where $\omega,\omega'$ are independent samples from $\hat \pi^\pl$. The analogous result holds for $\hat \pi^\mi$ for all $\beta\in (0,\beta_c+\delta]$.
 }
\begin{theorem}\label{thm:randomgraphRC}
\statelemrandomgraphRC
\end{theorem}
The proof of Theorem~\ref{thm:randomgraphRC} is given in Section~\ref{sec:randomgraphRC}. 

\subsection{A weak-Poincar\'e inequality}
Here we prove a weak-Poincar\'e inequality for the RC phase $\pi^\pl$, an analogous inequality holds for $\pi^\mi$. Note that while the inequality applies to the RC phases, the bad events are in terms of the lifted distributions $\hat\pi^\plm$.
\begin{lemma}\label{lem:poincareRC}
Let $d\ge3$, and real $q>2$, $B>1$, $\theta>0$. Let also $\ell,R,r\geq 2$ be integers  and set $k=d+d\lceil\log_2\ell\rceil+r$. The following holds for every max-degree $d$
 graph $G$ with edge activities $\beta\in (0,B]$. Let $\pi=\pi_{G,q,\beta}$ be the RC distribution on $G$, 
and let $\pi^\plm=\pi(\cdot\mid\cE^\plm_G)$ denote the $\pl/\mi$ phases. 

Let $P^\pl$ be the 
single-edge Glauber dynamics on $\cE^\pl_G$, with proposals
outside $\cE^\pl_G$ rejected. Then, 
\begin{equation}\label{eq:permanent-wpi}
 \mathrm{Var}_{\pi^\pl}(f)
 \le  A\,\mathcal E_{P^\pl}(f,f)+(b^\pl+\delta)(f_{\max}-f_{\min})^2
 \qquad (\forall f:\cE_G^\pl\to\mathbb R),
\end{equation}
where $ A=4q^{2(R+r+2)} n (3d\ell \emm^{2B})^{k+2}$, $\delta=4dnq^{2R+2}\exp(-\tfrac{\theta^2n}{16d^2\ell})$ and $b^\pl=\Pr[(\omega,\omega')\in\cA_{G,R}\cup \cB_{G,\ell,r}\cup \cH^\pl_{\theta n}]$ where $\omega,\omega'$ are independent samples
from $\hat\pi^\pl$.
The analogous inequality holds for $\pi^\mi$ and $P^\mi$.
\end{lemma}
We first reduce the comparison to switches of individual
disagreement components.  For the following definition, fix  arbitrary vertex and edge labellings. For a vertex set $C\subseteq V$ and an edge subset $F\subseteq E$, we denote by $C_F$ the edges in $F$ incident to \emph{some} vertex in $C$.
\begin{definition}\label{def:sweepingRC}
Let $\omega=(F,S)$ and $\omega'=(F',S')$ belong to $\Omega$.
Let $C$ be a connected component of $H(\omega,\omega')$
with $\ell$ vertices and tree excess $r$.
The component sweep $\gamma_{\omega,\omega',C}$ is a sequence
of single-edge moves $ F_0\rightarrow F_1\rightarrow\cdots\rightarrow F_m$,
where $F_0=F$ and  $F_m=(F\setminus C_F)\cup C_{F'}$.

Assume first that $C\subseteq S$ and $C\cap S'=\emptyset$ (cf. Observation~\ref{obs:actinact}). We maintain the current edge set $Y$ (initially equal to $F$), and the corresponding active set $S_Y$ (initially equal to $S$). We sweep through the
vertices of $C$ in the order given by~\Cref{lem:cutwidth}. 
When processing a vertex $v$, we:
\begin{enumerate}
 \item Delete all currently present edges incident to $v$,   one by one. 
 \item Once the first step is finished, replace $S_Y$ with $S_Y\backslash\{v\}$.
 \item Insert all edges of $F'$ incident to $v$ whose other endpoint has already been processed or lies outside $C$,
       one by one. Edges towards unprocessed vertices of $C$
       remain absent.
\end{enumerate}
The case $C\cap S=\emptyset$ and $C\subseteq S'$ is analogous, the only difference is that in Step 2 we add vertices into $S_Y$ (instead of deleting).

\end{definition}
\begin{lemma}\label{lem:sweepingRC}
Throughout
the sweep, the edge set $Y$  remains compatible with $S_Y$, i.e., $(Y,S_Y)\in \Omega$.
\end{lemma}
\begin{proof}
We consider the case $C\subseteq S\setminus S'$;
the other case is analogous.
Recall that compatibility means that the endpoints of every
edge in $Y$ either both belong to $S_Y$ or both lie outside it.

Step 1 preserves compatibility because it only deletes edges.
At Step 2, the vertex $v$ is isolated in $Y$, so removing it
from $S_Y$ also preserves compatibility.

It remains to check the edges inserted in Step 3.
Consider such an edge $\{v,w\}\in F'$.
At this point, $v\notin S_Y$.
The first case is when $w$ is a previously processed vertex of $C$: then $w$ has
also been removed from the active set, so $w\notin S_Y$. The remaining case is when  $w\notin C$. Since $\{v,w\}\in F\cup F'$,
we must have $w\notin S\oplus S'$ (otherwise, $w$ would
belong to the disagreement component $C$).
Moreover, compatibility of $(F',S')$ together with
$v\notin S'$, implies $w\notin S'$.
Consequently $w\notin S$.
Since $S_Y$ agrees with $S$ outside $C$, we obtain
$w\notin S_Y$.

Thus every inserted edge has both endpoints outside $S_Y$,
and compatibility is preserved.
\end{proof}

 For a pair $(\omega,\omega')$, even after performing all component switches in $H(\omega,\omega')$, we might still be left with some simple-edge disagreements, cf. Item~\ref{it:simpleedge} (note that a component sweep $\gamma_{\omega,\omega',C}$ handles all relevant edge disagreements incident to $C$, so any simple-edge disagreement remaining was present in the original pair $(\omega,\omega')$). We handle those simple-edge disagreements by simply flipping them one by one, and call them simple-edge flips.

Formally, for a simple-edge disagreement
$e\in(F\oplus F')$ whose endpoints both belong to $V\setminus(S\oplus S')$, see also Item~\ref{it:simpleedge} earlier, define the
individual coordinate path
$\gamma_{\omega,\omega',e}
 :=(F\rightarrow F\oplus\{e\})$,
with the active set $S$ unchanged.
Compatibility is preserved because the endpoints of $e$
have the same labels in the two configurations.

For $\omega=(F,S)$, let
$r_0(\omega)$ 
denote the total tree-excess of its inactive graph. Write also 
$\bar\pi^\plm=\hat\pi(\,\cdot\mid\Omega^\plm)$.
We use this unforced lift in the following lemma because
the configurations constructed in its proof need not
have their largest components active.
\begin{lemma}\label{lem:mappingRC}
Let $\ell,r\geq2$ be integers and set
$k=d+d\lceil\log_2\ell\rceil+r$.

Consider a single-edge Glauber  transition $Y\to Y'$ for edge subsets $Y,Y'\subseteq E$ with~$Y\neq Y'$.
Let $\mathcal P^{\plm}_{Y\to Y'}$ consist of the pairs
$\omega=(F,S),\omega'=(F',S')$ in
$(\Omega^\plm\times\Omega^\plm)\setminus(\cB_{G,\ell,r}\cup \cH^\plm_{2\ell})$
such that $Y\to Y'$ is traversed in the component switch $\gamma_{\omega,\omega',C}$
for some $C$ in the disagreement graph $H(\omega,\omega')$. Then
\begin{equation}\label{eq:RCcongestion}
 \sum_{(\omega,\omega')\in\mathcal P^{\plm}_{Y\to Y'}}
 \frac{\bar\pi^{\plm}(\omega)\bar\pi^{\plm}(\omega')}
      {(q-1)^{r_0(\omega)+r_0(\omega')}}
 \leq
 n(3d\ell\emm^{2\beta})^{k+1}
 \pi^{\plm}(Y)P^{\plm}(Y,Y').
\end{equation}
The same bound applies for the set $\hat{\mathcal P}^{\plm}_{Y\to Y'}$ of pairs $(\omega,\omega')\in (\Omega^\plm\times\Omega^\plm)\setminus(\cB_{G,\ell,r}\cup \cH^\plm_{2\ell})$ where the transition $Y\rightarrow Y'$ is traversed as a simple-edge flip. 
\end{lemma}
With this, we are ready to give the proof of Lemma~\ref{lem:poincareRC}.
\begin{proof}[Proof of Lemma~\ref{lem:poincareRC}]
Let $R_f=(f_{\max}-f_{\min})^2$.
We may assume $b^\pl+\delta<1/4$, where recall that 
$\delta=4dnq^{2R+2}\exp(-\tfrac{\theta^2n}{16d^2\ell})$,
since otherwise
the result follows from $\Var_{\pi^\pl}(f)\leq R_f/4$.
This implies  in particular $\theta<1/2$ and $\ell\leq\theta n/8$.

We first prove the result for $\pi^\pl$. Let
$ \mathcal Q
 :=
 (\Omega^\pl\times\Omega^\pl)
 \setminus
 (\cA_{G,R}\cup\cB_{G,\ell,r}\cup\cH^\pl_{\theta n})$.
Since the edge marginal of $\hat\pi^\pl$ is $\pi^\pl$,
the variance identity gives
\begin{align}
 \mathrm{Var}_{\pi^\pl}(f)\leq
 \frac12
 \mathbb E_{\omega,\omega'\sim\hat\pi^\pl}
 \big[
   \mathbf1_{\mathcal Q}
   (f(\omega_\edge)-f(\omega'_\edge))^2
 \big]
 +b^\pl R_f.
 \label{eq:RC-initial-pair-truncation}
\end{align}
We now group the remaining pairs in $\mathcal{Q}$ according to their
skeleton, analogously to the hard-core model.
For $\omega=(F,S)$ and $\omega'=(F',S')$, define the skeleton
\[
 O(\omega,\omega')
 :=
 \bigl(
 S\cap S',\,
 S\oplus S',\,
 F\cap F',\,
 F\oplus F',\,
 \omega_\inact\cup\omega'_\inact
 \bigr).
\]
The first two coordinates specify the common vertex
labels and the disagreement set $D=S\oplus S'$.
The next two specify which edges belong to both
configurations and which belong to exactly one.
The last coordinate records which edges are inactive
in at least one configuration. As we will see shortly, the skeleton of a pair of configurations determines the weight $\hat \pi^\pl(\omega) \hat \pi^\pl(\omega')$ modulo a small correction factor. 

To do this, we first look at the structure of paired-configurations that have a fixed skeleton $O$ (arising from a compatible pair).
The skeleton determines the disagreement graph $H$, since both
$D$ and $F\cup F'$ are determined. Write $\cC(\omega,\omega')$
for its collection of connected components.
Recall from Observation~\ref{obs:actinact} that every component $C$ of $H$ is active in
one configuration and inactive in the other.
We may therefore choose which configuration is active
on $C$. Once this choice is made, we can determine from which configuration every  edge $e$ in $F\oplus F'$ incident to a vertex in $C$ comes from:
$e$ belongs to the inactive configuration if it is
recorded in the last coordinate of the skeleton,
and to the active configuration otherwise.
Common edges remain in both configurations. There is one further choice for simple-edge disagreements (edges of
$(F\oplus F')[V\setminus D])$: we may choose independently
which configuration contains it.
Both endpoints are both active (or inactive) in the two
configurations, so either choice preserves compatibility.
These choices describe all pairs with skeleton $O$,
and hence identify its skeleton class with 
$\sigma\in\{-1,1\}^{h_O}$, where
\[
 h_O
 =
 |\cC(\omega,\omega')|
 +|(F\oplus F')[V\setminus D]|
 \leq n+|E|\leq dn.
\]
Write the configuration pair corresponding to $\sigma$ as
$(\omega_\sigma,\omega_{-\sigma})$, and let
$F_\sigma=(\omega_\sigma)_\edge$.
Indeed, reversing every coordinate exchanges the
two configurations.

Let $\mathcal O$ be the set of skeletons over all pairs in $\mathcal Q$.
For each $O\in\mathcal O$, define the skeleton classes
\[
 \mathcal F_O
 :=
 \big\{\sigma\in \{-1,+1\}^{h_O}:
     \omega_\sigma,\omega_{-\sigma}\in\Omega^\pl\big\},
 \qquad
 \cH_O
 :=
 \big\{\sigma\in \{-1,+1\}^{h_O}:
     (\omega_\sigma,\omega_{-\sigma})\in\mathcal Q\big\}.
\]
Observe that, inside a skeleton class $\mathcal F_O$, the total numbers of occupied
edge copies, inactive vertex copies, and inactive edge
copies are fixed. Using that $\kappa_0(\omega)
 =n-|\omega_\ver|-|\omega_\inact|+r_0(\omega)$,
it follows that $\frac{\bar\pi^\pl(\omega_\sigma)
       \bar\pi^\pl(\omega_{-\sigma})}
      {(q-1)^{r_0(\omega_\sigma)+r_0(\omega_{-\sigma})}}$ 
is independent of $\sigma\in\mathcal F_O$. Set
\[
 a_O:=\max_{\tau\in\cH_O}
 \hat\pi^\pl(\omega_\tau)\hat\pi^\pl(\omega_{-\tau}).
\]
The good orientations in $\cH_O$ have total inactive
excess at most $2R$. Moreover,
$\hat\pi^\pl\leq q\bar\pi^\pl$.
Consequently, for every $\sigma\in\mathcal F_O$,
\begin{equation}\label{eq:RC-skeleton-comparison}
 a_O
 \leq q^{2R+2}
 \frac{\bar\pi^\pl(\omega_\sigma)
       \bar\pi^\pl(\omega_{-\sigma})}
      {(q-1)^{r_0(\omega_\sigma)+r_0(\omega_{-\sigma})}}
 \leq q^{2R+2}
 \bar\pi^\pl(\omega_\sigma)
 \bar\pi^\pl(\omega_{-\sigma}).
\end{equation}
Summing~\eqref{eq:RC-skeleton-comparison} over the
phase-valid orientations of all skeletons gives
\begin{equation}
 \sum_{O\in\mathcal O}a_O|\mathcal F_O|
 \leq q^{2R+2}.
 \label{eq:RC-skeleton-mass}
\end{equation}
Indeed, the skeleton classes are disjoint, and the
products of the two unforced conditional weights sum
to at most one. Returning to~\eqref{eq:RC-initial-pair-truncation},
the definition of $a_O$ now yields
\begin{equation}
 \mathrm{Var}_{\pi^\pl}(f)
 \leq
 \frac12\sum_{O\in\mathcal O}a_O
 \sum_{\sigma\in\cH_O}
 \big(f(F_\sigma)-f(F_{-\sigma})\big)^2
 +b^\pl R_f.
 \label{eq:RC-grouped-pairs}
\end{equation}
We will apply the hypercube inequality separately
inside each skeleton class. For this purpose, for a skeleton $O$ define the larger set of orientations
\[
 \cH'_O
 :=
 \left\{
   \sigma:
   \min\{|F_\sigma|,|F_{-\sigma}|\}
   \geq \tfrac{dn}{2}\mathfrak{m}_c+d\theta n/4+2d\ell
 \right\}.
\]
Since $\ell\leq\theta n/8$, the threshold is at most
$\tfrac{dn}{2}\mathfrak m_c+d\theta n/2$, so
$\cH_O\subseteq\cH'_O\subseteq\mathcal F_O$.
The purpose of $\cH'_O$ is to ensure that each individual
coordinate switch can be implemented while remaining
in the phase. In particular, we impose no restriction
on the inactive excess or the largest components in
the definition of $\cH'_O$.

We next show that a uniformly random orientation $\sigma\in \{-1,+1\}^{h_O}$ belongs
to $\cH'_O$ with probability at least $1-\varepsilon$.
Each coordinate exchanges the ownership of a block
of edges in $F\oplus F'$. For a disagreement component,
this block consists of its incident edge disagreements;
the remaining blocks are the individual edge
disagreements outside $D$.
These blocks are disjoint and partition $F\oplus F'$.
Thus
\begin{equation}\label{eq:Deltai}
 |F_\sigma|-|F_{-\sigma}|
 =
 \sum_{i=1}^{h_O}\sigma_i\Delta_i,
\end{equation}
where $\Delta_i$ is the difference between the numbers
of edges assigned to the two configurations in block
$i$. Each block contains at most $d\ell$ edges, so $\sum_{i=1}^{h_O}\Delta_i^2
 \leq d\ell\,|F\oplus F'|
 \leq d^2n\ell$.
Also, $|F_\sigma|+|F_{-\sigma}|$ is fixed by the
skeleton. Since the skeleton is represented by a
pair in $\mathcal Q$, we have $\frac{1}{2}\big(|F_\sigma|+|F_{-\sigma}|\big)
 >\tfrac{dn}{2}\mathfrak{m}_c+d\theta n$.  So, using $\ell\leq\theta n/8$, Azuma's inequality applied to the sum in \eqref{eq:Deltai}
therefore gives that for a uniform $\sigma\in \{-1,+1\}^{h_O}$ it holds that $\Pr_\sigma[\sigma\notin\cH'_O]\leq \eps$ where $\eps:=2\emm^{-\theta^2n/(16d^2\ell)}$.
In particular,
\begin{equation}\label{eq:RC-valid-orientations}
 |\cF_O|  \geq|\cH'_O|  \geq(1-\varepsilon)2^{h_O}.
\end{equation}

Extend $f$ to all edge sets by setting $\hat f(F)=f(F)$
when $F\in\cE^\pl_G$, and $\hat f(F)=f_{\min}$ otherwise.
Apply Lemma~\ref{lem:cube-wpi} to
$\sigma\mapsto\hat f(F_\sigma)$ and the bad set
$\{-1,1\}^{h_O}\setminus\cH'_O$.
The expected squared difference between antipodal
orientations is at most four times the variance under
the uniform distribution on the cube. Hence
\begin{align*}
\sum_{\sigma\in\cH_O}
 \big(f(F_\sigma)-f(F_{-\sigma})\big)^2\leq
 \sum_{\sigma\in\{-1,1\}^{h_O}}
 \big(\hat f(F_\sigma)-\hat f(F_{-\sigma})\big)^2\leq
 \sum_{\sigma\in\cH'_O}\sum_{i=1}^{h_O}
 \big(f(F_\sigma)-f(F_{\sigma^{(i)}})\big)^2
 +h_O\varepsilon 2^{h_O}R_f.
\end{align*}
For $\sigma\in\cH'_O$, changing one coordinate alters
at most $d\ell$ edges, so both edge configurations in
each term of the first sum belong to the phase. Plugging this back into ~\eqref{eq:RC-grouped-pairs}, we see from~\eqref{eq:RC-skeleton-mass} and~\eqref{eq:RC-valid-orientations} that the total contribution of the exceptional term is $\leq \frac{dn\eps}{2(1-\eps)}
 \sum_{O\in\mathcal O}a_O|\mathcal F_O|\,R_f\leq dn\eps q^{2R+2}R_f
 \leq\delta R_f$, so
\begin{equation}\label{eq:RC-sc-reduction}
 \Var_{\pi^\pl}(f)
 \leq
 \frac12\sum_{O\in\mathcal O}a_O
 \sum_{\sigma\in\cH'_O}\sum_{i=1}^{h_O}
 \big(f(F_\sigma)-f(F_{\sigma^{(i)}})\big)^2
 +(b^\pl+\delta)R_f.
\end{equation}

It remains to implement each coordinate switch using
single-edge Glauber moves.
Let $\gamma_{O,\sigma,i}$ be the path from $F_\sigma$
to $F_{\sigma^{(i)}}$. 
For a disagreement-component
coordinate, use Definition~\ref{def:sweepingRC};
for an edge coordinate outside $D$, use the corresponding
single-edge move.
Each path has length at most $3d\ell$.
Moreover, throughout a component switch, each of the
two configurations differs from its initial edge set
only on edges incident to that component.
There are at most $d\ell$ such edges, so the buffer
in the definition of $\cH'_O$ ensures that the paths
remain in the phase. By Cauchy--Schwarz along each path,
\begin{equation}\label{eq:CS-RC}
 \big(f(F_\sigma)-f(F_{\sigma^{(i)}})\big)^2
 \leq
 3d\ell
 \sum_{(Y\rightarrow Y')\in\gamma_{O,\sigma,i}}
 \big(f(Y')-f(Y)\big)^2.
\end{equation}
Every orientation in $\cH'_O$ lies outside
$\cH^\pl_{2\ell}$ and hence satisfies the buffer
condition in Lemma~\ref{lem:mappingRC}.
For a fixed pair, the updated edge determines the
switched coordinate. Thus
\eqref{eq:RC-skeleton-comparison} and
Lemma~\ref{lem:mappingRC} give, for every directed
transition $(Y\to Y')$,
\begin{equation}\label{eq:RC-skcongestion}
 \sum_{O\in\mathcal O}a_O
 \sum_{\sigma\in\cH'_O}\sum_{i=1}^{h_O}
 \mathbf1\big\{(Y\to Y')\in\gamma_{O,\sigma,i}\big\}
 \leq
 2nq^{2R+2}(3d\ell\emm^{2B})^{k+1}
 \pi^\pl(Y)P^\pl(Y,Y').
\end{equation}
This includes both disagreement-component switches
and single-edge switches outside $D$.

Combining \eqref{eq:CS-RC},
\eqref{eq:RC-skcongestion}, and
\eqref{eq:RC-sc-reduction}, the coefficient of the
Dirichlet form is at most
\[
 6d\ell\,nq^{2R+2}(3d\ell\emm^{2B})^{k+1}
 \leq
 4q^{2(R+r+2)}n
 (3d\ell\emm^{2B})^{k+2}
 =A.
\]
Here we used $q>2$ and $B>1$.
This proves the claimed weak-Poincare inequality
for $\pi^\pl$.

For $\pi^\mi$, use the same skeletons and define
$\mathcal{Q}$ with $\Omega^\mi$ and $\cH^\mi_{\theta n}$.
There is no forced activation of the largest component,
so the weight comparison holds without the factor
$q^2$; in particular, the bounds above remain valid.
Replace the safe-orientation set by
$ \cH'_O
 :=
 \left\{
   \sigma:
   \max\{|F_\sigma|,|F_{-\sigma}|\}
   \leq \tfrac{dn}{2}\mathfrak m_c-d\theta n/4-2d\ell
 \right\}$.
Since the skeleton contains a good pair,
$\tfrac{1}{2}(|F_\sigma|+|F_{-\sigma}|) <\tfrac{dn}{2}\mathfrak m_c-d\theta n$, and the condition $\ell\leq\theta n/8$ therefore gives
$\Pr_\sigma[\sigma\notin\cH'_O]\leq\varepsilon$
by the same concentration argument.
The paths remain in $\cE^\mi_G$, and the
weighted congestion bound applies with
$\pi^\mi,P^\mi$ in place of $\pi^\pl,P^\pl$.
\end{proof}

\begin{proof}[Proof of~\Cref{lem:mappingRC}]
We give the proof for the $\pl$ phase; the $\mi$ phase
is analogous. Assume first that $Y'=Y\backslash \{e\}$, the other case $Y'=Y\cup \{e\}$ will follow with minor adaptation.
We establish the encoding for component sweeps first
and treat simple-edge flips at the end.
For a fixed pair, the updated edge determines the
disagreement component being swept, so there is no additional
multiplicity from the choice of component.
Each edge is deleted at most once and inserted at most
once during a component sweep.

As for the hard-core model, the key step is to define a function $f:=f_{Y,Y'}$
mapping $(\omega,\omega')\in\mathcal P^\pl_{Y\rightarrow Y'}$
to $(X,S_X,S_Y,L)$, where
$(Y,S_Y),(X,S_X)\in\Omega^\pl$ and $L$ is an edge set
with $|L|\leq k$, so that
\begin{enumerate}
 \item $\bar\pi^\pl(\omega)\bar\pi^\pl(\omega')
 \leq
 \emm^{2\beta k}(q-1)^{r_0(\omega)+r_0(\omega')}
 \bar\pi^\pl((Y,S_Y))\bar\pi^\pl((X,S_X))$.

 \item For each triple $(\hat X,\hat S_X,\hat S_Y)$,
 at most $8(3d\ell)^k$ pairs satisfy $f(\omega,\omega')=(\hat X,\hat S_X,\hat S_Y,L)$ for some~$L$.
\end{enumerate}
Since the edge marginal of $\bar\pi^\pl$ is $\pi^\pl$,
these properties imply
\[
 \sum_{(\omega,\omega')\in\mathcal P^\pl_{Y\to Y'}}\frac{\bar\pi^\pl(\omega)\bar\pi^\pl(\omega')}{(q-1)^{r_0(\omega)+r_0(\omega')}}
 \leq
 8(3d\ell)^k\emm^{2\beta k}\pi^\pl(Y)
 \leq
 n(3d\ell\emm^{2\beta})^{k+1}
 \pi^\pl(Y)P^\pl(Y,Y'),
\]
where in the last inequality we used that for the edge deletion $Y\rightarrow Y'$ we have $P^\pl(Y,Y')\geq\frac{1}{|E|}\emm^{-\beta}\geq\frac{2}{dn}\emm^{-\beta}$.

We now construct the map. Fix
$(\omega,\omega')\in\mathcal P^\pl_{Y\rightarrow Y'}$,
with $\omega=(F,S)$ and $\omega'=(F',S')$,
and let $C$ be the disagreement component whose sweep
traverses $Y\rightarrow Y'$.
Recall from Observation~\ref{obs:actinact} that $C$ is
active in exactly one of $\omega,\omega'$. Suppose that the transition is traversed while processing
$v\in C$, and let $U\subseteq C\setminus\{v\}$ be the
vertices processed before $v$.
Since the transition is a deletion, the label at $v$
has not yet been changed. Thus the current active set is $S_Y=S\oplus U$. By Lemma~\ref{lem:sweepingRC}, $Y$ is compatible with $S_Y$.

Let $\hat{F}_v=\{e\in F\cup F': e \text{ incident to } v\}$ 
be the set of all $F,F'$-union edges incident to $v$,
including those whose other endpoint lies outside $C$.
Define $L=E_C[U,C\setminus U]\cup\hat F_v$ 
where $E_C[U,C\setminus U]$ consists of the edges of
$C$ crossing the cut induced by $U$. Finally, set
\[
 \widetilde X
   =(F\cap F')\cup\big((F\oplus F')\setminus Y\big),
 \qquad
 X=\widetilde X\setminus L,
 \qquad
 S_X=S'\oplus U.
\]
Thus $\widetilde X$ retains every common edge of $F,F'$
and contains each edge of $F\oplus F'$ precisely when
that edge is absent from $Y$.
We then delete the cut edges and all edges incident to
the current vertex $v$.
The labels in $S_X$ are obtained from those in $S'$
by exchanging the two original labels on $U$.
In particular, $S_X$ agrees with $S$ on $U$ and with
$S'$ on $V\setminus U$.

We next verify that $X$ is compatible with $S_X$, i.e., that the endpoints of every edge in $X$ either both belong to $S_X$ or both do not. Note that, outside $L$, $Y$ agrees with  $F'$ on edges incident to $U$, and with $F$ on all remaining edges.  For $X$, the status of edges outside $L$ is flipped: outside $L$, it agrees with
$F$ on edges incident to $U$ and with $F'$ on all remaining edges. In particular, consider any edge $\{u,w\}\in X$.
\begin{itemize}
 \item If $u,w\notin U$, then $\{u,w\}\in F'$.
       Since $S_X$ agrees with $S'$ at both endpoints,
       compatibility follows from that of $(F',S')$. Similarly, if $u,w\in U$, then $\{u,w\}\in F$.        Since $S_X$ agrees with $S$ at both endpoints, compatibility follows from that of $(F,S)$.

 \item Suppose that $u\in U$ and $w\notin U$. Then, since we exclude the edges crossing the cut $(U,C\backslash U)$, we must have $w\notin C$. Since $\{u,w\}\in F\cup F'$, we also have   $w\notin S\oplus S'$ (otherwise, $w$ would belong
       to the disagreement component $C$), so
       $1_{w\in S}=1_{w\in S'}$.

       Moreover, we claim that $\{u,w\}\in F$. Indeed, first note that when $u$ was processed,
the edge $\{u,w\}$ was deleted if present, and then inserted
precisely when it belonged to $F'$. Since $w\notin C$,
no subsequent step of the sweep changes its status. Hence
       \[
 \{u,w\}\in Y \quad\Longleftrightarrow\quad \{u,w\}\in F'
\]

If $\{u,w\}\notin F$, its membership in
$X\subseteq F\cup F'$ would imply
$\{u,w\}\in F'\setminus F$, and therefore $\{u,w\}\in Y$.
But such an edge is excluded from $ \widetilde X
 =(F\cap F')\cup\big((F\oplus F')\setminus Y\big)$, contradicting $\{u,w\}\in X\subseteq\widetilde X$.
Thus $\{u,w\}\in F$ and $1_{u\in S}=1_{w\in S}$ from the compatibility of $(F,S)$.

Since $S_X$ agrees with
       $S$ at $u$ and with $S'$ at $w$; it follows that $1_{u\in S_X}=1_{u\in S}=1_{w\in S}=1_{w\in S'}=1_{w\in S_X}$.
       Compatibility again follows from that of $(F,S)$.
\end{itemize}
This proves that $(X,S_X)\in\Omega$. The cutwidth bound and the maximum-degree assumption give
\[
 |L|\leq d\lceil\log_2\ell\rceil+r+d=k.
\]

We next verify phase membership.
Both the sweep and the construction of $X$ change
only edges incident to $C$, so
\[
 |Y\oplus F|\leq d|C|\leq d\ell,
 \qquad
 |X\oplus F'|\leq d|C|\leq d\ell.
\]
The assumed buffer
$\min\{|F|,|F'|\}\geq \frac{dn}{2}\mathfrak{m}_c+2d\ell$
therefore gives $Y,X\in\cE_G^\pl$.
Together with compatibility, this proves
\[
 \omega_Y:=(Y,S_Y)\in\Omega^\pl,
 \qquad
 \omega_X:=(X,S_X)\in\Omega^\pl.
\]

We next prove the weight comparison. First, to bound the difference in the number of the edges, consider
\[
 h=|F|+|F'|-|Y|-|X|.
\]
For an edge $e\notin L$, we have $\mathbf{1}_{e\in F}+\mathbf{1}_{e\in F'}
=\mathbf{1}_{e\in Y}+\mathbf{1}_{e\in X}$.  Indeed, any common edge of $F\cap F'$ that is absent from $Y$
must cross the processed/unprocessed cut or be incident to $v$,
and hence belongs to $L$. Thus, outside $L$, common edges
belong to both $Y$ and $X$, while edges of $F\oplus F'$
belong to exactly one of them (by the definition of
$\widetilde X$).  On the other hand, for an edge of $L$, the original
pair has one or two copies, while the encoded pair retains
at most one of them. Consequently,
\[
 0\leq h\leq 2|L|\leq 2k.
\]
The construction preserves the total number of inactive
vertex copies. Outside $L$, it also preserves the total
number of inactive edge copies.
On $L$, only $\omega_Y$ can retain an edge, and each
retained copy has the activity status of its corresponding
original copy. Thus the total number of inactive edge
copies can only decrease.

Using
$\kappa_0(\zeta)
=n-|\zeta_\ver|-|\zeta_\inact|+r_0(\zeta)$
and nonnegativity of the output excesses, we obtain
\[
 \kappa_0(\omega)+\kappa_0(\omega')
 -\kappa_0(\omega_Y)-\kappa_0(\omega_X)
 \leq r_0(\omega)+r_0(\omega').
\]
All four configurations belong to $\Omega^\pl$,
so the normalizing constants cancel.
Since $q>2$ and $0\leq h\leq2k$, it follows that
\[
 \frac{\bar\pi^\pl(\omega)\bar\pi^\pl(\omega')}
      {(q-1)^{r_0(\omega)+r_0(\omega')}}
 \leq
 (\emm^\beta-1)^h
 \bar\pi^\pl(\omega_Y)\bar\pi^\pl(\omega_X)
 \leq
 \emm^{2\beta k}
 \bar\pi^\pl(\omega_Y)\bar\pi^\pl(\omega_X).
\]
This proves the required weight comparison.

It remains to bound the number of preimages of a fixed
triple $(X,S_X,S_Y)$.
The disagreement set is recoverable from this triple:
\[
 D=S_X\oplus S_Y=S\oplus S'.
\]
In addition to the triple, record:
\begin{itemize}
 \item the current vertex $v$, which is one of the two
       endpoints of the updated edge $e$;
 \item the original label $t\in\{0,1\}$ of $C$
       in $\omega$;
 \item the edges of $L$, each marked according to whether
       it belongs to $F\setminus F'$, $F'\setminus F$,
       or $F\cap F'$.
\end{itemize}
These data determine the original pair.
Indeed,
\[
 F\cup F'=X\cup Y\cup L,
\]
so $C$ is the component containing $v$ in
$(V,X\cup Y\cup L)[D]$.
The original labels on $C$ are $t$ in $\omega$
and $1-t$ in $\omega'$. Outside $C$, the original
labels agree with $S_Y,S_X$, respectively.
This recovers $S,S'$, and then
\[
 U=\{u\in C:\mathbf1_{\{u\in S_Y\}}\neq t\}.
\]
On $L$, the recorded marks recover the original edge
ownership. Outside $L$, the original configurations are
recovered by exchanging the ownership in $(Y,X)$ on
edges incident to $U$, and leaving all other ownership
unchanged. Thus $F,F'$ are also determined.

To count the records, explore $C$ from $v$, querying
ambient edges incident to discovered vertices in a fixed
deterministic order. Only vertices in $D$ are added to
the exploration. Membership in the original union is
known from $X\cup Y$, except for the missing edges in $L$,
which are supplied by the record.

There are at most $d|C|\leq d\ell$ edge queries.
Record the query positions of the edges in $L$, together
with their three possible ownership marks.
This record permits the same exploration to be reproduced
from $X,Y,D,v$.
For $|L|=j$, there are at most $(3d\ell)^j$ such records.
Including the choices of $v$ and $t$, the number of
preimages is therefore at most $
 4\sum_{j=0}^{k}(3d\ell)^j
 \leq 8(3d\ell)^k$, completing the construction of the mapping for pairs in $\mathcal{P}^\pl_{Y\rightarrow Y'}$.

For an insertion transition $Y'=Y\cup\{e\}$, use the
occupied endpoint $Y'$ in the encoding and weight
comparison. Define $U$ to be the vertices whose labels
have already been exchanged, so that $v\in U$ in this
case. The same formulas for $S_Y,S_X$ apply, and adding
$v$ to the prefix does not change $L$, since all union
edges incident to $v$ already belong to $L$.
The preceding compatibility, weight, and reconstruction
arguments therefore remain valid.
Apply the deletion-probability bound to the reverse
chain transition and use reversibility $ \pi^\pl(Y')P^\pl(Y',Y)
 =\pi^\pl(Y)P^\pl(Y,Y')$ to obtain the same congestion bound.

 Finally, we bound the congestion  for  a simple-edge flip  (outside $D=S\oplus S'$). We construct a mapping as before where for $\omega=(F,S)$ and $\omega'=(F',S')$, we take $(Y,S_Y)=(F,S)$ and $(X,S_X)=(F',S')$ for the edge deletion of $e$. For the edge addition we take  $(Y',S_Y)=(F\cup\{e\},S)$ and $(X,S_X)=(F'\setminus\{e\},S')$. Set $L=\emptyset$. These choices preserve the total numbers of occupied
edge copies, inactive vertex copies, and inactive edge copies. Compatibility follows because the endpoints of $e$ have the same labels in both $S,S'$,
and phase membership follows from the buffer.
The same weight comparison therefore applies,
and there is at most one preimage for a fixed output
triple. 
\end{proof}

\subsection{Completing the proof of
Theorem~\ref{thm:samplerRC1}}
Here, we complete the proof of Theorem~\ref{thm:samplerRC1}. The argument follows largely the argument for the hard-core model, so here we do the proof a bit quicker focusing on how to tie together the various pieces. We first state the following lemma that will be used as the initialisation for simulated annealing. The proof is given in Appendix~\ref{sec:initialization}.
\newcommand{\stateleminitRC}{Let $d\geq3$ be an integer,
$q>2$ be real, and $J>0$.
There exist constants $\beta_0,c,C>0$ such that,
whp over $G\sim\cG_{n,d}$, the following hold for any $\beta\geq \beta_0$.
\begin{enumerate}
    \item For every $\emm^{-cn}\leq\eta\leq1/4$,
    the RC Glauber dynamics at $\beta$, initialised
    from $X_0=E$ and with proposals outside
    $\cE_G^\pl$ rejected, produces an $\eta$-sample
    from $\pi^\pl_{G,\beta}$ after $t=\big\lceil Cn^{5/4}\log(\frac{1}{\eta})\big\rceil$    steps.
    \item There is an algorithm running
    in time $\widetilde O(n^{5/4})$ that computes
    $\widehat Z^\pl_{G,\beta}=(1\pm \frac{1}{n^J})Z^\pl_{G,\beta}$.
\end{enumerate}
}
\begin{lemma}\label{lem:initRC}
\stateleminitRC
\end{lemma}
We are now able to finish the proof of Theorem~\ref{thm:samplerRC1}.

\begin{proof}[Proof of Theorem~\ref{thm:samplerRC1}]
For the phase samplers, fix $\beta_0>\beta_c+1$ satisfying
Lemma~\ref{lem:initRC} with $J=1$. Choose $\delta>0$
sufficiently small for Theorem~\ref{thm:randomgraphRC} with
upper cutoff $\beta_0+1$. Now fix $K>0$, set
$\eps=n^{-K}$ and write $J=K+20$.
Write $M=|E|=dn/2$ and parameterise the edge activity by
$s=\log(\emm^\beta-1)$, so that
$Z^\plm_s=\sum_{F\in\cE_G^\plm}\emm^{s|F|}q^{\kappa(F)}$. Let $P_s^\plm$ denote the corresponding restricted Glauber dynamics.

We first prove the plus-phase assertion. We may assume that $\beta\leq \beta_0$ (the case $\beta\geq \beta_0$ is covered by Lemma~\ref{lem:initRC}). 
By Theorem~\ref{thm:randomgraphRC} and
Lemma~\ref{lem:poincareRC}, whp the restricted chains
along the annealing interval satisfy an $(A,b)$
weak-Poincar\'e inequality with $A=n^{1+o(1)}$ and $b\leq n^{-4J}$. We fix such a graph, also satisfying the initialization
lemma. Since a Glauber move changes $|F|$ by at most one,
\[
 \operatorname{Var}_{\pi_s^\pl}(|F|)
 \leq A/2+bM^2=:v=n^{1+o(1)}.
\]
These bounds also hold on a slightly extended interval,
as needed for the neighbouring-temperature comparisons.

Construct an auxiliary restricted chain $(Y_t)$,
starting from $E$. First run it at $\beta_0$ until its
distance from $\pi^\pl_{G,\beta_0}$ is at most $1/n^{J}$,
using Lemma~\ref{lem:initRC}. Then decrease $s$ to the
target value in increments of magnitude at most
$v^{-1/2}$, running $O(A\log n)$ steps at each temperature.
Exactly as in the proof of Lemma~\ref{lem:samplerbHC1},
the identity
 $\frac{d^2}{ds^2}\log Z_s^\pl
 =\operatorname{Var}_{\pi_s^\pl}(|F|)$ bounds the neighbouring $\chi^2$ distances by a constant.
Lemma~\ref{lem:anneal} therefore gives that for $T=n^{3/2+o(1)}$ it holds that
\[
 \| Y_T-\pi_G^\pl\|_{\TV}=O(1/n^{J}).
\]

To transfer this to the unrestricted chain $(X_t^\pl)$
at the target temperature, use the same sandwich coupling
as for hard-core, now with the partial order on edge set induced by set inclusion. Couple also a stationary restricted chain
$Z_t\sim\pi_G^\pl$ with the same randomness. Until its first rejected proposal,
\[
 Z_t\subseteq X_t^\pl\subseteq Y_t.
\]
Indeed, RC updates are monotone in both the configuration
and the temperature, and the plus-phase restriction
only rejects deletions.
Theorem~\ref{thm:randomgraphRC} bounds the stationary
mass of the phase boundary by $n^{-4J}$.
Using $|F|$ as the increasing rank function, the same
coupling calculation as in Lemma~\ref{lem:samplerbHC1}
gives
\[
 \|X_T^\pl-\pi_G^\pl\|_{\TV}
 \leq
 M\|Y_T-\pi_G^\pl\|_{\TV}
 +(M+1)Tn^{-4J}
 \leq\eps.
\]

The minus-phase argument is analogous, with the
inclusions reversed. Here initialize the auxiliary
chain at $\emptyset$ and choose starting value $s=s_0'$ where
$\emm^{s_0'}=n^{-J}/M$. Since
\[
 q^n\leq Z^\mi_{s_0'}
 \leq q^n(1+\emm^{s_0'})^M
 \leq q^n\emm^{n^{-J}},
\]
the empty configuration is already within $O(n^{-J})$
of $\pi^\mi_{s_0'}$.
The annealing interval has length $O(\log n)$, which
only contributes an additional logarithmic factor.
If the target activity is smaller than $\emm^{s_0'}$,
the empty configuration itself suffices.

We finally prove the full-sampling assertion.
After reducing $\delta$ if necessary, it suffices to
consider $|\beta-\beta_c|\leq\delta$.
Indeed, the phase coexistence at $\beta_c$ recalled
in the introduction gives
$\log(Z^\pl_{\beta_c}/Z^\mi_{\beta_c})=o(n)$.
Moreover, Theorem~\ref{thm:randomgraphRC} gives, throughout
a fixed neighbourhood of $\beta_c$,
\[
 \frac{d}{ds}\log\frac{Z_s^\pl}{Z_s^\mi}
 =
 \mathbb E_{\pi_s^\pl}|F|
 -\mathbb E_{\pi_s^\mi}|F|
 \geq \theta n.
\]
The ratio is increasing everywhere, since the phases
are separated by their edge counts. Thus, outside the
smaller critical window, the unwanted phase has
exponentially small mass and the corresponding phase
sampler suffices.

Inside the window, we follow exactly the same approach as in Section~\ref{sec:finalcombo}.
For the minus phase use the initial estimate
$\widehat Z^\mi_{s_0'}=q^n$.
For the plus-phase partition-function estimate, choose
$\beta_{\rm count}\geq\beta_0$ satisfying Lemma~\ref{lem:initRC}
with $J=K+20$. Applying Theorem~\ref{thm:randomgraphRC} with
upper cutoff $\beta_{\rm count}+1$ gives the additional ordered-phase
estimates on $[\beta_0,\beta_{\rm count}+1]$.
Lemma~\ref{lem:initRC} at $\beta_{\rm count}$ supplies the initial
sample and the estimate of $Z^\pl_{G,\beta_{\rm count}}$ with
relative error $n^{-J}$ in time $\widetilde O(n^{5/4})$.
Use schedules with increments
$O(1/(n\log n))$ in $s$, giving
$O(n\log^2 n)$ rounds, each of $O(A\log n)$ steps.

Lemmas~\ref{lem:anneal-product} and
\ref{lem:anneal-statistic}, applied with Hamiltonian
$-|F|$ on the decreasing-$s$ plus-phase schedule and
$|F|$ on the increasing-$s$ minus-phase schedule, provide the
phase partition-function estimators exactly as in
the hard-core proof. Here $|R_H|\leq M+1$, so the
statistic-coupling errors and the initial approximation
errors are negligible compared with $\eps$.
Lemma~\ref{lem:anneal-rejection} then selects the phase
with the required accuracy; a fresh sample from that
phase completes the algorithm.
The total running time, including the logarithmically many estimator repetitions, is $\widetilde O(nA)= n^{2+o(1)}$.
\end{proof}

\section{Disagreement Components for Hard-Core}\label{sec:disHC}

\subsection{Double Phase-Planting for Hard-Core and the Ising model}
We first explain the double phase-planting for the hard core model in more detail and then give the details for Ising a bit faster.

Let $\lambda>\lambda_c(d)$ and $s\in [0,s_0]$. We focus on the proofs for the $\pl$-phase, the case of $\mi$ is completely analogous. Consider the hard-core model on a $d$-regular bipartite graph $G=(V^\pl\sqcup V^\mi,E)$ where the activities on $V^\pl$ are equal to $\lambda \emm^{s}$ and the activities on $V^{\mi}$ are $\lambda \emm^{-s}$. For the purposes of this section, it will be convenient to view the hard-core model in its 2-spin representation.  So, for $\sigma\in \{0,1\}^V$, we have  \begin{equation}\label{eq:weightHC}
    w_{G,s}(\sigma)=\prod_{\{u,v\}\in E}\mathbf{1}_{\{\sigma(u)\sigma(v)=0\}}\prod_{v\in V^\pl; \sigma(v)=1} (\lambda \emm^{s})\prod_{v\in V^\mi; \sigma(v)=1} (\lambda \emm^{-s}).
\end{equation}
Note that the first product just captures that $\sigma$ encodes an independent set of $G$, whereas the remaining factors capture its weight whenever this is the case. 

We let $\mu_{G,s}$ be the distribution on $\{0,1\}^V$ induced by the weight function $w_{G,s}(\sigma)$. We also let $\cI^\pl$ consist of the configurations $\sigma\in \{0,1\}^V$ with $|\sigma^{-1}(1)\cap V^\pl|\geq |\sigma^{-1}(1)\cap V^\mi|$. Then, we can view the phase distribution $\mu^{\pl}_{G,s}$ as the distribution $\mu^{\pl}_{G,s}=\mu(\cdot\mid \cI^\pl)$; moreover $Z^\pl_{G,s}=\sum_{\sigma\in \cI^\pl} w_{G,s}(\sigma)$. For convenience, we will henceforth denote by $\bsigma^\pl_G$ a sample from $\mu^{\pl}_{G,s}$.

We denote by $\bG$  a random $d$-regular bipartite graph chosen uniformly at random from the configuration model; in the expressions below, expectations are taken with respect to the choice of~$\bG$.
\newcommand{\statelemZboundsHC}{Let $d\geq 3$ be an integer and $\lambda>\lambda_c(d)$. There are constants $C,\kappa>0$ such that the following holds for all $s\in [0,s_0]$ and vertex activities $(\lambda \emm^s,\lambda \emm^{-s})$. 

There is an increasing function $\Psi=\Psi(s)$ with $\Psi(s)\geq \Psi(0)+\kappa s$ so   that  
\begin{enumerate}
\item $Z_{G,s}\geq \emm^{n \Psi}/n^{C}$  for every $d$-regular bipartite graph $G$.
\item $\E[Z^{\pl}_{\bG,s}]\leq n^{C} \emm^{n \Psi}$ and $\E[(Z^{\pl}_{\bG,s})^2]\leq n^{C} \emm^{2n \Psi}$.
\end{enumerate}
}
\begin{lemma}\label{lem:ZboundsHC}
\statelemZboundsHC
\end{lemma}
Note that the lower bound is  deterministic and is supplied by a general result of Ruozzi. Note however that this applies to the full partition function, so we will need a small argument to transfer to the constrained partition function $Z^\pl_{G,s}$. The moment bounds, especially the second moment, typically require heavier technical work but in the models we study here follow quite easily by adapting previous techniques.

For $\sigma_1,\sigma_2\in \{0,1\}^{V}$, define a random graph $\hat\bG(\sigma_1,\sigma_2)$ by letting
\begin{equation}\label{eq:Gsigma}
\Pr\Big[\hat\bG(\sigma_1,\sigma_2)=G\Big]=\frac{\Pr[\bG=G]\, w_{G,s}(\sigma_1)w_{G,s}(\sigma_2)}{\mathbb{E}[w_{\bG,s}(\sigma_1)w_{\bG,s}(\sigma_2)]}.
\end{equation}
Furthermore,  we define a random pair of configurations $(\bsigma^\pl_\bone,\bsigma^\pl_\btwo)\in\{0,1\}^{V}\times\{0,1\}^{V} $ with distributions
\begin{align}\label{eq:plusphase}	\Pr\big[(\bsigma^\pl_\bone,\bsigma^\pl_\btwo)=(\sigma_1,\sigma_2)]=\frac{\mathbf{1}\{\sigma_1,\sigma_2\in \cI^\pl\}\mathbb{E}[w_{\bG,s}(\sigma_1)w_{\bG,s}(\sigma_2)]}{\mathbb{E}[(Z^\pl_{\bG,s})^2]}.
\end{align}
For convenience, we set $\bGp=\hat\bG(\bsigma^\pl_\bone,\bsigma^\pl_\btwo)$ to be the planted graph model in \eqref{eq:Gsigma} corresponding to the random pair of plus-configurations $(\bsigma^\pl_\bone,\bsigma^\pl_\btwo)$. We refer to $(\bGp,\bsigma^\pl_\bone,\bsigma^\pl_\btwo\big)$ as the double phase-planted distribution for the hard core model.

Assuming~\Cref{lem:ZboundsHC}, we next give the proof of~\Cref{lem:quietplanting}. Recall that $\cG^{\mathrm{bip}}_{n,d}$ denotes the set of all $n$-vertex regular (multi)graphs with degree $d$.  
 \setlist[enumerate,1]{leftmargin=10mm}
 \newcommand{\statelemquiteplanting}{Let $d\geq 3$ be integer and $\lambda>\lambda_c(d)$ be real. There exists a constant $K>0$ such that the following holds for all $s\in [0,s_0]$. Let $\mathcal{E}\subseteq \cG^{\mathrm{bip}}_{n,d}\times \{0,1\}^n\times \{0,1\}^n$ be an arbitrary event.
 
 Suppose that  for the hard-core double-planted model with activities $(\lambda \emm^s,\lambda \emm^{-s})$ it holds that $\Pr[(\bGp,\bsigma^\pl_\bone,\bsigma^\pl_\btwo)\in \mathcal{E}]\leq g(n)$ for some function $g(n)$. Then, 
 \[\mbox{whp over $\bG$, }\quad \Pr\Big[\big(\bG,\bsigma^{\pl}_{\bG,\bone},\bsigma^{\pl}_{\bG,\btwo}\big)\in \mathcal{E}\, \Big|\, \bG\Big]\leq g(n)\cdot n^K.\]
 }
\begin{lemma}\label{lem:quietplanting}
\statelemquiteplanting
\end{lemma}

\begin{proof}[Proof of~\Cref{lem:quietplanting}]
Set $D=\frac{2C+3}{\kappa}$ where $C,\kappa$ are the constants in Lemma~\ref{lem:ZboundsHC}. We first obtain a polynomial lower bound on
$Z^\pl_{\bG,s}$ with high probability, and then apply
the defining identity for double planting. 

For a configuration $\sigma$, recall that $M(\sigma)
=
|\sigma^{-1}(1)\cap V^\pl|
-
|\sigma^{-1}(1)\cap V^\mi|$,  and observe that $w_{G,s}(\sigma)=w_{G,0}(\sigma)\emm^{sM(\sigma)}$. Let 
\[Z^{<0}_{G,s}
=
\sum_{\sigma:M(\sigma)<0}w_{G,s}(\sigma) \mbox{ so that } Z_{G,s}=Z^\pl_{G,s}+Z^{<0}_{G,s}.\]
The graph distribution is invariant under interchanging
the two sides. Since $s\geq0$, this gives $\mathbb{E}[Z^{<0}_{\bG,s}]
\leq \mathbb{E}[Z^\pl_{\bG,s}]$, so by linearity of expectation 
$
\mathbb{E}[Z_{\bG,s}]
\leq 2\mathbb{E}[Z^\pl_{\bG,s}]$.
Using $\E[Z^{\pl}_{\bG,s}]\leq n^{C}\emm^{n\Psi(s)}$, we therefore obtain that
\begin{equation}\label{eq:planting-first-moment-rate}
\frac{\emm^{n\Psi(s)}}{2n^C}
\leq
\mathbb{E}[Z^\pl_{\bG,s}]
\leq
n^C\emm^{n\Psi(s)}.
\end{equation}
Consider the graph event $\mathcal{Q}
=
\big\{
G:Z_{G,0}\leq
2n^{C+1}\emm^{n\Psi(0)}
\big\}$. Since
$\mathbb{E}[Z_{\bG,0}]
\leq2n^C\emm^{n\Psi(0)}$,
Markov's inequality gives $
\Pr[\bG\notin\mathcal Q]\leq n^{-1}$.

Fix $G\in\mathcal Q$ and set $t_n=\frac{D}{n}\log n$ where recall that $D=(2C+3)/\kappa$. The weights of configurations with
negative imbalance decrease as $s$ increases, so $Z^{<0}_{G,s}
\leq Z_{G,0}
\leq2n^{C+1}\emm^{n\Psi(0)}$.
Using that $\Psi(s)\geq \Psi(0)+\kappa s$, we obtain that for all $s\geq t_n$ it holds that
\[
\frac{Z^{<0}_{G,s}}
{(\emm^{n\Psi(s)}/n^C)}
\leq
2n^{2C+1}\emm^{-n\kappa s}
\leq
2n^{-2}
\leq\frac12.
\]
and hence using the deterministic lower bound
on $Z_{G,s}$ we have that for $s\geq t_n$ it holds that 
\[Z^\pl_{G,s}
\geq\emm^{n\Psi(s)}/(2n^C).\] Similarly, for $0\leq s<t_n$, 
we can bound $Z^\pl_{G,s}\geq \emm^{-n(t_n-s)}Z^\pl_{G,t_n}$ using that  every configuration in $\cI^\pl$
has $0\leq M(\sigma)\leq n$. Using  the bound above on $Z^\pl_{G,t_n}$ and the fact that $\Psi(t_n)\geq \Psi(s)$, we conclude that for $0\leq s\leq t_n$ we have $Z^\pl_{G,s}\geq \emm^{n\Psi(s)}/(2n^{C+D})$.
It follows that, for graphs $G\in \mathcal Q$,
\begin{equation}\label{eq:planting-phase-lower}
Z^\pl_{G,s}\geq n^{-L}\emm^{n\Psi(s)}
\qquad\text{for every }s\in[0,s_0].
\end{equation}
where $L:=C+D+1$.

We now consider an arbitrary $s\in[0,s_0]$ and an event $\mathcal E$
as in the statement of the lemma. Write
\[
p_s(G)
=
\Pr\Big[
(G,\bsigma^\pl_{G,1},\bsigma^\pl_{G,2})
\in\mathcal E
\,\Big|\,G
\Big],
\]
where the two configurations $\bsigma^\pl_{G,1},\bsigma^\pl_{G,2}$ are independent samples
from $\mu^\pl_{G,s}$.
By the definition of double phase-planting,
\[
\mathbb{E}\big[
(Z^\pl_{\bG,s})^2p_s(\bG)
\big]=
\mathbb{E}\big[(Z^\pl_{\bG,s})^2\big]\,
\Pr\big[
(\bGp,\bsigma^\pl_\bone,\bsigma^\pl_\btwo)
\in\mathcal E
\big]\leq
n^C\emm^{2n\Psi(s)}g(n).
\]
By discarding graphs $G\notin \mathcal{Q}$ and using the lower bound  \eqref{eq:planting-phase-lower} on $G\in \mathcal{Q}$, we therefore obtain that
\[
\mathbb{E}\big[
p_s(\bG)\mathbf{1}_{\{\bG\in\mathcal Q\}}
\big]
\leq n^{C+2L}g(n).
\]
Taking $K=C+2L+1$ and applying Markov's inequality gives
\[
\Pr\big[p_s(\bG)>n^Kg(n)\big]
\leq
\Pr[\bG\notin\mathcal Q]+\frac1n
\leq\frac2n.
\]
The constant $K$ and the failure bound are uniform over
$s\in[0,s_0]$, proving the lemma.
\end{proof}

\subsection{Vertex-edge statistics in the double phase-planted model}
 To utilise Lemma~\ref{lem:quietplanting}, the key observation is that, for the planted models, once we condition on the vertex and edge statistics, the models become uniformly distributed. Formally, for a graph $G\in \cG_{n,d}^{\mathrm{bip}}$ and configurations $\sigma_1,\sigma_2\in \{0,1\}^V$, let $\gamma(\sigma_1,\sigma_2),\delta(\sigma_1,\sigma_2)$ be probability vectors capturing the vertex colour statistics on $V^\pl,V^\mi$ under the paired configurations $(\sigma_1,\sigma_2)$, i.e., for colours $i,j\in \{0,1\}$, the entries of $\gamma$ and $\delta$ are given by
\[\mbox{$\gamma_{ij}(\sigma_1,\sigma_2)=\frac{1}{n}|\sigma^{-1}_1(i)\cap\sigma^{-1}_2(j)\cap V^\pl|$ \mbox{ and } $\delta_{ij}(\sigma_1,\sigma_2)=\frac{1}{n}|\sigma^{-1}_1(i)\cap\sigma^{-1}_2(j)\cap V^\mi|$.}\]
Similarly, let $y^{G}(\sigma_1,\sigma_2)$ be the $4\times 4$ matrix with the following half-edge colour statistics under $(\sigma_1,\sigma_2)$, i.e., for colours $i,j,k,l\in \{0,1\}$,
\begin{equation}\label{eq:edgestat}
y^{G}_{ik,jl}(\sigma_1,\sigma_2)=\frac{1}{|E(G)|}\sum_{u\in V^{\pl},v\in V^\mi} \mathbf{1}\big\{\{u,v\}\in E(G)\} \ \mathbf{1}\left\{\begin{array}{l}\sigma_    1(u)=i,\sigma_1(v)=j\\ \sigma_    2(u)=k,\sigma_2(v)=l
\end{array}\right\}.
\end{equation}
Note that $|E(G)| = d n$.\footnote{For multigraphs, in \eqref{eq:edgestat}, \eqref{eq:xedgestat}
and \eqref{eq:yedgestat} we replace $\mathbf{1}\{\{u,v\}\in E(G)\}$
by the number of edges between $u$ and $v$ (counting each loop twice).}
So, for a colour class $(i,k)\in \{0,1\}^2$, out of the $d n \gamma_{ik}(\sigma_1,\sigma_2)$ half-edges incident to it, there are  $dny^{G}_{ik,jl}(\sigma_1,\sigma_2)$ ``going'' to a colour class $(j,l)\in \{0,1\}^2$.

To specify the typical statistics of $\gamma,\delta,y$ we first need the following lemma. Roughly, $\alpha,\beta$ correspond to the typical statistics for a single configuration.
\begin{lemma}\label{lem:alphabeta}
Let $d\geq 3$ and $\lambda>\lambda_c(d)$. There exists a constant $\zeta>0$ such that the following holds for all $s\geq 0$. The system of equations
\begin{equation}\label{eq:fixpoint}
\frac{\alpha}{1-\alpha}=\lambda \emm^{s} \bigg(\frac{1-\alpha-\beta}{1-\alpha}\bigg)^{d}, \quad \frac{\beta}{1-\beta}=\lambda \emm^{-s} \bigg(\frac{1-\alpha-\beta}{1-\beta}\bigg)^{d} \mbox{ for  $\alpha,\beta\geq 0$ with $\alpha+\beta\leq 1$.}
\end{equation}
There is a unique solution of \eqref{eq:fixpoint} with $1>\alpha>\beta>0$, and it satisfies  $(d-1)^2\frac{\alpha\beta}{(1-\alpha)(1-\beta)}<1-\zeta$ and $\alpha-\beta\geq \zeta$.
\end{lemma}

Let $\gamma^\pl=\gamma^\pl(s), \delta^\pl=\delta^\pl(s),y^\pl=y^\pl(s)$ be the vectors
\begin{equation}\label{eq:gdyoptimal}
\gamma^\pl=\big[\begin{smallmatrix}1-\alpha\\ \alpha\end{smallmatrix}\big]\otimes \big[\begin{smallmatrix}1-\alpha\\ \alpha\end{smallmatrix}\big], \quad \delta^\pl=\big[\begin{smallmatrix}1-\beta\\ \beta\end{smallmatrix}\big]\otimes \big[\begin{smallmatrix}1-\beta\\ \beta\end{smallmatrix}\big], \quad y^\pl=\big[\begin{smallmatrix} 1-\alpha-\beta& & \beta\\ \alpha & & 0\end{smallmatrix}\big]\otimes \big[\begin{smallmatrix} 1-\alpha-\beta& & \beta\\ \alpha & & 0\end{smallmatrix}\big].
\end{equation}

 Let $\gammab^\pl=\gamma(\bsigma^{\pl}_{\bone},\bsigma^{\pl}_{\btwo})$ and $\deltab^\pl=\delta(\bsigma^{\pl}_{\bone},\bsigma^{\pl}_{\btwo})$, $\yb^\pl=y^{\bGp}(\bsigma^{\pl}_{\bone},\bsigma^{\pl}_{\btwo})$. We have the following concentration statement overlap/edge statistics for the double-planted model.
 \newcommand{\statelemstatistics}{Let $d\geq 3$ be an integer
 and $\lambda>\lambda_c(d)$ be real. Then there exists a constant $\eps_0>0$ such that      for any constant $0<\eps<\eps_0$, there is $c>0$ such that uniformly over $s\in [0,s_0]$, it holds that
\[ \Pr\Big[\big\|\gammab^\pl-\gamma^\pl\big\|_1+\big\|\deltab^\pl-\delta^\pl\big\|_1+\big\|\yb^\pl-y^\pl\big\|_1\geq \eps\Big]\leq \emm^{-c  n}.\]
}
\begin{lemma}\label{lem:statistics}
\statelemstatistics
\end{lemma}

\subsection{Proof of ~\Cref{thm:randomgraphHC}}
We are now ready to prove~\Cref{thm:randomgraphHC} which we restate here for convenience. Recall that  $b_{G,\ell,r}(s)=\Pr\big[(I,J)\in\cB_{\ell,r}\big]$ where $I,J\sim \mu^\pl_{G,s}$ and $\cB_{\ell,r}$ is the event that $G[I\oplus J]$ contains a component with more than $\ell$ vertices or which has tree-excess $>r$.
\begin{thmrandomgraphHC}
\statelemrandomgraphHC
\end{thmrandomgraphHC}
\begin{proof}
Let $\zeta,\eps_0,c>0$ be the constants from ~\Cref{lem:statistics,lem:alphabeta} and  $0<\eps\leq\min\{\eps_0,\zeta\}/10$ be a sufficiently small constant satisfying~\Cref{lem:statistics} and let $\theta=\eps/10$. Consider arbitrary $s\in [0,s_0]$. 

We will mainly work in the planted model $(\bGp,\bsigma^\pl_\bone,\bsigma^\pl_\btwo)$. Recall that $\gammab^\pl=\gamma(\bsigma^{\pl}_{\bone},\bsigma^{\pl}_{\btwo})$ and $\deltab^\pl=\delta(\bsigma^{\pl}_{\bone},\bsigma^{\pl}_{\btwo})$ are the paired-spin overlaps   and $\yb^\pl=y^{\bGp}(\bsigma^{\pl}_{\bone},\bsigma^{\pl}_{\btwo})$ are the edge statistics between paired-color classes. By~\Cref{lem:statistics}, 
\begin{equation}\label{eq:vertexedg} \Pr\Big[\big\|\gammab^\pl-\gamma^\pl\big\|_1+\big\|\deltab^\pl-\delta^\pl\big\|_1+\big\|\yb^\pl-y^\pl\big\|_1\geq \eps \Big]\leq \emm^{-c \eps n},
\end{equation}
where $\gamma^\pl,\delta^\pl,y^\pl$ are the vectors in \eqref{eq:gdyoptimal}. Let $\cE$ be the event in \eqref{eq:vertexedg}.

We first show the second item. Consider arbitrary vectors $\gamma,\delta,y$ that are within $\eps$ distance from $\gamma^\pl,\delta^\pl,y^\pl$. Conditioned on $(\gammab^\pl,\deltab^\pl,\yb^\pl)=(\gamma,\delta,y)$, we have that $M(\bsigma_\bone^{\pl})\geq \theta n$:   
the number of occupied vertices in $V^\pl$ is $(\gamma_{10}+\gamma_{11})n\geq (\alpha-2\eps) n$,  the number of occupied vertices in $V^\mi$ is $(\delta_{10}+\delta_{11})n\leq (\beta+2\eps)n$ and $\alpha-\beta\geq \zeta\geq 10\eps $ from~\Cref{lem:alphabeta}. We therefore conclude that  $M(\bsigma_\bone^{\pl})\geq \theta n$, i.e., $\bsigma_\bone^{\pl}\notin \cI_G^{\mathrm{bal}}(\theta)$. Using the transfer of~\Cref{lem:quietplanting}, we obtain that, whp over $ \bG$,
\[\mu^\pl_{\bG,s}(\cI_{\bG}^{\mathrm{bal}}(\theta))=\Pr\big[\bsigma^\pl_{\bG,\bone}\in \cI^{\mathrm{bal}}_{\bG}(\theta)\mid \bG\big]\leq  \Pr\big[(\bG,\bsigma^\pl_{\bG,\bone},\bsigma^\pl_{\bG,\btwo})\in \cE\mid \bG\big]\leq \emm^{- c \eps n/2}.\]

For the first item, we will use that, conditioned on the vertex-edge statistics prescribed by $(\gamma,\delta,y)$, the graph $\bGp$ is uniformly random. Let $V^{\plm}_{ik}$ be the vertices in $V^\plm$ with colour $i$  under $\bsigma_\bone^{\pl}$ and colour $k$ under $\bsigma_\btwo^{\pl}$. To understand the disagreement graph we only need to look at the disjoint graphs $\boldsymbol{H}=\bGp[V^\pl_{01}\cup V^\mi_{10}]$ and $\boldsymbol{H'}=\bGp[V^\pl_{10}\cup V^\mi_{01}]$.

In particular,  we  start  from a vertex $u\in  V^{\pl}_{01}$ and aim to explore the component $C_u$ of $u$ in $\boldsymbol{H}$ (the case $u\in V^\pl_{10}$ is completely analogous). To do this, we maintain an ``active'' set $A_t\subseteq V^{\pl}_{01}$. Initially $t=0$ and $A_0=\{u\}$. Mark $u$ as discovered, and mark every other vertex of $\boldsymbol H$ as soon as it is first reached.
While $|A_t|\neq 0$ and $t<\ell$: pick $w\in A_t$. Reveal the unmatched half-edges of $w$ one by one and let $B_t$ consist of the newly discovered neighbours in $V^\mi_{10}$. Once finished revealing the neighbours of $w$, do the same with the vertices in $B_t$ (revealing their unmatched half-edges one by one). Let $N_t$ consist of the newly discovered neighbours in $V^\pl_{01}$ and set $A_{t+1}=(A_t\setminus\{w\})\cup N_t$. Repeat the loop.

Let $A_{\leq t}=\cup^t_{t'=0} A_{t'}$ be all the vertices that have been in the active set up to time $t$. Note that each  vertex $w\in A_{\leq t}$ other than the starting vertex $u$ can add at most $(d-1)^2$ to $A_{t+1}$, and $u$ itself can add at most $d(d-1)$. Since we stop the process at $t=\ell$ (at the latest), we conclude that by the end of the process we have revealed $d^2\ell\leq \eps n/2$ half-edges. In particular, at every stage the probabilities that a fresh half-edge
from $V^\pl_{01}$ reaches $V^\mi_{10}$ and conversely are, respectively,
\[
 \frac{y_{01,10}}{\gamma_{01}}+O(\eps)
 =\frac{\beta}{1-\alpha}+O(\eps),\qquad
 \frac{y_{01,10}}{\delta_{10}}+O(\eps)
 =\frac{\alpha}{1-\beta}+O(\eps)
\] (noting that $\gamma_{01}=\sum_{j,l}y_{01,jl}$ and $\delta_{10}=\sum_{i,k}y_{ik,10}$).

Using the inequality $(d-1)^2 \frac{\alpha \beta}{(1-\alpha)(1-\beta)}<1-\zeta$ from Lemma~\ref{lem:alphabeta}, it follows that for small $\epsilon>0$ there are constants $a,b$ with $(d-1)^2ab\leq 1-\zeta/2$ such that, for $t\geq 1$ and $|A_t|\neq 0$, $|A_{t+1}|-|A_t|$ is stochastically dominated above by $X_t-1$ where  $X_t$ are independent with distribution $Z=\sum^{d-1}_{i=1}D_i Y_i$ and $D_i\sim \mathrm{Ber}(a), Y_i\sim \mathrm{Bin}(d-1,b)$. Since  $\E[Z]=(d-1)^2ab<1$,  the expected drift is negative and standard concentration arguments give that $\Pr[|C_u|>\ell]\leq \Pr[|A_{\ell/d^2}|\neq 0]\leq \emm^{-\rho\ell}$ for some constant $\rho>0$ (depending only on $d,a,b$).

To bound the tree-excess of $C_u$, note that an excess edge is created exactly when an edge of $\boldsymbol H$ is revealed whose other endpoint has already been discovered, including earlier in the current step. Since at most $O_d(\ell)$ vertices have been discovered, the conditional probability of an excess edge at each exposure is at most $\frac{\ell}{d^2\rho n}$ for small enough $\rho>0$. After at most $d^2\ell$ trials, the probability of seeing $\geq r$ excess edges is at most $\binom{d^2\ell}{r}(\frac{\ell}{d^2\rho n})^{r}\leq (\frac{\ell^2}{\rho n})^{r}$.

A union bound over the starting vertices gives the planted
bounds. Applying the argument of Lemma~\ref{lem:quietplanting},
and increasing $T$ and decreasing $\rho$ if necessary,
we obtain
\[
 b_{G,\ell,r}(s)\leq
 \tfrac12 n^T\big(\emm^{-\rho\ell}+(L\ell^2/n)^r\big),
 \qquad
 \mu^\pl_{G,s}(\cI_G^{\mathrm{bal}}(\theta))
 \leq\emm^{-2\rho n},
\]
with failure probability at most $n^{-6}$ for each fixed
$s,\ell,r$, conditionally on the graph event $\mathcal Q$
used there. Recall that $\Pr[\bG\in\mathcal Q]=1-o(1)$
and that its partition-function lower bound holds
simultaneously for all $s\in[0,s_0]$.

To obtain the conclusions uniformly, take a set
$\mathcal S$ of $O(n^2)$ values of $s$ spread uniformly
over $[0,s_0]$. There are only $O(n^2)$ relevant integer
pairs $(\ell,r)$: it suffices to consider
$\ell<\sqrt{n/L}$ and $r\leq d\ell$, together with the
event that some component has size greater than $\ell$.
For larger $\ell$ the claimed bound is trivial, and
for $r>d\ell$ the bad event reduces to this size event.
A union bound therefore gives both estimates whp
simultaneously for all $s\in\mathcal S$ and all $\ell,r$.

Finally, the pointwise ratio of $\mu^\pl_{G,s}$ and
$\mu^\pl_{G,s'}$ lies between $\emm^{-n|s-s'|}$ and
$\emm^{n|s-s'|}$, so the corresponding ratio for two
samples is at most $\emm^{2n|s-s'|}$.
Comparison with the nearest value in $\mathcal S$
proves both assertions for every $s\in[0,s_0]$.
\end{proof}

\section{Disagreement components for Random Cluster}\label{sec:randomgraphRC}
\subsection{From Random-Cluster to Ising}
To study $\hat\pi_G$ and its phase restrictions, we introduce
a reference distribution on the same space~$\Omega$.
Write $B=\emm^\beta$ and $Q=q-1$, and recall that 
\[\hat\pi(\omega)\propto (B-1)^{|\omega_\edge|}Q^{\kappa_0(\omega)}
  \mbox{ for all } \omega\in \Omega\] 
To remove the dependence on the number of components and obtain a more conventional vertex-edge model, we will remove a factor depending on the total tree-excess of the inactive graph.  

More precisely, recall that for $\omega=(F,S)$ the set of inactive vertices is $V\backslash S$ and the set of edges in inactive components is given by $F[V\backslash S]$; then, $r_0(\omega)$ is the aggregate excesses over the components of the graph $(V\backslash S,F[V\backslash S])$. Define $\hat\nu_G$ by removing the inactive-excess factor
from $\hat\pi_G$:
\begin{equation}\label{eq:RC-Ising-reference}
 \hat\nu_G(\omega)
 \propto
 Q^{-r_0(\omega)}\hat\pi_G(\omega)\propto
 (B-1)^{|F|}Q^{|V\setminus S|-|F[V\setminus S]|}.
\end{equation}

We define the corresponding phase restrictions using the
same events as for $\hat\pi_G^\mi$ and $\hat\pi_G^\pl$:
\begin{equation}\label{eq:RC-Ising-ref2}
 \hat\nu_G^\mi
 =\hat\nu_G(\,\cdot\mid\Omega^\mi),
 \qquad
 \hat\nu_G^\pl
 =\hat\nu_G\bigl(\,\cdot\mid
     \Omega^\pl\cap\{C_1(\omega) \text{ is  active}\}\bigr).
\end{equation}
\begin{lemma}\label{lem:RCtoIsing}
Let $q>2$, $\beta>0$, and $R\geq 0$.
For any graph $G$ and any event $\cE\subseteq\Omega\times \Omega$,
we have
\[
\Pr_{\omega,\omega'\sim\hat\pi_G^\pl}
 [(\omega,\omega')\in\cE]
\leq
2\mathbb{E}_{\omega\sim\hat\nu_G^\pl}
 \big[q^{c(\omega)}
       \mathbf1_{\{r(\omega)>R\}}\big]
+
Q^{2R}
\Pr_{\omega,\omega'\sim\hat\nu_G^\pl}
 [(\omega,\omega')\in\cE],
\]
where $c(\omega)$ and $r(\omega)$ denote the number of cyclic components and the total tree excess in  $\omega$, respectively, excluding the contribution of the largest component $C_1(\omega)$ from the counts.
\end{lemma}

\begin{proof}
From the definitions \eqref{eq:RC-Ising-reference} and \eqref{eq:RC-Ising-ref2}, observe that we have 
\begin{equation}\label{eq:RC-Ising-phase-comparison}
\hat\pi_G^\pl(\omega)
=
\frac{Q^{r_0(\omega)}}
     {\mathbb E_{\hat\nu_G^\pl}[Q^{r_0}]}
\hat\nu_G^\pl(\omega)
\leq Q^{r_0(\omega)}\hat\nu_G^\pl(\omega),
\end{equation}
where the inequality follows from $Q>1$ and $r_0\geq0$. Let $
\mathcal D_R
=\{(\omega,\omega'):
       r_0(\omega)\leq R,\ r_0(\omega')\leq R\}$.
By independence and \eqref{eq:RC-Ising-phase-comparison},
\begin{equation}\label{eq:455hy5}
\Pr_{\omega,\omega'\sim\hat\pi_G^\pl}
 [(\omega,\omega')\in\cE\cap \mathcal D_R]
\leq
Q^{2R}\Pr_{\omega,\omega'\sim\hat\nu_G^\pl}
 [(\omega,\omega')\in\cE\cap \mathcal D_R].
\end{equation}
On the other hand, the union bound gives
\begin{equation}\label{eq:455hy5b}
\Pr_{\omega,\omega'\sim\hat\pi_G^\pl}
 [(\omega,\omega')\notin\mathcal D_R]
\leq
2\Pr_{\omega\sim\hat\pi_G^\pl}
 [r_0(\omega)>R]
\leq
2\mathbb E_{\omega\sim \hat\nu_G^\pl}
 \big[Q^{r_0(\omega)}\mathbf1_{\{r_0(\omega)>R\}}\big].
\end{equation}
where in the last inequality we used again \eqref{eq:RC-Ising-phase-comparison}. The final bound we are going to show shortly is 
\begin{equation}\label{eq:expectationcomp}
\mathbb E_{\omega\sim \hat\nu_G^\pl}
 \big[Q^{r_0(\omega)}\mathbf1_{\{r_0(\omega)>R\}}\big]\leq  \mathbb E_{\omega\sim \hat\nu_G^\pl}
 \big[q^{c(\omega)}\mathbf1_{\{r(\omega)>R\}}\big].
\end{equation}
Assuming this for the moment, we obtain the desired claim using the bound in~\eqref{eq:455hy5} and~\eqref{eq:455hy5b}.

To prove \eqref{eq:expectationcomp}, let us observe that the definitions of $c(\omega)$ and $r(\omega)$ extend naturally to edge configurations $F$. That is, for an edge configuration $F$, let $\mathcal{C}(F)$ denote the components of the graph $(V,F)$ and, for $C\in \mathcal{C}(F)$, write
$r_C=\mathtt{exc}(C)$ for the tree excess of $C$. With $C_1(F)$ denoting the largest component in $\mathcal{C}(F)$, note that
\[
 c(F)
 =\sum_{C\in \mathcal{C}(F); C\neq  C_1(F)}\mathbf{1}\{r_C\geq 1\},
 \qquad
 r(F)
 =\sum_{C\in \mathcal{C}(F); C\neq  C_1(F)} r_C.
\]

Let $\tilde\nu^\pl_G(F)$ denote the edge marginal of $\hat\nu_G^\pl$. We can lift $F\sim \tilde\nu^\pl_G$ to $\omega=(F,S)\sim \hat\nu^\pl_G$ by declaring the largest
component $C_1(F)$ to be active, while the other components become inactive independently  with probability $\frac{Q^{1-r_C}}{1+Q^{1-r_C}}$. Note that from the lifting we have that  $r_0(\omega)\leq r(F)$, so 
\begin{align}
\mathbb{E}_{\omega\sim \hat\nu_G^\pl(\cdot \mid F)} \big[Q^{r_0}\mathbf1_{\{r_0>R\}}\mid F\big]
\leq
 \mathbf1_{\{r(F)>R\}}
 \prod_{C\ne C_1(F)}
 \frac{1+Q^{1-r_C}Q^{r_C}}{1+Q^{1-r_C}}\leq 
 q^{c(F)}\mathbf1_{\{r(F)>R\}}.
 \label{eq:inactive-label-average}
\end{align}
Averaging over $F\sim \tilde\nu^\pl_G$ gives the desired inequality \eqref{eq:expectationcomp}, completing the proof.
\end{proof}

\subsection{Double Phase-Planting for the reference Ising model}\label{sec:refIsing}

In light of Lemma~\ref{lem:RCtoIsing}, we can henceforth focus on understanding the distributions $\hat\nu_G,\hat\nu^\pl_G$. The key here is that vertex marginals of $\hat\nu_G$ have an Ising type distribution on $\sigma\in \{0,1\}^V$ with interaction matrix $\big[\begin{smallmatrix} 1+\frac{B-1}{Q}& 1\\ 1& B\end{smallmatrix}\big]$  and external field $Q$ on spin 0, where recall that $B=\emm^\beta$, $Q=q-1$ and $\hat B=1+\frac{B-1}{Q}$. Namely define the Ising distribution
\[\nu_G(\sigma)=w_G(\sigma)/Z_G \mbox{ where } w_G(\sigma)=Q^{|\sigma^{-1}(0)|} \hat B^{|M_{0}(\sigma)|}B^{|M_{1}(\sigma)|}\]
and $M_0(\sigma), M_1(\sigma)$ are the sets of edges whose both endpoints have spin $0,1$ under $\sigma$.

\begin{lemma}[Ising lifting]\label{lem:Ising-lifting}
For every graph $G$, $q>1$, and $\beta>0$, the following holds. 

A sample $\sigma\sim \nu_G$ can be lifted to a vertex-edge configuration $\omega=(F,S)\sim \hat\nu_G$  by setting $S=\sigma^{-1}(1)$ and setting $F=\mathrm{Bin}(M_0(\sigma), p_0)\cup \mathrm{Bin}(M_1(\sigma), p_1)$ where 
$p_0=\frac{\hat B-1}{\hat B}$ and $p_1=\frac{B-1}{B}$.

Conversely, we obtain a sample $\sigma\sim \nu_G$ from a vertex-edge configuration $\omega=(F,S)\sim \hat\nu_G$ by setting $\sigma(v)=\mathbf{1}\{v\in S\}$.
\end{lemma}
\begin{proof}
Fix $S\subseteq V$. Compatibility requires that every
edge between $S$ and $V\setminus S$ be absent. Summing
independently over the remaining edges gives
\begin{align*}
 \sum_{F:(F,S)\in\Omega}w_{\hat\nu,G}(F,S)
 &=
 Q^{|V\setminus S|}
 \prod_{e\in E[S]}(1+B-1)
 \prod_{e\in E[V\setminus S]}
 \bigg(1+\frac{B-1}{Q}\bigg)=w_{\nu,G}(S).
\end{align*}
Summing over $S$ proves equality of the partition
functions. Dividing the joint weight by $w_{\nu,G}(S)$
shows that the conditional edges are independent
with the stated probabilities. This proves the
factorisation and the expectation identity.
\end{proof}
Now that we have paved the path to study the random-cluster model via the reference Ising model, we are left with the double-planting argument for the Ising model. The details are analogous to those for the hard-core model though some extra care is needed to handle the tree-excess.

We begin with the definition of the phases, and then the relevant vertex-edge statistics. Define 
\[\Sigma^\mi=\{\sigma\in \{0,1\}^V:\ \  |\sigma^{-1}(1)|< n/2\}, \qquad \Sigma^\pl=\{\sigma\in \{0,1\}^V:\ \ |\sigma^{-1}(1)|\geq n/2\},\]
and let $\nu_G^\mi(\cdot)=\nu_G(\cdot \mid \Sigma^\mi)$ and $\nu_G^\pl(\cdot)=\nu_G(\cdot \mid \Sigma^\pl)$ be the corresponding phase models. We let $Z^\plm_{G}, Z_{G}$ denote the partition functions corresponding to the distributions $\nu^\plm,\nu$. For convenience, we will henceforth denote by $\bsigma^\pl_G$ a sample from $\nu^{\pl}_{G}$, and similarly for $\bsigma^\mi_G,\bsigma_G$.

We can now define the planted models analogously to the case of the hard-core model.  Recall that $\bG$ denotes a graph chosen uniformly at random from the configuration model; in the expressions below, expectations are taken with respect to the choice of $\bG$.

For $\sigma,\sigma_1,\sigma_2\in \{0,1\}^{V}$, define random graphs $\hat\bG_\bone(\sigma)$ and $\hat\bG_\btwo(\sigma_1,\sigma_2)$ by letting
\begin{equation}\label{eq:Gsigmab}
\Pr\Big[\hat\bG_\bone(\sigma)=G\Big]=\frac{\Pr[\bG=G]\, w_{G}(\sigma)}{\mathbb{E}[w_{\bG}(\sigma)]} \mbox{ and }\Pr\Big[\hat\bG_\btwo(\sigma_1,\sigma_2)=G\Big]=\frac{\Pr[\bG=G]\, w_{G}(\sigma_1)w_{G}(\sigma_2)}{\mathbb{E}[w_{\bG}(\sigma_1)w_{\bG}(\sigma_2)]}.
\end{equation}
Furthermore,  we define a random configuration $\bsigma^\pl\in \{0,1\}^V$ and a random pair of configurations $(\bsigma_\bone^\pl,\bsigma_\btwo^\pl)\in\{0,1\}^{V}\times\{0,1\}^{V} $ with distributions
\begin{align}	\Pr\big[\bsigma^\pl=\sigma]=\frac{\mathbf{1}\{\sigma\in \Sigma^\pl\}\mathbb{E}[w_{\bG}(\sigma)]}{\mathbb{E}[Z^\pl_{\bG}]} \mbox{ and }\Pr\big[(\bsigma^\pl_\bone,\bsigma^\pl_\btwo)=(\sigma_1,\sigma_2)]=\frac{\mathbf{1}\{\sigma_1,\sigma_2\in \Sigma^\pl\}\mathbb{E}[w_{\bG}(\sigma_1)w_{\bG}(\sigma_2)]}{\mathbb{E}[(Z^\pl_{\bG})^2]}.\notag
\end{align}
We set $\bGp_\bone=\hat\bG_\bone(\bsigma^\pl)$ to be the planted graph model in \eqref{eq:Gsigmab} corresponding to the random configuration $\bsigma^\pl$ and analogously $\bGp_\btwo=\hat\bG_\btwo(\bsigma^\pl_\bone,\bsigma^\pl_\btwo)$ to be the planted graph model in \eqref{eq:Gsigmab} corresponding to the random pair of plus-configurations $(\bsigma^\pl_\bone,\bsigma^\pl_\btwo)$. We refer to $(\bGp_\bone,\bsigma^\pl)$ as the single phase-planted distribution and $(\bGp_\btwo,\bsigma^\pl_\bone,\bsigma^\pl_\btwo\big)$ as the double phase-planted distribution for the Ising model. 

We will show the following bounds on the partition functions in~\Cref{sec:Isinglower}. 
\newcommand{\statelemZboundsRC}{Let $d\geq 3$ be an integer, $q>2$ and $\beta_0>0$ be reals. For small $\delta>0$, there are constants $C>0$  and continuous functions $\Psi^\pl=\Psi^\pl(\beta)$ and $\Psi^\mi=\Psi^\mi(\beta)$ with $\Psi^\mi(\beta_c)=\Psi^\pl(\beta_c)$ such that the following hold. For the plus phase, for all $n$ sufficiently large and any $\beta\in [\beta_c-\delta,\beta_0]$:
\begin{enumerate}
\item It holds that $\E[Z^\pl_{\bG}]\leq n^{C} \emm^{n \Psi^\pl}$ and $\E\big[(Z^\pl_{\bG})^2\big]\leq n^{C} \emm^{2n \Psi^\pl}$.
\item for any $G\in \cG_{n,d}$ it holds that $Z^\pl_G\geq \frac{1}{2}\emm^{n \Psi^\pl}$
\end{enumerate}
The analogous statements also hold for the minus phase for all $\beta\in [1/\beta_0,\beta_c+\delta]$.}
\begin{lemma}\label{lem:ZboundsRC}
\statelemZboundsRC
\end{lemma}
With this at hand, we can next show the following. Recall that for a graph $G$, $\bsigma^\pl_G$ denotes a sample from $\nu^\pl_G$ (and similarly $\bsigma^\pl_{G,1},\bsigma^\pl_{G,2}$).
 \setlist[enumerate,1]{leftmargin=10mm}
 \newcommand{\statelemquiteplantingRC}{Let $d\geq 3$ be integer, $q>2$ and $\beta_0>0$ be reals. There exist $\delta,T>0$ such that the following holds for all large enough  $n$ and any $\beta\in[\beta_c-\delta,\beta_0]$, for any   $g:\mathbb{N}\rightarrow \mathbb{R}_{> 0}$.

 \begin{enumerate}[noitemsep]
 \item  Suppose that for an event $\cE\subseteq\cG_{n,d}\times \{0,1\}^n$ it holds that  $\Pr\Big[
 \big(\bGp_\bone,\bsigma^\pl\big)
 \in\mathcal E
\Big]\leq g(n)$ in the single-planted model. Then, whp over $\bG$, $\Pr\Big[\big(\bG,\bsigma^{\pl}_{\bG}\big)\in \mathcal{E}\, \Big|\, \bG\Big]\leq g(n)\cdot n^T$.
 \item Suppose that for an event $\cE\subseteq\cG_{n,d}\times \{0,1\}^n \times \{0,1\}^n$  it holds that $\Pr\Big[
 (\bGp_\btwo,\bsigma^\pl_\bone,\bsigma^\pl_\btwo)
 \in\mathcal E
\Big]\leq g(n)$ in the double-planted model. Then, whp over $\bG$, $\Pr\Big[\big(\bG,\bsigma^{\pl}_{\bG,\bone},\bsigma^{\pl}_{\bG,\btwo}\big)\in \mathcal{E}\, \Big|\, \bG\Big]\leq g(n)\cdot n^T$.
\end{enumerate}
The analogous statements hold for the minus phase for all
$\beta\in[1/\beta_0,\beta_c+\delta]$.
 }
\begin{lemma}\label{lem:quietplantingRC}
\statelemquiteplantingRC
\end{lemma}
\begin{proof}[Proof of Lemma~\ref{lem:quietplantingRC}]
 Consider $\beta\in[\beta_c-\delta,\beta_0]$ and write $\Psi^\pl=\Psi^\pl(\beta)$ where $\Psi^\pl$ is the function from Lemma~\ref{lem:ZboundsRC}. We give the proof for the double-planted plus-phase model; the single-planted and minus-phase proofs are analogous. 
 
 For a graph $G$, define $p_G^\pl=\Pr[(G,\bsigma_{G,1}^\pl,\bsigma_{G,2}^\pl)\in \mathcal{E}]$. The double-planted distribution gives
\[
\Pr\Big[
 (\bGp_\btwo,\bsigma^\pl_\bone,\bsigma^\pl_\btwo)
 \in\cE
\Big]
=
\frac{
 \mathbb E[(Z^\pl_{\bG})^2p^\pl_{\bG}]
}{
 \mathbb E[(Z^\pl_{\bG})^2]
}.
\]
Thus the hypothesis and Item~1 of
Lemma~\ref{lem:ZboundsRC} imply $\mathbb E[(Z^\pl_{\bG})^2p_{\bG}^\pl]
\leq g(n)n^C\emm^{2n\Psi^\pl}$. Using the deterministic lower bound from Item~2,
we obtain
\[\mathbb E[p^\pl_{\bG}]
\leq
4\emm^{-2n\Psi^\pl}
\mathbb E[(Z^\pl_{\bG})^2p^\pl_{\bG}]
\leq 4n^C g(n).\] So, by Markov's inequality, with probability 
$\geq 1-4/n^{2}$ over $\bG$, $p^\pl_{\bG}\leq n^{C+2}g(n)$, proving the lemma with $T=C+2$.
\end{proof}

\subsection{Vertex-edge statistics and the transfer lemma}
For a graph $G\in \cG_{n,d}$ and configuration $\sigma\in \{0,1\}^V$, let 
$\alpha(\sigma)=\frac{1}{n}|\sigma^{-1}(1)|$ be the fraction of vertices with spin 1. Let also $x^{G}(\sigma)$ be the $2\times 2$ matrix with the following half-edge colour statistics under $\sigma$, i.e., for colours $i,j\in \{0,1\}$,
\begin{equation}\label{eq:xedgestat}
x^{G}_{i,j}(\sigma)=\frac{1}{2|E(G)|}\sum_{u,v} \mathbf{1}\big\{\{u,v\}\in E(G)\} \ \mathbf{1}\left\{\sigma(u)=i,\sigma(v)=j\right\}.
\end{equation} 
Define $\alpha^\mi=a^\mi=1/q$ and $\alpha^\pl,a^\pl\in (\tfrac{1}{q},1)$ by $\alpha^\pl=\frac{t^{d}}{t^{d}+q-1}$ and  $a^\pl=\frac{t^{d-1}}{t^{d-1}+q-1}$ where $t$ is the largest solution to $\emm^{\beta}-1=\frac{(t-1)(t^{d-1}+q-1)}{t^{d-1}-t}$. Let $x^\pl$ be the vector
\begin{equation}\label{eq:xoptimal}
 x^\pl=\frac{1}{X^\pl}\Big[\begin{array}{cc} \hat B(1-a^\pl)^2 &  a^\pl(1-a^\pl)\\ a^\pl(1-a^\pl)  & B(a^\pl)^2\end{array}\Big] \mbox{ where } X^\pl=\hat B(1-a^\pl)^2+2a^\pl(1-a^\pl)+B(a^\pl)^2.
\end{equation}

Similarly, for a graph $G\in \cG_{n,d}$ and configurations $\sigma_1,\sigma_2\in \{0,1\}^V$, let $\gamma(\sigma_1,\sigma_2)$ be a probability vector capturing the vertex colour statistics on $V$ under the paired configurations $(\sigma_1,\sigma_2)$, i.e., for colours $i,j\in \{0,1\}$, the entries of $\gamma$ are given by $\gamma_{ij}(\sigma_1,\sigma_2)=\tfrac{1}{n}|\sigma^{-1}_1(i)\cap\sigma^{-1}_2(j)|$. Let $y^{G}(\sigma_1,\sigma_2)$ be the $4\times 4$ matrix with the following half-edge colour statistics under $(\sigma_1,\sigma_2)$, i.e., for colours $i,j,k,l\in \{0,1\}$,
\begin{equation}\label{eq:yedgestat}
y^{G}_{ik,jl}(\sigma_1,\sigma_2)=\frac{1}{2|E(G)|}\sum_{u,v} \mathbf{1}\big\{\{u,v\}\in E(G)\} \ \mathbf{1}\left\{\begin{array}{l}\sigma_    1(u)=i,\sigma_1(v)=j\\ \sigma_    2(u)=k,\sigma_2(v)=l
\end{array}\right\}.
\end{equation}
 Let $\gamma^\pl,y^\pl$ be the vectors
\begin{equation}\label{eq:gdyoptimalb}
\gamma^\pl=\big[\begin{smallmatrix}1-\alpha^{\pl}\\ \alpha^\pl\end{smallmatrix}\big]\otimes \big[\begin{smallmatrix}1-\alpha^{\pl}\\ \alpha^\pl\end{smallmatrix}\big], \quad y^\pl=x^\pl \otimes x^\pl.
\end{equation}
 Define analogously $\gamma^\mi,y^\mi,x^\mi$.

Let $\alphab^\pl=\alpha(\bsigma^\pl)$, $\xb^\pl=x^{\bGp_\bone}(\bsigma^\pl)$ be the vertex-edge statistics in the single-planted model and $\gammab^\pl=\gamma(\bsigma^\pl_\bone,\bsigma^\pl_\btwo)$, $\yb^\pl=y^{\bGp_\btwo}(\bsigma^\pl_\bone,\bsigma^\pl_\btwo)$ denote the vertex-edge statistics in the double-planted model.
\newcommand{\statelemstatisticsRC}{Let $d\geq 3$ be an integer
 and $q>2$ and $\beta_0>0$ be reals. There exists $\delta>0$ such that for all sufficiently small $\eps>0$, there is $c>0$ such that the following holds for all sufficiently large $n$. 
 
 For any $\beta\in[\beta_c-\delta,\beta_0]$, for the single-planted and double-planted models, we have respectively
\[ \Pr\Big[\big\|\alphab^\pl-\alpha^\pl\big\|_1+\big\|\xb^\pl-x^\pl\big\|_1\geq \eps\Big]\leq \emm^{-c  n}\quad \mbox{and}\quad\Pr\Big[\big\|\gammab^\pl-\gamma^\pl\big\|_1+\big\|\yb^\pl-y^\pl\big\|_1\geq \eps\Big]\leq \emm^{-c  n}.\]
The analogous statements hold for the minus phase for all
$\beta\in[1/\beta_0,\beta_c+\delta]$.}
\begin{lemma}\label{lem:statisticsRC}
\statelemstatisticsRC
\end{lemma}
The proof is given in~\Cref{sec:Isinglower}. We will use the following bounds for the mass of $\Sigma^\pl,\Sigma^\mi$ outside the critical window.
\begin{lemma}\label{lem:offcritical}
Let $d\geq 3$ be an integer, and let $q>2$ and $\beta_0>1$ be reals.
For any $\xi>0$, there exist $\delta,\chi>0$ such that, whp over
$G\sim\cG_{n,d}$, the following hold for all
$\beta\in[1/\beta_0,\beta_0]$. 
\begin{enumerate}
\item We have $Z_G^\mi/Z_G^\pl\leq \emm^{-\chi n}$ for $\beta>\beta_c+\delta$ and $Z_G^\pl/Z_G^\mi\leq \emm^{-\chi n}$ for $\beta<\beta_c-\delta$.
\item For $|\beta-\beta_c|\leq \delta$, we have $\emm^{-\xi n}\leq Z_G^\pl/Z_G^\mi\leq \emm^{\xi n}$.
\end{enumerate}
\end{lemma}
\begin{proof}
Set $h(\beta)=\frac{1}{Q}\big(B/\hat B\big)^{d/2}$. Note that $h$ is strictly increasing and $h(\beta_c)=1$. Fix $\delta\in(0,\beta_c)$ sufficiently small so  that
$ \emm^{- \xi/4}\leq h(\beta)\leq \emm^{ \xi/4} $ whenever
$|\beta-\beta_c|\leq\delta$.

For $\sigma\in \{0,1\}^V$, let $\overline{\sigma}$ be the configuration obtained from $\sigma$ by flipping $0$s and $1$s. Since $G$ is $d$-regular, we have $|M_1(\sigma)|-|M_0(\sigma)| =d\big(|\sigma^{-1}(1)|-n/2\big)$, so $\frac{w_G(\sigma)}{w_G(\overline\sigma)}
 =h(\beta)^{2|\sigma^{-1}(1)|-n}$.

We first give the proof of the first item  for odd $n$ where $\sigma\mapsto \overline{\sigma}$ is a bijection between $\Sigma^\pl$ and $\Sigma^\mi$.  Consider  the case $\beta>\beta_c+\delta$.
By Lemmas~\ref{lem:statisticsRC} and
\ref{lem:quietplantingRC}, and since the plus-phase
occupation density is bounded away from $1/2$,
there are constants $\kappa,c>0$ such that, whp,
\[
 \nu_G^\pl\big(|\sigma^{-1}(1)|<(1/2+\kappa)n\big)
 \leq \emm^{-c n}.
\]
The spin flip operation maps $\Sigma^\pl$ bijectively onto
$\Sigma^\mi$, so $
 \frac{Z_G^\mi}{Z_G^\pl}
 =
 \mathbb E_{\sigma\sim\nu_G^\pl}
 \big[h(\beta)^{n-2|\sigma^{-1}(1)|}\big]$. Since $h(\beta)>1$, for $\sigma\in \Sigma^\pl$ we have $h(\beta)^{n-2|\sigma^{-1}(1)|}\leq 1$ so 
splitting the expectation according to whether $|\sigma^{-1}(1)|<(1/2+\kappa)n$  gives
\[
 \nu_G(\Sigma^\mi)
 \leq \frac{Z_G^\mi}{Z_G^\pl}
 \leq \emm^{-c n}
      +\big(h(\beta_c+\delta)\big)^{-2\kappa n}
 \leq \emm^{-\chi n},
\]
for a constant $\chi>0$ and all sufficiently large $n$. A similar argument for $\beta<\beta_c-\delta$ gives that   $\nu_G^\mi\big(|\sigma^{-1}(1)|>(1/2-\kappa)n\big)
 \leq \emm^{-c n}$ and analogously $\nu_G(\Sigma^\pl)\leq\frac{Z_G^\pl}{Z_G^\mi}\leq \emm^{-\chi n}$.  

For even $n$, the argument is identical, except that the spin flip operation fixes the set of configurations $\sigma$ with $\{|\sigma^{-1}(1)|=n/2\}$. For $\beta>\beta_c+\delta$, the argument above works after replacing the expectation equality for $\frac{Z_G^\mi}{Z_G^\pl}$ by a $\leq $ inequality. So, we focus on the argument for $\beta<\beta_c-\delta$, where  we next show a bound on the weight of
these balanced configurations by $\emm^{-c n/2} Z^\mi_G$. Changing one spin from
$1$ to $0$ maps a balanced configuration to one with
$|\sigma^{-1}(1)|=n/2-1$, with at most $n$ preimages and weight ratio
at most $B^{d}$. Consequently, for $\beta<\beta_c-\delta$, we have that $B$ is upper bounded by $\emm^{\beta_c}$ so
\[
\sum_{\sigma:\,|\sigma^{-1}(1)|=n/2}w_G(\sigma)
 \leq n B^d Z_G^\mi\,
       \nu_G^\mi\big(|\sigma^{-1}(1)|=n/2-1\big)
 \leq \emm^{-cn/2}Z_G^\mi.
\]
Thus $\nu_G(\Sigma^\pl)
 \leq \frac{Z_G^\pl}{Z_G^\mi}
 \leq
 \mathbb E_{\sigma\sim\nu_G^\mi}
 \big[h(\beta)^{n-2|\sigma^{-1}(1)|}\big]
 +\emm^{-c n/2}
 \leq \emm^{-\chi n}$,  where the expectation is bounded as in the odd-$n$ case. Since $\log(Z_G^\pl/Z_G^\mi)$ is $(dn/2)$-Lipschitz in $\beta$,
applying these bounds on a sufficiently fine fixed grid in each
off-critical interval gives Item~1 uniformly.

 For the second item, our choice of $\delta$ and
the spin flip operation give $
 Z_G^\mi/Z_G^\pl\leq\emm^{\xi n/4}$. 
For odd $n$, the reciprocal ratio satisfies the same
bound. For even $n$, the one-spin comparison above,
without using concentration, bounds the weight of
balanced configurations by $nB^d Z_G^\mi$. Thus,
$Z_G^\pl/Z_G^\mi  \leq\emm^{\xi n/4}+nB^d
 \leq\emm^{\xi n}$, for all sufficiently large $n$, since $B\leq\emm^{\beta_c+\delta}$ in this window. These inequalities prove the second item.
\end{proof}

\subsection{The final probabilistic bounds}
For a graph $G$ and $\sigma\in \{0,1\}^V$, we denote by $\omegab_{G,\sigma}$ a random lifted configuration $(\bF_\sigma,S)\in \Omega$ obtained by setting $S=\sigma^{-1}(1)$ and  $\bF_\sigma=\mathrm{Bin}(M_0(\sigma), p_0)\cup \mathrm{Bin}(M_1(\sigma), p_1)$.

\begin{lemma}\label{lem:tranfRCtoIsing}
Let $d\geq 3$ be an integer, and let $q>2$ and $\beta_0>1$ be reals. There exist $\delta,c>0$ such that, whp over $G\sim \cG_{n,d}$, the following hold for any function $f:\Omega\times \Omega\rightarrow \mathbb{R}_\geq 0$
\begin{enumerate}
    \item For any $\beta\in[\beta_c-\delta,\beta_0]$, $\E_{\omega_1,\omega_2\sim\hat\nu_G^\pl}
 [f(\omega_1,\omega_2)]\leq 4q^2 \E_{\sigma_1,\sigma_2\sim \nu_G^\pl}[f(\omegab_{G,\sigma_1},\omegab_{G,\sigma_2})]+\emm^{-cn}f_{\max}$. 
 \item For any $\beta\in[1/\beta_0,\beta_c+\delta]$, $\E_{\omega_1,\omega_2\sim\hat\nu_G^\mi}
 [f(\omega_1,\omega_2)]\leq  \E_{\sigma_1,\sigma_2\sim \nu_G^\mi}[f(\omegab_{G,\sigma_1},\omegab_{G,\sigma_2})]+\emm^{-cn}f_{\max}$. 
\end{enumerate}
\end{lemma}
\begin{proof}
We prove the plus-phase assertion first.
Recall that $\omegab_{G,\sigma}$ is obtained by setting
$S=\sigma^{-1}(1)$ and retaining monochromatic edges
independently with probabilities $p_0,p_1$.
By Lemma~\ref{lem:Ising-lifting}, when $\sigma\sim\nu_G$,
this produces a sample from $\hat\nu_G$.

By Lemmas~\ref{lem:statisticsRC},
\ref{lem:quietplantingRC} and
\ref{lem:RC-edge-statistics}, together with a Chernoff
bound for the independent edge selections, there exists
$\xi>0$ such that, whp over $G$,
\[
 \Pr_{\sigma\sim\nu_G^\plm}
 [\omegab_{G,\sigma}\notin\Omega^\plm]
 \leq\emm^{-2\xi n},
\]
uniformly on the respective intervals by the grid argument of
Lemma~\ref{lem:offcritical} (since the log-weights of the lifted
configurations are $O(n)$-Lipschitz there).
Apply Lemma~\ref{lem:offcritical} with this $\xi$ to obtain
$\delta,\chi>0$, and let $\varepsilon=\emm^{-\min\{\xi,\chi\}n}$.

For $\beta\in[\beta_c-\delta,\beta_0]$, we claim that
\[
 \Pr_{\sigma\sim\nu_G}
 [\sigma\in\Sigma^\mi,\,
   \omegab_{G,\sigma}\in\Omega^\pl]
 \leq\varepsilon\,\nu_G(\Sigma^\pl).
\]
Indeed, dividing the left-hand side by
$\nu_G(\Sigma^\pl)$ gives $ \frac{Z_G^\mi}{Z_G^\pl}
 \Pr_{\sigma\sim\nu_G^\mi}
 [\omegab_{G,\sigma}\in\Omega^\pl]$. For $|\beta-\beta_c|\leq\delta$, the two factors are at
most $\emm^{\xi n}$ and $\emm^{-2\xi n}$, respectively, since $\Omega^\pl$ and $\Omega^\mi$ are disjoint.
For $\beta>\beta_c+\delta$, the first factor is at most
$\emm^{-\chi n}$, while the second is at most $1$.
This proves the claim.

Let $
 \mathcal D^\pl
 =\Omega^\pl\cap\{C_1(F)\text{ is active}\}$,
so that $\hat\nu_G^\pl=\hat\nu_G(\cdot\mid\mathcal D^\pl)$. Conditional on $F$ under $\hat\nu_G$, the probability
that $C_1(F)$ is active is $\geq\frac1q$, so 
\[
 \hat\nu_G(\mathcal D^\pl)
 \geq\tfrac1q\hat\nu_G(\Omega^\pl)
 \geq\tfrac{1-\varepsilon}{q}\nu_G(\Sigma^\pl).
\]

For any nonnegative function $g:\Omega\to\mathbb R$,
split its unnormalised expectation over $\mathcal D^\pl$
according to whether $\sigma\in\Sigma^\pl$.
Using the preceding bounds, we obtain
\begin{align*}
 \mathbb E_{\hat\nu_G^\pl}[g]
 &\leq
 \frac{\nu_G(\Sigma^\pl)}
      {\hat\nu_G(\mathcal D^\pl)}
 \left(
 \mathbb E_{\sigma\sim\nu_G^\pl}
 [g(\omegab_{G,\sigma})]
 +\varepsilon g_{\max}
 \right)\leq
 2q 
 \left(
 \mathbb E_{\sigma\sim\nu_G^\pl}
 [g(\omegab_{G,\sigma})]
 +\varepsilon g_{\max}
 \right).
\end{align*}
Applying this inequality successively to the two
coordinates of $f$ gives
\[
 \mathbb E_{\omega_1,\omega_2\sim\hat\nu_G^\pl}
 [f(\omega_1,\omega_2)]
 \leq
 4 q^2
 \mathbb E_{\sigma_1,\sigma_2\sim\nu_G^\pl}
 [f(\omegab_{G,\sigma_1},\omegab_{G,\sigma_2})]
 +O_q(\varepsilon)f_{\max},
\]
proving the first assertion. The minus-phase argument is identical, with the roles
of the phases interchanged. Here the conditioning event
is simply $\Omega^\mi$, and $
 \hat\nu_G(\Omega^\mi)
 \geq(1-\varepsilon)\nu_G(\Sigma^\mi)$.
Thus the same argument gives the second assertion.
\end{proof}

We will also use the following standard percolation bounds for the configuration model. Define first the following generalisation.
\begin{definition}\label{def:partial-configuration}
For integers $N,d,M$ with $0\leq2M\leq dN$, let
$\CM_{d}(N,M)$ denote the following random
multigraph on $N$ vertices.
Assign $d$ labelled half-edges to each vertex, choose
$2M$ of the $dN$ half-edges uniformly, and pair the
chosen half-edges uniformly.
\end{definition}
When $dN$ is even, this is equivalently obtained by
sampling a $d$-regular configuration graph and
retaining a uniformly chosen set of exactly $M$
of its edges.

 For a graph $H$, we let $C_1(H)$ denote its largest component, and $H^*$ to denote the graph obtained from $H$ by removing $C_1(H)$. We use $\mathcal{C}(H)$ to denote the collection of components of $H$ and $\mathtt{exc}(H)$ to denote the total tree-excess of $H$. Let $W(H)$ denote the number of cyclic components in $H$.

\begin{lemma}\label{lem:percsub}
Fix $d\geq3$ and $\eta>0$. There exist constants $\kappa,K>0$ such that the following hold for  $H\sim\mathsf{CM}_{d}(N,M)$ whenever $p=\frac{2M}{dN}$ satisfies
$|(d-1)p-1|\geq\eta$ and $p\leq 1-\eta$.  

For integers $\ell,r\geq 1$, let $t:= KN\emm^{-\kappa\ell}  +N\left(\frac{K\ell^2} {N}\right)^r+\emm^{-\kappa N}$. 
\begin{enumerate}
 \item For any $z\geq 1$ with $z\leq \exp\big(\sqrt{\log N}\big)$, it holds that $\mathbb E\big[
   z^{W(H)}  \big]
 \leq K\emm^{K (z-1)}$.
 \item if $(d-1)p<1-\eta$, it holds that $\Pr[\mathtt{exc}(C)\geq r \mbox{ or } |C|\geq \ell \text{ for some } C\in \mathcal{C}(H)]\leq  t$.
 \item if $(d-1)p>1+\eta$, it holds that $\Pr[\mathtt{exc}(C)\geq r \mbox{ or } |C|\geq \ell \text{ for some } C\in \mathcal{C}(H^*)]\leq  t$.
\end{enumerate}
\end{lemma}

To utilise Lemma~\ref{lem:percsub}, we will need the following inequalities that establish the relevant subcritical/supercritical conditions. The proof is given in Section~\ref{sec:remainingproofsb}.
\begin{lemma}\label{lem:RC-edge-statistics}
Fix $d\geq3$ and $q>2$. There exist constants $\delta,\eta,\theta>0$ such that the following hold. Let $p_0=\frac{\hat B-1}{\hat B}$ and $p_1=\frac{B-1}{ B}$ where $B=\emm^\beta$ and $\hat B=1+\frac{\emm^\beta-1}{q-1}$. Let $p=p_0+p_1-p_0p_1$.  

\begin{enumerate}
 \item For every $\beta>\beta_c-\delta$, we have $p_0 \frac{x_{00}^\pl}{1-\alpha^\pl},\  p \frac{x_{00}^\pl x_{11}^\pl}{\alpha^\pl(1-\alpha^\pl)} 
       \leq \frac{1-\eta}{d-1}$ and  $p_1 \frac{x_{11}^\pl}{\alpha^\pl}\geq \frac{1+\eta}{d-1}$.
 \item For every $\beta<\beta_c+\delta$, we have $p_0 \frac{x_{00}^\mi}{1-\alpha^\mi},\ p_1 \frac{x_{11}^\mi}{\alpha^\mi}, \  p \frac{x_{00}^\mi x_{11}^\mi}{\alpha^\mi(1-\alpha^\mi)} \leq \frac{1-\eta}{d-1}$.
\end{enumerate}
Moreover, the functions $\mathfrak{m}^\plm(\beta)=p_0x^\plm_{00}+p_1x^\plm_{11}$ are increasing in $\beta$, and for all $\beta\in (\beta_c-\delta,\beta_c+\delta)$ it holds that  $\mathfrak{m}^\pl(\beta)-\mathfrak{m}^\mi(\beta)\geq 10\theta$.
\end{lemma}
\begin{remark}\label{rem:pcvalue}
The functions $\mathfrak{m}^\mi(\beta),\mathfrak{m}^\pl(\beta)$ capture the most likely density of edges in $\pi^\mi_G,\pi^\pl_G$. In light of the final part of Lemma~\ref{lem:RC-edge-statistics}, the threshold value of $\mathfrak{m}_c$ that separates the phases (cf. \eqref{eq:phasesRCdef})  can be chosen as $\mathfrak{m}_c=\frac{1}{2}(\mathfrak{m}^\pl(\beta_c)+\mathfrak{m}^\mi(\beta_c))$.
\end{remark}
Finally, we will use the following standard fact about random regular graphs.
\begin{lemma}[{\cite[Eq.~(2.2)]{MWW04}}]\label{lem:numcycles}
Let $d\geq 3$ be an integer. Then, there are constants $\rho,\zeta\in(0,1/2)$ such that, whp over $G\sim\cG_{n,d}$, there are $\leq n^\zeta$ cycles with length $\leq\rho\log n$.
\end{lemma}

We are now ready to prove Theorem~\ref{thm:randomgraphRC}, which we restate here for convenience. Recall that $\mathcal{A}_{G,R}$ is the event in $\Omega^2$ that at least one of the two inactive graphs has total tree-excess greater than $R$, $\mathcal{B}_{G,\ell, r}$ is the event in $\Omega\times \Omega$ that the disagreement graph has a component bigger than $\ell$ or  tree-excess bigger than $r$.

\begin{thmrandomgraphRC}
\statelemrandomgraphRC
\end{thmrandomgraphRC}
\begin{proof}[Proof of Theorem~\ref{thm:randomgraphRC}]
Let $\delta,\theta>0$ be small constants satisfying Lemma~\ref{lem:RC-edge-statistics}, $K,\kappa>0$ be the constants from Lemma~\ref{lem:percsub} and $\eps,c>0$ be sufficiently small constants satisfying Lemma~\ref{lem:statisticsRC}; we require in particular that $\theta\gg \eps$. Let $\tau=\frac{1}{10}\min\{\kappa,c,\eps^2\}$.
Lemma~\ref{lem:RC-edge-statistics} and
Remark~\ref{rem:pcvalue} ensure that $\mathfrak{m}^\pl(\beta)\geq \mathfrak{m}_c+4\theta$ for $\beta>\beta_c-\delta$.

We focus first on the plus-phase assertion; in fact, it suffices to prove it for
each fixed $\beta\in[\beta_c-\delta,\beta_0]$. To obtain from this the conclusion for all $\beta$,
use a union bound over $O(n)$ events corresponding to values of $\beta$ spaced out by $1/n$ in the interval. We can ensure that each such event has  failure probability $\leq 1/n^{D+3}$, so for any intermediate $\beta$  the probability of the desired event is within a factor $O(1)$ from that of the closest $\beta$-value.

So consider $\beta>\beta_c-\delta$. For a graph $G$, define $f:\Omega\rightarrow \mathbb{R}$ and the event $\mathcal{D}_G$ by 
\[
 f(\omega)
 =
 q^{c(\omega)}
 \mathbf1_{\{r(\omega)>R\}}
 \mathbf1_{\{C_1(\omega)\text{ is active}\}},
 \qquad
 \mathcal D_G=\cB_{G,\ell,r}\cup\cH^\pl_{\theta n},
\]
and write $b_G^\pl$ for $\Pr[(\omega,\omega')\in \cA_{G,R}\cup \mathcal{B}_{G,\ell,r}\cup \cH^{\pl}_{\theta n}]$. By
Lemma~\ref{lem:RCtoIsing}, for every graph $G$
\[
 b_G^\pl
 \leq
 2\mathbb E_{\omega\sim\hat\nu_G^\pl}[f(\omega)]
 +Q^{2R}
 \Pr_{\omega,\omega'\sim\hat\nu_G^\pl}
 [(\omega,\omega')\in\mathcal D_G].
\]
We say that a graph $G$ is \emph{cycle-good} if it satisfies the small cycle-counts in Lemma~\ref{lem:numcycles}.  A good graph $G$ has at most $n^{\zeta}$ cycles of length $\leq \rho \log n$, so for every $F\subseteq E$ there at most $n^\zeta+\frac{n}{\rho \log n}=o(n)$ cyclic components in $(V,F)$ so $c(\omega)=o(n)$ for any $\omega\in \Omega$ and $f_{\max}=\emm^{o(n)}$. 

We say that a graph $G$ \emph{lifts well} if it satisfies the  lifting-transfer inequality  of Lemma~\ref{lem:tranfRCtoIsing}.  Recall in particular that for $\sigma\in \{0,1\}^V$, the lifted configuration $\omegab_{G,\sigma}$ is obtained by keeping each edge in $M_0(\sigma)$ with probability $p_0$ and each  edge in $M_1(\sigma)$ with probability $p_1$. 
Applying  the inequality in Lemma~\ref{lem:tranfRCtoIsing} twice (for 
the function $(\omega,\omega')\mapsto f(\omega)$ and the event $\mathcal D_G$) yields that, for graphs $G$ that lift well, 
\begin{equation}\label{eq:RC-final-reduction}
\begin{aligned}
 b_G^\pl
 \leq{}&
 8q^2\mathbb E_{\sigma\sim\nu_G^\pl}
       [f(\omegab_{G,\sigma})]+4q^2Q^{2R}
 \Pr_{\sigma_1,\sigma_2\sim\nu_G^\pl}
 [(\omegab_{G,\sigma_1},\omegab_{G,\sigma_2})
       \in\mathcal D_G]
 +Q^{2R}\emm^{-cn/2}.
\end{aligned}
\end{equation}

We estimate the expectation using the single-planted
model $(\bGp_\bone,\bsigma^\pl)$ and the probability using the double-planted
model $(\bGp_\btwo,\bsigma^\pl_\bone,\bsigma^\pl_\btwo)$.  Set for simplicity  $\omegab^\pl=\omegab_{\bGp_\bone,\sigma^\pl}$ for the lifted vertex-edge configuration in the single-planted model and $\omegab^\pl_\bone=\omegab_{\bGp_\btwo,\bsigma^\pl_\bone}$, $\omegab^\pl_\btwo=\omegab_{\bGp_\btwo,\bsigma^\pl_\btwo}$ for the lifted vertex-edge configurations in the double-planted model.

Let $\tilde{f}(G,\sigma)=\mathbb{E}[f(\omegab_{G,\sigma})]\mathbf{1}\{G\text{ is cycle-good}\}$ where the expectation is just over the randomness of the lifting $\omegab_{G,\sigma}$; note that the indicator for cycle-goodness ensures  that $\tilde{f}_{\max}=\emm^{o(n)}$. Let also $g(G,\sigma_1,\sigma_2)=\Pr[(\omegab_{\sigma_1},\omegab_{\sigma_2})\in \mathcal{D}_G]$ and note that $g_{\max}\leq 1$. We will show  that  for an arbitrary large constant $L>0$ it holds that
\begin{equation}\label{eq:boundsplant12}
\begin{aligned}
\mathbb E
       [\tilde{f}(\bGp_\bone,\bsigma^\pl)]&\leq  n^{-L}+ K^2\Big(\frac{\emm^{10q KL}}{R}\Big)^{R/(5L)}+\emm^{-\tau n},\\
\mathbb E[g(\bGp_\btwo,\bsigma^\pl_\bone,\bsigma^\pl_\btwo)]&\leq  K n\emm^{-\kappa\ell}
 +n\left(\frac{K\ell^2}{n}\right)^r+\emm^{-\tau n}.
\end{aligned}
\end{equation}
Note that the planting Lemma~\ref{lem:quietplantingRC} also extends to expectations (by integrating for example over appropriate events). So, we conclude that whp over $\bG$, 
\[\mathbb E_{\sigma\sim\nu_{\bG}^\pl}
       [\tilde{f}(\bG,\sigma)]\leq n^T \mathbb  E
       [\tilde{f}(\bGp_\bone,\bsigma^\pl)] \text{ and }\mathbb E_{\sigma_1,\sigma_2\sim\nu_{\bG}^\pl}[g(\bG,\sigma_1,\sigma_2)]\leq n^T \mathbb E[g(\bGp_\btwo,\bsigma^\pl_\bone,\bsigma^\pl_\btwo)].\]
       Let's call the graphs satisfying these bounds ``tame''. By a union-bound $\bG$ is whp cycle-good, tame and lifts well. On that intersection, from  cycle-goodness and tameness, we have 
       \[\mathbb E_{\sigma\sim\nu_{\bG}^\pl}
       [f(\omegab_{\bG,\sigma})]=\mathbb E_{\sigma\sim\nu_{\bG}^\pl}
       [\tilde{f}(\bG,\sigma)]\leq n^T \mathbb E
       \big[\tilde{f}(\bGp_\bone,\bsigma^\pl)\big].\]
       Note also $\Pr_{\sigma_1,\sigma_2\sim\nu_{\bG}^\pl}
 [(\omegab_{G,\sigma_1},\omegab_{G,\sigma_2})
       \in\mathcal D_{\bG}]=\mathbb E_{\sigma_1,\sigma_2\sim\nu_{\bG}^\pl}[g(\bG,\sigma_1,\sigma_2)]\leq n^T \mathbb E[g(\bGp_\btwo,\bsigma^\pl_\bone,\bsigma^\pl_\btwo)]$. So,  using that $G$ lifts well, plugging these bounds into \eqref{eq:RC-final-reduction} and using $\tilde{f}_{\max}=\emm^{o(n)}$, we can bound $b_G^\pl\leq 1/n^D$ for $R=C\frac{\log n}{\log \log n}$ and $\ell=C'\log n$ (provided $C,C'$ are large enough).
       
To obtain the bounds in \eqref{eq:boundsplant12}, we will need to consider their typical vertex-edge statistics. Fix a sufficiently small constant $\eps>0$ and let
\begin{equation*}
\mathcal{T}_1=\big\{\big\|\alphab^\pl-\alpha^\pl\big\|_1+\big\|\xb^\pl-x^\pl\big\|_1\leq \eps\} \mbox{ and }
\mathcal{T}_2=\big\{\big\|\gammab^\pl-\gamma^\pl\big\|_1+\big\|\yb^\pl-y^\pl\big\|_1\leq \eps\big\}
\end{equation*}
be the events that the vertex-edge statistics in the two models are close to the values $\alpha^\pl,x^\pl$ and $\gamma^\pl,y^\pl$, respectively. By Lemma~\ref{lem:statisticsRC}, $\Pr\big[\neg\mathcal{T}_1\big]\leq \emm^{-cn}\leq \emm^{-10\tau n}$ and $\Pr[\neg \mathcal{T}_2]\leq \emm^{-cn}\leq \emm^{-10\tau n}$.

We first bound $\mathbb E
       [\tilde{f}(\bGp_\bone,\bsigma^\pl)]$ for the single-planted model. Condition on the spin configuration $\bsigma^\pl$, its edge
statistics, and the remaining edge counts after the percolation step. Vertex-edge statistics that do not satisfy $\mathcal{T}_1$  have probability at most $\emm^{-cn}$ from Lemma~\ref{lem:statisticsRC}, so they contribute at most $\tilde{f}_{\max} \emm^{-10\tau n}$ to $\mathbb E[\tilde{f}(\bGp_\bone,\bsigma^\pl)]$. 

For vertex-edge statistics that do satisfy $\mathcal{T}_1$,    let $H_0,H_1$ be the  graphs on the two spin
classes after the percolation steps in $\omega^\pl$. 
Uniformity of the half-edge matchings gives that 
$H_i\sim\CM_d(N_i,m_i)$ where $N_i=|(\bsigma^\pl)^{-1}(i)|$ and $m_i=|E(H_i)|$. Standard Chernoff bounds give that the number of remaining edges $m_0,m_1$ satisfy  with probability $1-\emm^{-c n}$ the density bounds
$\frac{2m_0}{dN_0}
 =p_0\frac{x^\pl_{00}}{1-\alpha^\pl}+O(\eps)$ and $ \frac{2m_1}{dN_1}
 =p_1\frac{x^\pl_{11}}{\alpha^\pl}+O(\eps)$ (those that do not, contribute at most $\emm^{o(n)}\emm^{-10\tau n}\leq \emm^{-5\tau n}$ to the target expectation). By Lemma~\ref{lem:RC-edge-statistics}, $H_0$ is
uniformly subcritical and $H_1$ is uniformly
supercritical.
Items~2 and~3 of Lemma~\ref{lem:percsub} therefore
apply to $H_0$ and $H_1^*$, respectively.

Let $W_0$ count the cyclic components of $H_0$ and $W_1$ those of $H_1^*$, and note that 
\[f(\omega)=q^{c(\omega)}
 \mathbf1_{\{r(\omega)>R\}}
 \mathbf1_{\{C_1(\omega)\text{ is active}\}}\leq q^{W_0+W_1}\mathbf{1}_{\mathtt{exc}(H_0)+\mathtt{exc}(H_1^*)>R}.\] For a (large) constant $L>0$, let also $\mathcal F_L$ be the event that some component
of $H_0$ or $H_1^*$ has excess greater than $5L$. Outside $\mathcal F_L$, observe that there are at most $W:=W_0+W_1$ components with non-zero tree excess (which must have excess $\leq 5L$), so $\mathtt{exc}(H_0)+\mathtt{exc}(H_1^*)\leq 5L W$. It follows that
\[\mathbb E
       \big[\tilde{f}(\bGp_\bone,\bsigma^\pl)\big]\leq \mathbb E\big[q^{W}\mathbf1_{\mathcal F_L}\big]+\mathbb E\big[q^{W}\mathbf1_{W>R/(5L)}\big].\]
Independence of the graphs $H_0,H_1$ (conditioned on the edge counts) and  Lemma~\ref{lem:percsub} (for $z=q^2$) imply the uniform bounds (over typical statistics)
\[
 \mathbb E[q^{2W}]=\mathbb E[q^{2W_0}] \mathbb E[q^{2W_1}] \leq K^2\emm^{2Kq^2 },\quad \Pr[\mathcal F_L]\leq n^{-3L},
\]
Consequently,
\[
 \mathbb E[q^{W}\mathbf1_{\mathcal F_L}]
 \leq
 \sqrt{\mathbb E[q^{2W}]\Pr[\mathcal F_L]}
 \leq n^{-L}.
\]
Applying in an analogous manner Lemma~\ref{lem:percsub} for $z=q R$ gives that
\[ K^2 \emm^{2qKR}\geq \mathbb E\big[(qR)^{W}\big]\geq R^{R/(5L)}\,  \mathbb E\big[q^{W}\mathbf1_{\{W>R/(5L)\}}\big].
\]
Combining these bounds give the first inequality in \eqref{eq:boundsplant12}.

To show the second inequality in \eqref{eq:boundsplant12} for $\Pr[(\omegab^\pl_\bone,\omegab^\pl_\btwo)\in \mathcal{D}_{\bGp_\btwo}]$  the proof is analogous, a bit simpler in fact since we only have to care about the probability of an event instead of an expectation of a potentially large function. In particular, the disagreement graph $H(\omegab^\pl_\bone,\omegab^\pl_\btwo)$ is the disjoint union of its
restrictions to the paired-spin classes $(0,1)$ and $(1,0)$.
Within either class, an edge is kept in the union
with probability $p=p_0+p_1-p_0p_1$ and  each restriction is a partial configuration
model.

Conditional on the event $\mathcal{T}_2$ of typical vertex-edge statistics, the density is $p\frac{x^\pl_{00}x^\pl_{11}}
        {\alpha^\pl(1-\alpha^\pl)}
 +O(\eps)$ with probability $\geq 1-\emm^{-10\tau n}$. Since the density is uniformly subcritical by
Lemma~\ref{lem:RC-edge-statistics}, applying Lemma~\ref{lem:percsub} to the two paired-spin classes
bounds the probability of $\cB_{G,\ell,r}$ by
$K
 n\emm^{-\kappa\ell}
 +n\left(\frac{K\ell^2}{n}\right)^r+\emm^{-\kappa n}$.
For the event $\cH^\pl_{\theta n}$, conditional on $\mathcal T_2$, we have that, simultaneously for both configurations,  the number of remaining edges is $\frac{dn}{2}\mathfrak{m}^\pl(\beta)+O(\eps n)$ with probability $\geq 1-\emm^{-10\tau n}$ (from a Chernoff bound and a union bound).
Since $\mathfrak{m}^\pl(\beta)\geq \mathfrak{m}_c+4\theta$ and $\theta\gg \eps$, we have  that the conditional probability of $\cH^\pl_{\theta n}$ is $\leq \emm^{-10\tau n}$.

For the minus phase with $\beta$ bounded away from zero,
the proof is analogous and simpler.
The percolated spin-class graphs are subcritical by
Item~2 of Lemma~\ref{lem:RC-edge-statistics}
(here we don't need the largest-active indicator).
The remaining argument uses the corresponding minus-phase
statistics and transfer bounds.

To cover small $\beta$, choose a fixed $\beta_1>0$ such that
$p_1:=1-\emm^{-2\beta_1}
<\min\{1/(4(d-1)),\mathfrak m_c/2-\theta\}$.
For every $\beta\leq\beta_1$, each edge has conditional
inclusion probability at most $1-\emm^{-\beta_1}$
under $\pi^\mi_{G,\beta}$: the phase restriction either
leaves the usual RC conditional probability unchanged
or forces the edge to be absent.
Consequently, two independent samples $F,F'$ can be
coupled so that $F\cup F'\subseteq U$, where $U$ is
Bernoulli percolation on $G$ with parameter $p_1$.
Let $\mathcal E$ be the event that $U$ has more than
$p_1dn$ edges, total tree-excess greater than $R$,
or a component with more than $\ell$ vertices or
tree-excess greater than $r$.
Outside $\mathcal E$, neither inactive graph has
tree-excess greater than $R$, and every disagreement
component has size at most $\ell$ and tree-excess
at most $r$.
Moreover,
$|F|,|F'|\leq p_1dn
<\frac{dn}{2}\mathfrak m_c-d\theta n$.
Thus the pair avoids
$\mathcal A_{G,R}\cup\mathcal B_{G,\ell,r}
 \cup\mathcal H^\mi_{\theta n}$,
regardless of its component labels.

For $\bG$ sampled from the configuration model,
conditional on $|U|=m\leq p_1dn$,
the graph $(V,U)$ has distribution $\CM_d(n,m)$
with density at most $2p_1$, which is uniformly
subcritical.
The probability that $|U|>p_1dn$ is exponentially
small by a Chernoff bound.
Apply the subcritical bounds of Lemma~\ref{lem:percsub}
and the same cyclic-component argument used above:
outside the event that some component has excess
greater than $r$, we have
$\mathtt{exc}(U)\leq rW(U)$, and taking $z=R/r$
in Item~1 gives, conditional on the retained count,
a bound
\[
 \exp\left\{-\frac Rr\log\frac Rr+O(R/r)\right\}
\]
for the remaining total-excess tail.
Choosing $r,C',C$ sufficiently large therefore gives
$\Pr[\mathcal E]\leq n^{-D-2}$.
By Markov's inequality, whp over $\bG$,
$\Pr[\mathcal E\mid\bG]\leq n^{-D}$.
Since the dominating percolation law does not depend
on $\beta$, this gives the desired conclusion
simultaneously for all $0<\beta\leq\beta_1$.
Combining this with the compact-interval argument
on $[\beta_1,\beta_c+\delta]$ completes the proof.
\end{proof}

\section{Disagreement components for the ferromagnetic Ising model}
\label{sec:randomgraphIS}

We prove Lemma~\ref{lem:randomgraphISb} using double planting,
analogously to the hard-core and random-cluster models, so we give the relevant details a bit faster this time.
Throughout this section, fix $d\geq3$, $\beta>\beta_c(d)$ and
$s_0>0$, and let $s\in[0,s_0]$. Define
\[
 w_{G,s}(\sigma)=\emm^{\beta m(\sigma)+sM(\sigma)},
 \qquad
 Z^\pl_{G,s}=\sum_{\sigma\in\Sigma^\pl}w_{G,s}(\sigma).
\]
The double phase-planted distribution is defined by
\[
 \Pr[(\bGp,\bsigma^\pl_1,\bsigma^\pl_2)=(G,\sigma_1,\sigma_2)]
 =
 \frac{\Pr[\bG=G]\,
       w_{G,s}(\sigma_1)w_{G,s}(\sigma_2)
       \mathbf1_{\{\sigma_1,\sigma_2\in\Sigma^\pl\}}}
      {\E[(Z^\pl_{\bG,s})^2]}.
\]
Thus, conditional on the planted graph, the two configurations
are independent samples from its plus-phase distribution.
We suppress the dependence of the planted variables on $s$.

We will use the following partition-function bounds, proved in
Section~\ref{sec:Isingfieldbounds}.
\begin{lemma}\label{lem:ZboundsIS}
There are a constant $C_0>0$ and a continuous function $\Psi(s)$
such that, uniformly for $s\in[0,s_0]$ and all sufficiently large $n$,
\[
 \E[Z^\pl_{\bG,s}]\leq n^{C_0}\emm^{n\Psi(s)},
 \qquad
 \E[(Z^\pl_{\bG,s})^2]\leq n^{C_0}\emm^{2n\Psi(s)}.
\]
Moreover, every $d$-regular graph $G$ on $n$ vertices satisfies
$Z^\pl_{G,s}\geq\frac12\emm^{n\Psi(s)}$.
\end{lemma}

As in Lemma~\ref{lem:quietplantingRC}, these bounds yield a
planting comparison. It is convenient here to retain its
expectation form, before applying Markov's inequality. Recall that for a graph $G$, we use $\bsigma_G^\pl$ to denote a sample from $\nu^\pl_{G,s}$. We use $\bsigma_{\bG,1}^\pl,\bsigma^\pl_{\bG,2}$ to denote (independent) samples from $\nu^\pl_{\bG,s}$
\begin{lemma}\label{lem:quietplantingIS}
For every nonnegative function
$f:\cG_{n,d}\times\{0,1\}^n\times\{0,1\}^n\to\mathbb R_{\geq0}$,
\[
 \E_{\bG}
       [f(\bG,\bsigma_{\bG,1}^\pl,\bsigma^\pl_{\bG,2})]
 \leq
 4n^{C_0}\E[f(\bGp,\bsigma^\pl_1,\bsigma^\pl_2)].
\]
The same comparison holds if, under both laws, we additionally
sample $F_{\sigma_1,\sigma_2}$ conditional on the graph and configurations,
and allow $f$ to depend on this edge set.
\end{lemma}
\begin{proof}
The definition of the planted law gives
\[
 \E[f(\bGp,\bsigma^\pl_\bone,\bsigma^\pl_\btwo)]
 =
 \frac{\E_{\bG}\big[(Z^\pl_{\bG,s})^2
       \E_{\sigma_1,\sigma_2\sim\nu^\pl_{\bG,s}}
       f(\bG,\sigma_1,\sigma_2)\big]}
      {\E[(Z^\pl_{\bG,s})^2]}.
\]
The deterministic lower bound and second-moment upper bound
in Lemma~\ref{lem:ZboundsIS} prove the claim.
For the additional edge sampling, apply this comparison to
the conditional expectation of $f$ given the graph and configurations.
\end{proof}

We next specify the typical vertex-edge statistics.
Write $B=\emm^\beta$, and let $a^\pl=a^\pl(s)\in(1/2,1)$
be the solution of the equation
\begin{equation}\label{eq:IS-BP}
 \frac{a^\pl}{1-a^\pl}
 =
 \emm^s
 \left(\frac{1-a^\pl+Ba^\pl}{B(1-a^\pl)+a^\pl}\right)^{d-1}.
\end{equation}
Its existence and uniqueness are verified in
Section~\ref{sec:Isingfieldbounds}.
The corresponding occupation probability $\alpha^\pl=\alpha^\pl(s)\in (1/2,1)$
satisfies
$\frac{\alpha^\pl}{1-\alpha^\pl}
 =
 \emm^s
 \left(\frac{1-a^\pl+Ba^\pl}{B(1-a^\pl)+a^\pl}\right)^d$.

Define the edge statistics by
\[
 x^\pl=\frac{1}{X^\pl}
 \begin{pmatrix}
 B(1-a^\pl)^2&a^\pl(1-a^\pl)\\
 a^\pl(1-a^\pl)&B(a^\pl)^2
 \end{pmatrix}, \mbox{ where } 
X^\pl=B(1-a^\pl)^2+2a^\pl(1-a^\pl)+B(a^\pl)^2.
\]
We also set $\tau^\pl=1-x^\pl_{01}/(\alpha^\pl(1-\alpha^\pl))$. Let $\gamma^\pl$ and $y^\pl$ be defined as
\begin{equation}\label{eq:gdyoptimalbbb}
\gamma^\pl=\big[\begin{smallmatrix}1-\alpha^{\pl}\\ \alpha^\pl\end{smallmatrix}\big]\otimes \big[\begin{smallmatrix}1-\alpha^{\pl}\\ \alpha^\pl\end{smallmatrix}\big], \quad y^\pl=x^\pl \otimes x^\pl.
\end{equation}

Let $\gammab^\pl=\gamma(\bsigma^\pl_\bone,\bsigma^\pl_\btwo)$, $\yb^\pl=y^{\bGp}(\bsigma^\pl_\bone,\bsigma^\pl_\btwo)$ denote the vertex-edge statistics in the double-planted model.
As before, conditioning the planted model on the configurations
and these statistics gives a uniform half-edge matching with the
prescribed edge-type counts.

\begin{lemma}\label{lem:statisticsIS}
There is a constant $\zeta>0$ such that, for every $s\in[0,s_0]$,
\[
 \alpha^\pl\geq\tfrac12+\zeta,
 \qquad
 0\leq(d-1)\tau^\pl\leq1-\zeta.
\]
Moreover, for every fixed $\eps>0$, there is $c>0$ such that,
uniformly for $s\in[0,s_0]$ and  large enough~$n$,
\[
 \Pr\left[
 \big\|\gammab^\pl-\gamma^\pl\big\|_1+
 \big\|\yb^\pl-y^\pl\big\|_1
 \geq\eps
 \right]\leq\emm^{-cn}.
\]
\end{lemma}
The proof is given together with Lemma~\ref{lem:ZboundsIS} in
Section~\ref{sec:Isingfieldbounds}.

\begin{proof}[Proof of Lemma~\ref{lem:randomgraphISb}]
Choose $0<\theta<\zeta/2$ and a sufficiently small constant
$\eps>0$ in Lemma~\ref{lem:statisticsIS}. Since $\gamma^\pl_{10}+\gamma^\pl_{11}
=\alpha^\pl\geq1/2+\zeta$, taking $\eps$ sufficiently
small in Lemma~\ref{lem:statisticsIS} gives an
exponentially small double-planted probability that
$M(\bsigma_1^\pl)\leq\theta n$.
Applying Lemma~\ref{lem:quietplantingIS} to the event
$\{M(\sigma_1)\leq\theta n\}$, which depends only on
the first configuration, therefore gives for some $c>0$ that
\begin{equation}\label{eq:IS-balanced-average}
 \E_{\bG}\big[\nu^\pl_{\bG,s}(M\leq\theta n)\big]
 \leq\emm^{-cn}.
\end{equation}

For the disagreement components, sample $F_{\sigma,\sigma'}$
in the double-planted model. Its edges lie within the two
classes of vertices with joint spins $(0,1)$ and $(1,0)$.
Conditional on the configurations, their edge statistics, and the
number $M$ of remaining edges within either class, the resulting graph
on its $N$ vertices has the distribution $\CM_d(N,M)$ (see Definition~\ref{def:partial-configuration}).
Indeed, the remaining half-edges form a uniform subset of the $dN$
half-edges in the class and are paired uniformly.

Since $\alpha^\pl(s)\in(1/2,1)$ is continuous on
$[0,s_0]$, the quantity
$\alpha^\pl(s)(1-\alpha^\pl(s))$ is uniformly bounded
away from zero. Thus, for sufficiently small $\eps$,
each disagreement class has $N=\Theta(n)$ vertices
on the statistics-concentration event of Lemma~\ref{lem:statisticsIS}.
By Chernoff concentration of the remaining edge count, its half-edge density
is, except with exponentially small probability, arbitrarily close to
\[
 (1-B^{-2})\frac{x^\pl_{00}x^\pl_{11}}
                    {\alpha^\pl(1-\alpha^\pl)}
 =
 \frac{x^\pl_{00}x^\pl_{11}-(x^\pl_{01})^2}
      {\alpha^\pl(1-\alpha^\pl)}
 =\tau^\pl.
\]
Here $x^\pl_{00}x^\pl_{11}=B^2(x^\pl_{01})^2$ follows directly
from the definition of $x^\pl$.
Lemma~\ref{lem:statisticsIS} therefore verifies the subcriticality
assumption of Lemma~\ref{lem:percsub}, uniformly in $s$.
Applying its subcritical bound to the two classes and then
Lemma~\ref{lem:quietplantingIS}, we obtain constants $K,\kappa>0$
such that
\begin{equation}\label{eq:IS-disagreement-average}
 \E_{\bG}[b_{\bG,\ell,r}(s)]
 \leq
 4n^{C_0}\left[
 K n\emm^{-\kappa\ell}
 +n\left(K\ell^2/n\right)^r
 +\emm^{-\kappa n}\right].
\end{equation}

Choose $C$ and then the integer $r$ sufficiently large that,
with $\ell=\lceil C\log n\rceil$, the right-hand side of
\eqref{eq:IS-disagreement-average} is at most $n^{-(D+6)}$.
Markov's inequality, \eqref{eq:IS-balanced-average}, and a union
bound over a set $\mathcal{S}$ of $n^2$ values of $s$ spread uniformly over $[0,s_0]$ give that, whp, for all $s\in \mathcal{S}$, 
\[
 b_{G,\ell,r}(s)\leq n^{-D}/2,
 \qquad
 \nu^\pl_{G,s}(M\leq\theta n)\leq\emm^{-cn/2}.
\]
Finally, the pointwise ratio of $\nu^\pl_{G,s}$ and
$\nu^\pl_{G,s'}$ lies between
$\emm^{-n|s-s'|}$ and $\emm^{n|s-s'|}$, while the
conditional law of $F_{\sigma,\sigma'}$ is independent
of $s$. Thus the corresponding ratio for two samples
is at most $\emm^{2n|s-s'|}$.
Comparison with the nearest $s\in \mathcal{S}$ proves both
assertions for every $s\in[0,s_0]$, with $\rho=c/3$.
\end{proof}

\section{Partition function  and vertex-edge statistics estimates}
Here we prove the partition function bounds used in the planting transfer lemmas. We start with the hard-core model, and then with the two Ising models (for random-cluster and Ising respectively).

Throughout this section, $\bG$ denotes a random multigraph sampled
from the $d$-regular configuration model. We take $n$ vertices on
each side in the bipartite case (see Footnote~\ref{foo:configmodel})
and $n$ vertices in total for the two Ising models.

\subsection{First Moment Bounds via Matrix Norms}
Let $q\geq 2,d\geq 3$ be integers. Let $\Bb=\{B_{ij}\}_{i,j\in [q]}$ be a $q\times q$   matrix with nonnegative entries. Consider a $d$-regular bipartite graph $G=(V^\pl\sqcup V^\mi,E)$ with $n$ vertices on each side. We use $Z_G=Z_G(\Bb)$ to denote the partition function of $G$.

For vectors $\alphab=\{\alpha_i\}_{i\in [q]},\betab=\{\beta_j\}_{j\in [q]}$ and a matrix $\xb=\{x_{ij}\}_{i,j\in [q]}$, let $\mathcal{M}=\mathcal{M}(\Bb)$ denote the space of all vectors  $\alphab=\{\alpha_i\}_{i\in [q]},\betab=\{\beta_j\}_{j\in [q]}$ and matrix $\xb=\{x_{ij}\}_{i,j\in [q]}$ such that
\begin{equation*}
\begin{aligned}
\mbox{$\sum_{i}$}\,\alpha_i=1,&\quad \mbox{$\sum_{j}$}\, \beta_j=1,\quad \alpha_i,\beta_j\geq 0,\\
\mbox{$\sum_{j}$}\, x_{ij}=\alpha_i\ \ (i\in [q]),& \quad \mbox{$\sum_{i}$}\, x_{ij}=\beta_j \ \ (j\in [q]), \quad x_{ij}\geq 0, x_{ij}=0 \text{ if } B_{ij}=0.
\end{aligned}
\end{equation*}
For a probability vector $\mathbf{p}=\{p_i\}$, we  use $H(\mathbf{p})=-\sum_{i}p_i\log p_i$ to denote its entropy, and similarly for a matrix $\mathbf{W}=\{W_{ij}\}_{ij}$ with nonnegative entries that sum to 1. Throughout we use the conventions $0\log 0\equiv 0, \log 0\equiv -\infty$.

Let
$Z_G^{\alphab,\betab,\xb}$ be the total weight of configurations
having $n\alpha_i$ vertices of colour $i$ in $V^\pl$,
 $n\beta_j$ vertices of colour $j$ in $V^\mi$ and $dn x_{ij}$ edges from colour $i$ vertices in $V^\pl$ to colour $j$ in $V^\mi$.
An edge from colour $i$ in $V^\pl$ to colour $j$ in
$V^\mi$ contributes weight $B_{ij}$, and parallel edges
are counted with multiplicity. 

\begin{lemma}\label{lem:first-moment-bipartite}
Fix integers $q\geq 2,d\geq 3$ and a nonzero matrix
$\Bb\in\mathbb{R}_{\geq 0}^{q\times q}$.
Let $C=10q^3 d$, and $A= R^{-5d q^2}$ where $R$ is  the ratio of the max over the min positive entries of $\Bb$.

For all  $(\alphab,\betab,\xb)\in \mathcal{M}$ with $n\alphab,n\betab,dn\xb$ integral, it holds that
\begin{equation}\label{eq:first-moment-polynomial}
 n^{-C}\emm^{n\Phi(\alphab,\betab,\xb)}
\leq
\mathbb{E}\big[Z_{\bG}^{\alphab,\betab,\xb}\big]
\leq
n^C\emm^{n\Phi(\alphab,\betab,\xb)}.
\end{equation}
where $\Phi(\alphab,\betab,\xb)
=-(d-1)\bigl(H(\alphab)+H(\betab)\bigr)
+dH(\xb)
+d\sum_{i,j:B_{ij}>0}x_{ij}\log B_{ij}$. 

Moreover, for any real $(\hat \alphab,\hat\betab,\hat\xb)\in \mathcal{M}$ and all sufficiently large $n$, there are  $(\alphab,\betab,\xb)\in \mathcal{M}$ with $n\alphab,n\betab,dn\xb$ integral and $\|(\alphab,\betab,\xb)-(\hat\alphab,\hat\betab,\hat \xb)\|_{\infty}\leq 1/n$ such that $\mathbb{E}\big[Z_{\bG}^{\alphab,\betab,\xb}\big]\geq A n^{-C}\emm^{n\Phi(\hat\alphab,\hat\betab,\hat\xb)}$.
\end{lemma}

GSV reformulated the first moment via matrix norms. Following their notation, for nonzero vectors $\rb=(R_1,\hdots,R_q)$ and $\cb=(C_1,\hdots,C_q)$ define 
\begin{equation}\label{eq:defPsi}
\Psi(\rb,\cb)=d\ln\frac{\rb^T \Bb \cb}{\|\rb\|_p \|\cb\|_p}, \quad \|\Bb\|_{p\rightarrow d}=\max_{\cb\geq 0}\frac{\|\Bb \cb\|_d}{\|\cb\|_p}
\end{equation}
The quantity $\|\Bb\|_{p\rightarrow d}$ is known as the induced matrix norm of $\Bb$. 
Using H\"older's inequality, it is not hard to see that $\max_{\rb,\cb} \Psi(\rb,\cb)=d\ln\|\Bb\|_{p\rightarrow d}$. We refer to the vectors $\cb$ which achieve the maximum in \eqref{eq:defPsi} as the maximizers of $\|\Bb\|_{p\rightarrow d}$.

\begin{lemma}[{\cite[Lemma 3.1\& Theorem 4.1]{AndreasBIP}}]\label{le:gsvl}
Fix integers $q\geq 2,d\geq 3$, and let $p=d/(d-1)$. Consider a matrix
$\Bb\in\mathbb{R}_{\geq 0}^{q\times q}$ whose bipartite support is connected. Then, 
\[\max_{(\alphab,\betab,\xb)\in \mathcal{M}} \Phi(\alphab,\betab,\xb) =\max_{\rb,\cb\geq 0}  \Psi(\rb,\cb)=d\ln \|\Bb\|_{p\rightarrow d},\] and  the maxima of the functions $\Phi$ and $\Psi$  are in  correspondence via 
\[\alpha_i=\frac{R_i\big(\sum_{j} B_{ij}C_j\big)}{\rb^T \Bb \cb}  \quad (i\in [q]), \quad \beta_j=\frac{C_j\big(\sum_{i} B_{ij}R_i\big)}{\rb^T \Bb \cb} \quad (j\in [q]), \quad x_{ij}=\frac{B_{ij}R_i C_j}{\rb^T \Bb \cb} \quad (i,j\in [q])\]
Further, the maxima of $\Psi_B(\rb,\cb)$ occur at fixpoints of the tree recursions, and satisfy in particular
\[R_i \propto \bigg(\sum_{j} B_{ij}C_j\bigg)^{d-1} \quad (i\in [q]),\qquad  C_j \propto \bigg(\sum_{i} B_{ij}R_i\bigg)^{d-1} \quad (j\in [q]).\]
\end{lemma}

\begin{lemma}[{\cite[Lemma~3.2]{AndreasBIP}}]
Let $\Bb$ be a matrix with nonnegative entries with $\mathrm{det}\Bb\neq 0$. 

Then, for any $1\leq p < r<\infty$, it holds that $\|\Bb\otimes  \Bb\|_{p\rightarrow r}=(\|\Bb\|_{p\rightarrow r})^2$ and the maximizers of $\|\Bb\otimes  \Bb\|_{p\rightarrow r}$ are of the form $\mathbf{z}_1\otimes \mathbf{z}_2$ where $\mathbf{z}_1, \mathbf{z}_2$ are maximizers of $\|\Bb\|_{p\rightarrow r}$ .
\end{lemma}

\subsection{Hard-core: Proof of Lemmas~\ref{lem:ZboundsHC} and~\ref{lem:statistics}}
Here we prove the remaining lemmas that we used for the hard-core model, namely Lemmas~\ref{lem:ZboundsHC} and~\ref{lem:statistics}, that we restate here for convenience.

Recall that, for a bipartite graph $G=(V^\pl \sqcup V^\mi,E)$,  we use $Z_{G,s}$ for the full partition function with vertex activities $\lambda \emm^{\pm s}$ on $V^\plm$, whereas $Z^\pl_{G,s}$ for the plus-phase partition function. 
\begin{lemZbound}
\statelemZboundsHC
\end{lemZbound}

For the vertex-edge statistics lemma, recall that the random vectors $\gammab^\pl,\deltab^\pl, \yb^\pl$ are the statistics of the double-planted model, whereas $\gamma^\pl=\gamma^\pl(s), \delta^\pl=\delta^\pl(s),y^\pl=y^\pl(s)$ are the vectors
\begin{equation*}\tag{\ref{eq:gdyoptimal}}
\gamma^\pl=\big[\begin{smallmatrix}1-\alpha\\ \alpha\end{smallmatrix}\big]\otimes \big[\begin{smallmatrix}1-\alpha\\ \alpha\end{smallmatrix}\big], \quad \delta^\pl=\big[\begin{smallmatrix}1-\beta\\ \beta\end{smallmatrix}\big]\otimes \big[\begin{smallmatrix}1-\beta\\ \beta\end{smallmatrix}\big], \quad y^\pl=\big[\begin{smallmatrix} 1-\alpha-\beta& & \beta\\ \alpha & & 0\end{smallmatrix}\big]\otimes \big[\begin{smallmatrix} 1-\alpha-\beta& & \beta\\ \alpha & & 0\end{smallmatrix}\big].
\end{equation*}
where $1>\alpha>\beta>0$ satisfy the equations $\frac{\alpha}{1-\alpha}=\lambda \emm^{s} \bigg(\frac{1-\alpha-\beta}{1-\alpha}\bigg)^{d}, \quad \frac{\beta}{1-\beta}=\lambda \emm^{-s} \bigg(\frac{1-\alpha-\beta}{1-\beta}\bigg)^{d}$.
\begin{lemstatistics}
\statelemstatistics
\end{lemstatistics}
\begin{proof}
On a $d$-regular bipartite graph, the hard-core model with vertex activities $\lambda \emm^s$ on one side and $\lambda \emm^{-s}$ on the other (cf. ~\eqref{eq:weightHC}) can be viewed as the two-spin model with interaction matrix
$\Bb_s=\big[\begin{smallmatrix}
        1 &(\lambda \emm^{-s})^{1/d}\\
        (\lambda \emm^s)^{1/d}&0
      \end{smallmatrix}\big]$.
Let $\Psi(s)=d\log \|\Bb_s\|_{p\rightarrow d}$ be the maximum of the function in \eqref{eq:defPsi} corresponding to the matrix $\Bb_s$. Similarly let  $\Phi_s$ be the function $\Phi$ corresponding to the interaction matrix $\Bb_s$, and note that the hard-core interaction matrix gives the equality  $\Phi_s(\alphab,\betab,\xb)=\Phi_{0}(\alphab,\betab,\xb)
                   +s(\alpha_1-\beta_1)$.
 It follows that $\Psi(s)$ is convex in $s$ since, by Lemma~\ref{le:gsvl},  it is  the maximum of
\[
\max_{\alphab,\betab,\xb}\Phi_s(\alphab,\betab,\xb)=\max_{\alphab,\betab,\xb}\{\Phi_{0}(\alphab,\betab,\xb)
                   +s(\alpha_1-\beta_1)\},
\]
and pointwise maximum of convex functions is convex. Note that $\Psi(s)$ is even (by swapping the sides) and hence it is nondecreasing for $s\geq0$.

Let  $\alphab_0=\big[\begin{smallmatrix}
    1-\alpha_0\\ \alpha_0
\end{smallmatrix}\big],\betab_0=\big[\begin{smallmatrix}
    1-\beta_0\\ \beta_0
\end{smallmatrix}\big]$ be the maximizing fixed point at $s=0$ (it exists since $\lambda$ is in the non-uniqueness region). 
Using corresponding $\alpha,\beta$  in $\Phi_s$ we
have for $s\geq 0$
\begin{equation}\label{eq:hc-psi-growth}
 \Psi(s)\geq\Psi(0)+\kappa s,
 \qquad \kappa:=\alpha_0-\beta_0>0.
\end{equation}

We next prove the deterministic lower bound by using Ruozzi's theorem~\cite{Ruozzi} that bounds the partition function for log-supermodular interaction matrices by the Bethe entropy. We give the relevant definitions first.  A Bethe assignment $\mathbf{b}$ consists of a probability vector
$b_v$ at each vertex and a probability matrix $b_{uv}$
at each edge, whose marginals are $b_u,b_v$ and which
is supported on the allowed spin pairs.
Its Bethe free entropy is
\[
 \Phi_{\mathrm{Bethe},G,s}(\mathbf b)
 =
 \sum_{\substack{uv\in E;u\in V^\pl,\ v\in V^\mi}}
 \bigg(
  H(b_{uv})
  +\sum_{i,j:\,(B_s)_{ij}>0}
       b_{uv}(i,j)\log(B_s)_{ij}
 \bigg)
 -(d-1)\sum_{v\in V}H(b_v),
\]
and $Z_{\mathrm{Bethe}}(G,s)
 =
 \exp\left(
   \max_{\mathbf b}\Phi_{\mathrm{Bethe},G,s}(\mathbf b)
 \right)$.
These beliefs are required to be locally consistent;
they need not be marginals of a distribution on
configurations. Given any feasible $(\alphab,\betab,\xb)$, assign
$b_u=\alphab$ on $V^\pl$, $b_v=\betab$ on $V^\mi$,
and $b_{uv}=\xb$ on every edge.
The marginal constraints make this a feasible Bethe
assignment. Since there are $n$ vertices on each side
and $dn$ edges, its value is exactly
\[
 \Phi_{\mathrm{Bethe},G,s}(\mathbf b)
 =
 n\Phi_s(\alphab,\betab,\xb).
\]
In particular,
$Z_{\mathrm{Bethe}}(G,s)\geq\emm^{n\Psi(s)}$. To apply Ruozzi's theorem, complement the spin on
$V^\mi$. This exchanges the columns of $\Bb_s$,
giving
$\widetilde\Bb_s=
 \big[\begin{smallmatrix}
  (\lambda\emm^{-s})^{1/d}&1\\
  0&(\lambda\emm^s)^{1/d}
 \end{smallmatrix}\big]$. This interaction is log-supermodular, since
 $(\widetilde B_s)_{00}(\widetilde B_s)_{11}
 \geq
 (\widetilde B_s)_{01}(\widetilde B_s)_{10}$.
The spin complementation is a bijection on
configurations and on Bethe assignments, preserving
both their weights and their Bethe free entropies.
Ruozzi's theorem~\cite{Ruozzi} therefore gives
\begin{equation}\label{eq:hc-ruozzi-lower}
 Z_{G,s}\geq Z_{\mathrm{Bethe}}(G,s)
          \geq\emm^{n\Psi(s)}.
\end{equation}

The first moment upper bound follows directly from Lemma~\ref{lem:first-moment-bipartite} (summing over $\alphab,\betab,\xb$)
\[
 \mathbb E_G[Z_{G,s}]
           \leq  n^{100d} \emm^{n\Psi(s)}.
\]
For the second moment, we can view $Z_{G,s}^2=\sum_{\sigma_1,\sigma_2} w_{G,s}(\sigma_1) w_{G,s}(\sigma_2)$ as the partition function of a spin system with interaction matrix  $\Bb_s\otimes \Bb_s$. So we can write $(Z_{G,s})^2=\sum_{\gammab,\deltab,\yb} Z^{\gammab,\deltab,\yb}_G$ where $\gammab=\gamma_{ik},\deltab=\delta_{jl},\yb={y_{ik,jl}}$ range over all possible vertex-edge statistics of $(\sigma_1,\sigma_2)$. In particular, we have that $Z_G(\Bb_s\otimes \Bb_s)=\big(Z_{G,s}\big)^2$. Using therefore the tensorisation identity $\|\Bb_s\otimes \Bb_s\|_{p\rightarrow d}=(\| \Bb_s\|_{p\rightarrow d})^2$ and applying Lemma~\ref{lem:first-moment-bipartite} to the corresponding $4$-state system with 
$\Bb_s\otimes \Bb_s$ yields therefore that
\[
 \mathbb E_G[Z_{G,s}^2]
           \leq n^{600d}  \emm^{2 n\Psi(s)}.
\]
Finally, using that $0\leq Z_{G,s}^\plm\leq Z_{G,s}$, we observe that these moment bounds extend to $Z_{G,s}^\plm$. 

To obtain the desired concentration of $\gammab,\deltab,\yb$ we need to focus a bit more closely on the vectors $\gammab,\deltab,\yb$ that contribute to $(Z^\pl_{G,s})^2$. Note that the row marginals of $\gammab,\deltab$ correspond to the vertex-statistics of $\sigma_1$ on $V^\pl, V^\mi$; the column marginals correspond to the vertex-statistics of $\sigma_2$ on $V^\pl, V^\mi$, so to ensure that $\sigma_1\in \cI^\pl$ we need to impose that 
$\sum_{k}\gamma_{1k}\geq \sum_{l}\delta_{1l}$
and similarly for $\sigma_2\in \cI^\pl$ we need the constraint $\sum_{i}\gamma_{i1}\geq \sum_{j}\delta_{j1}$.

By the final part of Lemma~\ref{le:gsvl} the maximizers for $\Psi(\rb,\cb)$ satisfy 
\[
 \frac{R_1}{R_0}
 =
 \left(
  \frac{(\lambda\emm^s)^{1/d}C_0}
       {C_0+(\lambda\emm^{-s})^{1/d}C_1}
 \right)^{d-1},
 \qquad
 \frac{C_1}{C_0}
 =
 \left(
  \frac{(\lambda\emm^{-s})^{1/d}R_0}
       {R_0+(\lambda\emm^s)^{1/d}R_1}
 \right)^{d-1}.
\]
Writing $\alpha=\alpha_1$ and $\beta=\beta_1$, we have $x_{00}=1-\alpha-\beta$, so 
the correspondence in~\Cref{le:gsvl} gives
\[
 \frac{\alpha}{1-\alpha}=\frac{R_1}{R_0}
 \frac{(\lambda\emm^s)^{1/d}C_0}
      {C_0+(\lambda\emm^{-s})^{1/d}C_1}
 \qquad
  \frac{\beta}{1-\beta}=
 \frac{C_1}{C_0}
 \frac{(\lambda\emm^{-s})^{1/d}R_0}
      {R_0+(\lambda\emm^s)^{1/d}R_1},
 \qquad
 \frac{1-\alpha-\beta}{1-\alpha}=\frac{C_0}{C_0+(\lambda\emm^{-s})^{1/d}C_1}.
\]
From these, it is not hard to conclude that the maximizing occupation probabilities satisfy
the equations in~\Cref{lem:alphabeta}.

We next identify which solutions can be maximizers.
For $s>0$, exchanging the two sides changes
$(\alphab,\betab,\xb)$ to
$(\betab,\alphab,\xb^T)$, and
\[
 \Phi_s(\alphab,\betab,\xb)
 -\Phi_s(\betab,\alphab,\xb^T)
 =2s(\alpha-\beta).
\]
Hence a maximizer cannot have $\alpha<\beta$.
The fixed-point equations also exclude
$\alpha=\beta$ when $s>0$.
By~\Cref{lem:alphabeta}, there is a unique solution
with $\alpha>\beta$, so this is the unique global
maximizer. At $s=0$, it is well-known (see for example \cite{DFJ02,Sly}) that there are exactly two global
maximizers, the plus solution and the solution
obtained by exchanging the two sides (there is also a symmetric solution with $\alpha=\beta$ which is not a local maximum).

We now apply the tensor-maximizer lemma to
$\Bb_s\otimes\Bb_s$.
Every maximizing input vector is a tensor product
of two maximizing input vectors for $\Bb_s$.
The corresponding left vector is also a tensor
product, by equality in H\"older's inequality.
Therefore, by~\Cref{le:gsvl}, every maximizing
statistics triple has the form
\[
 \gamma=\alphab^{(1)}\otimes\alphab^{(2)},
 \qquad
 \delta=\betab^{(1)}\otimes\betab^{(2)},
 \qquad
 y=\xb^{(1)}\otimes\xb^{(2)},
\]
where each
$(\alphab^{(j)},\betab^{(j)},\xb^{(j)})$
is a maximizing triple for $\Bb_s$.

For $s>0$, both factors must be the plus maximizer.
For $s=0$, there are four possibilities, according
to whether each factor is the plus or minus
maximizer. The two plus-phase inequalities imposed
above require each factor to be the plus maximizer.
Thus, for every $s\geq0$, the unique maximizer
among statistics satisfying these inequalities is
$ (\gamma^\pl(s),\delta^\pl(s),y^\pl(s))$, and its value is $2\Psi(s)$.

We can now conclude the concentration statement
directly from the first-moment estimates.
The maximizing triple is continuous in $s$,
by continuity of the objective and uniqueness
on the closed plus-phase feasible set.
Fix $\eps>0$.
Compactness gives a constant $c_\eps>0$ such that,
uniformly over $s\in[0,s_0]$, 
\[
 \Phi_{\Bb_s\otimes\Bb_s}(\gamma,\delta,y)
 \leq 2\Psi(s)-c_\eps
\]
for every feasible triple satisfying both
plus-phase inequalities and $\|\gamma-\gamma^\pl\|_1+
 \|\delta-\delta^\pl\|_1+
 \|y-y^\pl\|_1\geq\eps.$
By~\Cref{lem:first-moment-bipartite}, and since
there are polynomially many possible statistics
triples, the total expected weight of these pairs is at most 
$n^{O(1)}\emm^{n(2\Psi(s)-c_\eps)}$. 

For the normalizing constant, apply the rounding
in~\Cref{lem:first-moment-bipartite}
directly to
$(\gamma^\pl,\delta^\pl,y^\pl)$ for the matrix
$\Bb_s\otimes\Bb_s$.
Each of the two phase inequalities has slack
$\alpha(s)-\beta(s)\geq\zeta$ by~\Cref{lem:alphabeta},
so the rounded triple still corresponds to two
plus-phase configurations for sufficiently large $n$.
Its expected weight therefore gives
$\mathbb E[(Z^\pl_{\bG,s})^2]
 \geq n^{-O(1)}\emm^{2n\Psi(s)}$, uniformly over $s\in[0,s_0]$ (note that the prefactor from the rounding of~\Cref{lem:first-moment-bipartite} is uniformly bounded below because $s$ ranges over a compact interval). By definition of the double phase-planted law,
the probability of exceptional statistics is
their total expected weight divided by
$\mathbb E[(Z^\pl_{\bG,s})^2]$.
Consequently,
\[
 \Pr\left[
 \|\gammab^\pl-\gamma^\pl\|_1+
 \|\deltab^\pl-\delta^\pl\|_1+
 \|\yb^\pl-y^\pl\|_1\geq\eps
 \right]
 \leq n^{O(1)}\emm^{-c_\eps n}
 \leq\emm^{-cn}
\]
for some $c>0$, uniformly over $s\in[0,s_0]$.
This proves~\Cref{lem:statistics}.
\end{proof}
\stopmark

\subsection{Partition function bounds for the reference Ising model}\label{sec:Isinglower}
Here we prove Lemmas~\ref{lem:ZboundsRC} and~\ref{lem:statisticsRC} for the reference Ising model of Section~\ref{sec:refIsing} which was used in the random-cluster arguments.
\begin{lemZboundsRC}
\statelemZboundsRC
\end{lemZboundsRC}
\begin{lemstatisticsRC}
\statelemstatisticsRC
\end{lemstatisticsRC}
\begin{proof}
Recall that $Q=q-1$, $B=\emm^\beta$ and
$\hat B=1+(B-1)/Q$. We have  $Z_G=\sum_\sigma w_G(\sigma)$ where $w_G(\sigma)=Q^{|\sigma^{-1}(0)|}\hat B^{|M_0(\sigma)|} B^{|M_1(\sigma)|}$.  On a $d$-regular graph, this Ising model  can be viewed as the two-spin model with interaction matrix
$\Bb=\big[\begin{smallmatrix}
        Q^{2/d} \hat B & Q^{1/d}\\
        Q^{1/d}&  B
      \end{smallmatrix}\big]$. Note that $\Bb$ is positive definite, since
$\det\Bb=Q^{2/d}(B\hat B-1)>0$. This will allow us to use the first/second moment ingredients from \cite{BIShard}, suitably applied. We first review the most important pieces for a general ferromagnetic matrix $\Bb$ and then explain how to apply them in our setting.

Analogously to Lemma~\ref{lem:first-moment-bipartite}, we decompose $Z_G=\sum_{\alphab,\xb} Z_G^{\alphab,\xb}$. From \cite[Section 4]{BIShard}, using  the same multinomial estimates as in
Lemma~\ref{lem:first-moment-bipartite}, we have 
\begin{equation}\label{eq:g65g543r4}
\E[Z_{\bG}^{\alphab,\xb}]
 =n^{\Theta(1)}
   \emm^{n\Phi_{\Bb}(\alphab,\xb)} \mbox{ where } \Phi_\Bb(\alphab,\xb)
 =-(d-1)H(\alphab)
   +\frac d2H(\xb)
   +\frac d2\sum_{i,j}x_{ij}\log B_{ij}.
\end{equation}
Throughout, the sums over statistics range over
admissible integral tables. The constants in the
polynomial factors can be chosen uniformly when
$\beta$ is bounded above.

Let $p=d/(d-1)$. By the variational correspondence
in \cite[Section~5.1]{BIShard}, the maximum of
$\Phi_\Bb$ equals the maximum of $\Psi_\Bb(\rb)
 =\frac d2\log
   \frac{\rb^{\mathsf T}\Bb\rb}{\|\rb\|_p^2}$ in the space $\rb\geq 0$ with $\|\rb\|=1$. In fact,  their maximizers occur at stationary points, which  are in correspondence via
\begin{equation}\label{eq:rfvvrte345t}
 \alpha_i
 =\frac{R_i\sum_{j}B_{ij}R_j}{\rb^{\mathsf T}\Bb\rb},
 \qquad
 x_{ij}
 =\frac{B_{ij}R_iR_j}{\rb^{\mathsf T}\Bb\rb}
\end{equation} 
 In \cite[Section~6 \& Lemma 40]{BIShard}, it is further shown that the stationary points of $\Psi$ satisfy  the tree recursion
$R_i\propto(\sum_{j} B_{ij}R_j)^{d-1}$ and that   a stationary point of $\Psi$ has negative definite Hessian iff the fixpoint of the tree recursions $\rb$ has Jacobian with eigenvalues in $(-1,1)$ iff the matrix $\mathbf{M}=\{\frac{x_{ij}}{\alpha_i}\}$ has its nontrivial eigenvalues strictly less than $1/(d-1)$ (note that $\mathbf{M}$ is stochastic so its maximum eigenvalue is equal to 1, and hence we refer to the other eigenvalues as being nontrivial).

We now apply these results in the setting here. Recall that $Z_G^\pl$ is the contribution of configurations $\sigma$ to the partition function from $\sigma$ with at least as many pluses as minuses, so we have $Z_G^\pl=\sum_{\alphab,\xb} Z_G^{\alphab,\xb}$ where the sum ranges over $\alphab,\xb$ such that $\alpha_1\geq \alpha_0$. For convenience, below we will refer to the set of such $\alpha$'s as $\Sigma^\pl$ as well. We decompose $Z_G^\mi$ analogously.

From the fixpoint equations for $\rb=(R_0,R_1)$, we obtain by setting $R_1/R_0=(t/Q^{1/d})^{d-1}$ that 
\begin{equation}\label{eq:Ising-reference-BP}
 t=\frac{Bt^{d-1}+Q}{t^{d-1}+B+Q-1}.
\end{equation}
Thus $t=1$ gives a fixed point, while for $t\ne1$
the equation is equivalent to $B-1
 =\frac{(t-1)(t^{d-1}+Q)}{t^{d-1}-t}$.
This gives the  vertex marginal $\alpha=\frac{t^d}{t^d+Q}$ and, with $a=\frac{t^{d-1}}{t^{d-1}+Q}$, the edge statistics 
\[
 x=\frac1X
 \begin{pmatrix}
 \hat B(1-a)^2 & a(1-a)\\
 a(1-a) & Ba^2
 \end{pmatrix},
 \qquad
 X=\hat B(1-a)^2+2a(1-a)+Ba^2.
\]
that were given back in \eqref{eq:xoptimal}. The matrix $\mathbf{M}$ capturing stability is therefore given by $\mathbf{M}=\big[\begin{smallmatrix} \frac{x_{00}}{1-\alpha} & \frac{x_{01}}{1-\alpha}\\ \frac{x_{01}}{\alpha} & \frac{x_{11}}{\alpha}\end{smallmatrix}\big]$ with eigenvalues $1$ and $\tau=1-\frac{x_{01}}{\alpha(1-\alpha)}$.

We next identify the relevant vertex-edge statistics that have the maximum contribution to each of $\E[Z_{\bG}^\pl],\E[Z_{\bG}^\mi]$. Equation~\eqref{eq:Ising-reference-BP} is equivalent to
\[
 t^d-Bt^{d-1}+(B+Q-1)t-Q=0,
\]
which has at most three positive roots by Descartes'
rule of signs.  We have that $t=1$ is always a solution, and it is a double root when $\beta=\beta_s$; the corresponding fixpoint is Jacobian attractive for $\beta<\beta_s$. In \cite[Lemma 17]{BIShard}, it is shown that there are solutions with $t\neq 1$ iff $\beta\geq \beta_u$. In fact,  for $\beta\in [\beta_u,\beta_s)$ there are two solutions with $t>1$ (technically, at $\beta=\beta_u$ the solution is a double root), and for $\beta>\beta_s$  there is one solution with $t>1$ and one solution with $t<1$; of these, only the one with $t>1$ gives $\alpha>1/2$. 
For $\beta\in (\beta_u,\beta_s)$, only the one with the larger $t$ corresponds to a Jacobian stable fixpoint.  For $\beta>\beta_s$ the solution with $t<1$ is stable (but as we will see soon its $\Phi$-value is less than that of the solution with $t>1$). At  $B_c=\emm^{\beta_c}
 =\frac{Q-1}{Q^{1-2/d}-1}$, the interaction matrix $\Bb$ has equal diagonal entries and the solutions are by inspection $t=1$, $t=Q^{1/d}$ and $t=Q^{2/d}$ which give $\alpha=1/q$, $\alpha=1/2$ and $\alpha=(q-1)/q$,
respectively.

To determine which fixed point is dominant, note that on an $n$-vertex $d$-regular graph $G$,  we have  for $\sigma\in \{0,1\}^V$  that $|\sigma^{-1}(0)|=n-|\sigma^{-1}(1)|$ and $|M_1(\sigma)|-|M_0(\sigma)|=d(|\sigma^{-1}(1)|-n/2)$. So, we can rewrite the weight function $w_G(\sigma) =Q^{|\sigma^{-1}(0)|}\hat B^{|M_0(\sigma)|} B^{|M_1(\sigma)|}$ as \[A\, h^{|\sigma^{-1}(1)|} W^{|M_0(\sigma)|+|M_1(\sigma)|} \mbox{ where } A=Q^n (\hat B/B)^{dn/4},\ \ h=h(\beta)=\tfrac{1}{Q}(B/\hat B)^{d/2}, \mbox{ and } W=(B \hat B)^{1/2}.\] 
As a function of $\beta$, $h(\beta)$ is strictly increasing and is equal to 1 at $\beta=\beta_c$.
Hence spin flips show that every global maximizer needs to align
with the field. Since the field-aligned solution is
unique, $(\alphab^\pl,\xb^\pl)$ is the unique global
maximizer for $\beta>\beta_c$, the $(\alphab^\mi,\xb^\mi)$
solution is the unique global maximizer for
$\beta<\beta_c$; at $\beta=\beta_c$, the only global maximizers are $(\alphab^\pl,\xb^\pl)$ and $(\alphab^\mi,\xb^\mi)$ and hence  $\Psi^\mi(\beta_c)=\Psi^\pl(\beta_c)$. We can further conclude that, for small $\delta>0$, $(\alphab^\mi,\xb^\mi)$ is the unique maximizer within $\Sigma^\mi$ for $\beta<\beta_c+\delta$ and similarly  $(\alphab^\pl,\xb^\pl)$ is the unique maximizer within $\Sigma^\pl$ for $\beta>\beta_c-\delta$.

For the second moment, we have $(Z^\plm_G)^2=\sum_{\sigma_1,\sigma_2\in \Sigma^\plm}w_G(\sigma_1)w_G(\sigma_2)$. For  a pair of configurations, the product of their weights is the same as that of the weight in a paired spin model  with interaction matrix
$\Bb\otimes\Bb$. Analogously to before, we decompose $(Z^\plm_G)^2=\sum_{\gammab,\yb} Z^{\gammab,\yb}_G$ where the sum ranges over vertex-edge statistics $\gammab,\yb$ with $\gammab\in \Sigma^\plm_2$ (here,  $\Sigma^\pl_2,\Sigma^\mi_2$ denote the sets of all vectors $\gammab$ whose column and row marginals lie inside $\Sigma^\pl,\Sigma^\mi$ respectively). We therefore have that $\E[Z^{\gammab,\yb}_G]=n^{\Theta(1)}\emm^{n\Phi_{\Bb\otimes\Bb}(\gammab,\yb)}$. 
The tensor-norm argument in
\cite[Section~5.2]{BIShard}  shows
that all maximizers for $\Phi_{\Bb\otimes\Bb}$ are tensor products of maximizers of
$\Phi_{\Bb}$. Then, using the condition that $\gammab$ lies in $\Sigma_2^\pl$, we obtain that for $\beta\geq \beta_c$ the only maximizer over $\gammab,\yb$ with $\gammab\in \Sigma^\pl_2$ is $(\gammab^\pl,\yb^\pl)=(\alphab^\pl\otimes\alphab^\pl,\,
  \xb^\pl\otimes\xb^\pl)$. Similarly, for $\beta\leq \beta_c$ the only maximizer over $\gammab,\yb$ with $\gammab\in \Sigma^\mi_2$ is $(\gammab^\mi,\yb^\mi)=(\alphab^\mi\otimes\alphab^\mi,
  \xb^\mi\otimes\xb^\mi)$.

 We next extend these second-moment conclusions to a neighbourhood
of $\beta_c$. Consider the plus phase for $\beta\in (\beta_c-\delta,\beta_c)$ for some small $\delta>0$ to be specified shortly, the case of the minus phase is analogous.
We use  $\rb^\pl$ to denote the fixpoint of the tree recursions corresponding to $(\alphab^\pl,\xb^\pl)$.  Since the phase fixpoint $\rb^\pl$ is strictly stable for the $\Bb$-fixed-point equations, we conclude that the same is true for any $\beta$ with $|\beta-\beta_c|< \delta'$ where $\delta'>0$ is a sufficiently small constant (depending on $\eps'$). It follows that for $\beta>\beta_c-\delta'$ the $\Bb\otimes \Bb$-fixed-point equation  is also a contraction in an $\eps'$-neighbourhood of $\rb^\pl\otimes \rb^\pl$.  Indeed, the relevant matrix for the Jacobian stability is $\mathbf{M}\otimes \mathbf{M}$ and hence has eigenvalues $1,\tau,\tau,\tau^2$; since $(d-1)\tau<1$ all the nontrivial eigenvalues are less than $1/(d-1)$. It follows that the $\Bb\otimes \Bb$-tree-recursion has the unique fixed point $\rb^\pl\otimes \rb^\pl$ in that neighbourhood. From this, we conclude that, for some small constant $\eps>0$, the function $\Phi_{\Bb\otimes \Bb}(\gammab,\yb)$ has a unique maximum in an $\eps$-neighbourhood of $(\gammab^\pl,\yb^\pl)=(\alphab^\pl\otimes \alphab^\pl,\xb^\pl\otimes \xb^\pl)$ for any $\beta>\beta_c-\delta'$; this maximum equals $2\Psi^\pl(\beta)$ and is attained at $(\gammab^\pl,\yb^\pl)$. So it only remains to exclude maximizers outside of that $\eps$-neighbourhood.

 At $\beta=\beta_c$ we have $(\gammab_c^\pl,\yb^\pl_c)$  is the unique maximum of $\Phi_{\Bb_c\otimes \Bb_c}$ over all $\gammab,\yb$ with $\gamma\in  \Sigma^\pl_2$. By uniform continuity and compactness there are $\delta'', \zeta>0$ so that, for any $\beta\in [\beta_c-\delta'',\beta_c]$, it holds that $\Phi_{\Bb\otimes \Bb}(\gammab,\yb)\leq\Phi_{\Bb_c\otimes \Bb_c}(\gammab^\pl_c,\yb^\pl_c)-\zeta$ for all $\gammab,\yb$ with $\gammab\in \Sigma^\pl_2$ which are $(\eps/2)$-far from $(\gammab_c^\pl,\yb^\pl_c)$. Choose now $\delta<\delta',\delta''$ so that  $(\gammab^\pl,\yb^\pl)$ is $(\eps/2)$-close to $(\gammab_c^\pl,\yb^\pl_c)$ and $\Phi_{\Bb\otimes\Bb}(\gammab^\pl,\yb^\pl)$ is $(\zeta/2)$-close to $\Phi_{\Bb_c\otimes\Bb_c}(\gammab^\pl_c,\yb^\pl_c)=2\Psi^\pl(\beta_c)$. Then, for any $\beta\in (\beta_c-\delta,\beta_c)$, we conclude that $\Phi_{\Bb\otimes \Bb}(\gammab,\yb)\leq 2\Psi^\pl(\beta)-\zeta/2$ for all $(\gammab,\yb)$ which are $\eps$-far from $(\gammab^\pl,\yb^\pl)$. This completes the proof that there are no maximizers other than $(\gammab^\pl,\yb^\pl)$.

To conclude the bounds on $\E[Z^\plm_{\bG}]$ and $\E[(Z^\plm_{\bG})^2]$, we observe that there are only $n^{O(1)}$ many vertex and edge
statistics,
so  using
\eqref{eq:g65g543r4}, we obtain for some constant $C>0$ and any $\beta\in[\beta_c-\delta,\beta_0]$ that
\[
 \mathbb E[Z^\pl_{\bG}]
 \leq n^C\emm^{n\Psi^\pl},
 \qquad
 \mathbb E[(Z^\pl_{\bG})^2]
 \leq n^C\emm^{2n\Psi^\pl}.
\]

For the deterministic lower bound, consider first the plus
phase, and let $\alphab=\alphab^\pl$ and $\xb=\xb^\pl$. From~\eqref{eq:rfvvrte345t} and the fixed-point equations, we have $x_{ij}=\frac{B_{ij}R_iR_j}{\rb^T\Bb\rb}$ and  $\alpha_i=R_i^{p}/\|\rb\|^p_p$ where $p=d/(d-1)$. Using $d$-regularity, we therefore have for every $\sigma\in\{0,1\}^V$ that
\begin{equation}\label{eq:helpxhelp}
 \prod_{\{u,v\}\in E}x_{\sigma(u),\sigma(v)}
 =
 (\rb^T\Bb\rb)^{-dn/2}
 w_G(\sigma)\prod_{v\in V}R_{\sigma(v)}^d
 =
 \emm^{-n\Psi^\pl}w_G(\sigma)
 \prod_{v\in V}\alpha_{\sigma(v)}^{d-1}.
\end{equation}
Rearranging, and using $d$-regularity once more, yields
\[
 w_G(\sigma)
 =
 \emm^{n\Psi^\pl}
 \frac{\prod_{\{u,v\}\in E}x_{\sigma(u),\sigma(v)}}
      {\prod_{v\in V}\alpha_{\sigma(v)}^{d-1}}
 =
 \emm^{n\Psi^\pl}
 \prod_{v\in V}\alpha_{\sigma(v)}
 \prod_{\{u,v\}\in E}
 \frac{x_{\sigma(u),\sigma(v)}}
      {\alpha_{\sigma(u)}\alpha_{\sigma(v)}}.
\]
For $\tau=1-x_{01}/(\alpha_0\alpha_1)$, we can further write 
\[
 x_{ij}=(1-\tau)\alpha_i\alpha_j
       +\tau\alpha_i\mathbf1_{\{i=j\}}.
\]
Note that $\tau\geq0$ (due to the ferromagnetic $\Bb$), while Jacobian stability of the fixed point gives $(d-1)\tau<1$. We therefore obtain that
\[
 w_G(\sigma)=\emm^{n\Psi^\pl}
 \prod_{v\in V}\alpha_{\sigma(v)}
 \prod_{\{u,v\}\in E}
 \left(1-\tau+
 \tau\frac{\mathbf1_{\{\sigma(u)=\sigma(v)\}}}
               {\alpha_{\sigma(u)}}\right).
\]

We can now expand the product over the edges to obtain the following. For a subset $F\subseteq E$, let $\mathcal{C}(F)$ denote the components of the graph $(V,F)$. We say that $\sigma$ is compatible with $F$ if, for every component $C\in\mathcal{C}(F)$, the vertices in $C$  have the same colour $\sigma_C$ under $\sigma$; let $G(\sigma)$ be the set of compatible edge sets. For compatible $F$ and $\sigma$, let $\hat w_F(\sigma)=\prod_{C\in \mathcal{C}(F)}(\alpha_{\sigma_C})^{1-\mathtt{exc}(C)}$ where recall that $\mathtt{exc}(C)$ is the tree excess of $C$. Then,
\[w_G(\sigma)=\emm^{n\Psi^\pl}\sum_{F\in G(\sigma)}(1-\tau)^{|E|-|F|}\tau^{|F|}\hat w_F(\sigma)\]
For $F\subseteq E$, let $\hat Z_F=\sum_{\sigma; F\in G(\sigma)} \hat w_F(\sigma)$ and $\hat  Z_F^\pl=\sum_{\sigma\in \Sigma^\pl; F\in G(\sigma)} \hat w_F(\sigma)$. Then, we can write 
\[
 \frac{Z_G^\pl}{\emm^{n\Psi^\pl}}
 = 
 \mathbb E_F\big[ \hat Z_F^\pl\big] \mbox{ where } F\sim \mathrm{Bin}(E,\tau).
\]

Fix $F\subseteq E$.  Note that, in $\hat w_F(\sigma)=\prod_{C}(\alpha_{\sigma_C})^{1-\mathtt{exc}(C)}$, cyclic components contribute a factor of at least 1, while the others a factor of $\alpha_{\sigma_C}$. Then, we can 
write 
\[
 \hat Z_F^\pl
 \geq
 \Pr[\bsigma\in\Sigma^\pl],
\]
where $\bsigma$ is the random configuration obtained by assigning
spin $1$ to every cyclic component, and spin $1$ to each
tree component with probability $\alpha_1$ (independently).
Let $N_1=|\bsigma^{-1}(1)|$. Then  $\E[N_1]\geq\alpha_1 n$ and $\Var[N_1]=\alpha_0\alpha_1
 \sum_{C\in \mathcal{C}(F); \mathtt{exc}(C)=0}|C|^2$. Choose $\theta>0$ with $\alpha_1-\theta>1/2$. Then, Chebyshev's inequality gives
\[
 \hat Z_F^\pl
 \geq \Pr[N_1\geq (\alpha_1-\theta) n]\geq 
 1-\frac{\alpha_0\alpha_1}{\theta^2n^2}
       \sum_{C\in\mathcal C(F)}|C|^2.
\]
For a vertex $v$, let $C_F(v)$ denote its component
in $(V,F)$. Then $\sum_{C}|C|^2=\sum_{v} |C_F(v)|$. Bounding the size of $C_F(v)$ by the number of paths with start-point $v$ and using $(d-1)\tau<1$,  we can bound 
\[
 \mathbb E_F\sum_{C\in\mathcal C(F)}|C|^2
 =
 \sum_{v\in V}\mathbb E_F|C_F(v)|
 \leq
 n\bigg(1+\sum_{k\geq1}d(d-1)^{k-1}\tau^k\bigg)
 =
 n\left(1+\tfrac{d\tau}{1-(d-1)\tau}\right).
\]
Averaging the preceding bound on $\hat Z_F^\pl$
therefore yields $Z_G^\pl\geq \frac{1}{2}\emm^{n\Psi^\pl}$
for all sufficiently large $n$.

For the minus phase, use its corresponding BP beliefs
and assign spin $0$ to every cyclic component.
Then $\mathbb E[N_1\mid F]\leq\alpha_1n$.
Choosing $\theta>0$ with $\alpha_1+\theta<1/2$,
the same argument bounds
$\Pr[N_1>(\alpha_1+\theta)n]$ and gives
$Z_G^\mi\geq\frac12\emm^{n\Psi^\mi}$.
The occupation and stability gaps established above
make these estimates uniform over the respective
parameter ranges.

Finally, the proof of the statistics lemma is
completely analogous to the hard-core model:
within either phase, the maximum of $\Phi_\Bb$
is achieved uniquely at
$(\alphab^\plm,\xb^\plm)$, and the maximum of
$\Phi_{\Bb\otimes\Bb}$ is achieved uniquely at
$(\gamma^\plm,y^\plm)$.
By compactness, the strict gaps away from these maximizers
are uniform over the respective parameter intervals.
Summing over the polynomial number of relevant vectors gives
the claimed exponential concentration
in the single- and double-phase-planted models.
\end{proof}

\subsection{Partition-function bounds for the ferromagnetic Ising model}
\label{sec:Isingfieldbounds}

\begin{proof}[Proof of Lemmas~\ref{lem:ZboundsIS}
and~\ref{lem:statisticsIS}]
The proofs are very close to those of Section~\ref{sec:Isinglower}, so we give the relevant arguments a bit more quickly, focusing on highlighting the main differences. Recall that $B=\emm^\beta$. We use the variational and BP
calculations from Section~\ref{sec:Isinglower} with interaction matrix
$\Bb_s=\big[\begin{smallmatrix}
 B\emm^{-s/d}&1\\
 1&B\emm^{s/d}
 \end{smallmatrix}\big]$.
On $d$-regular graphs, its configuration weight is precisely
$\emm^{\beta m(\sigma)+sM(\sigma)}$.
Moreover, $\det\Bb_s=B^2-1>0$, so the ferromagnetic
matrix-norm arguments apply.

We first identify the maximizers and verify their stability.
Writing $r=\emm^{s/d}R_1/R_0$, the BP equations
$R_i\propto(\sum_j(B_s)_{ij}R_j)^{d-1}$ become
$ r=\emm^s\left(\frac{1+Br}{B+r}\right)^{d-1}$, and the substitution $r=a/(1-a)$ gives the definition of $a$ in~\eqref{eq:IS-BP}.
To analyse its solutions with $r>1$, set
\[
 g(r)=\log r-(d-1)\log\frac{1+Br}{B+r},
 \qquad
 \tau(r)=\frac{(B^2-1)r}{(B+r)(1+Br)}.
\]
We have $ rg'(r)=1-(d-1)\tau(r)$. Observe that $\tau(r)=\frac{B^2-1}{B(r+r^{-1})+B^2+1}$ is strictly decreasing for $r>1$. 
Since $\beta>\beta_c(d)$, we have
$(d-1)\tau(1)=(d-1)(B-1)/(B+1)>1$.
Thus $g$ first decreases and then increases to infinity.
As $g(1)=0$, it has a unique further zero $r_0>1$.
Consequently, for every $s\geq0$, the equation $g(r)=s$
has a unique solution $r_s>1$, with $r_s\geq r_0$.
On this increasing branch, $r_s$ is continuous and increasing in $s$.

The corresponding message is $a^\pl=r_s/(1+r_s)$.
Substituting in the BP formulas gives the occupation probability
$\alpha^\pl$ and edge statistics $x^\pl$ defined in
Section~\ref{sec:randomgraphIS}, with
$ \frac{\alpha^\pl}{1-\alpha^\pl}
 =r_s\frac{1+Br_s}{B+r_s}$ and 
 $ \tau^\pl=\tau(r_s).$
Using that $r\frac{1+Br}{B+r}$ and $-\tau(r)$ are strictly increasing in $r$, we obtain that
$\alpha^\pl(s)\geq\alpha^\pl(0)>1/2$ and
$(d-1)\tau^\pl(s)\leq(d-1)\tau^\pl(0)<1$, proving the two uniform
gaps in Lemma~\ref{lem:statisticsIS}.

Every global maximizer of the first-moment functional is
interior and corresponds to a BP fixed point, as in
Section~\ref{sec:Isinglower}.
For $s>0$, spin reversal increases this functional whenever
the spin-$1$ occupation is less than $1/2$, while a balanced
maximizer cannot satisfy the BP equations.
Hence every global maximizer corresponds to $r_s>1$.
At $s=0$, the balanced fixed point $r=1$ has a strictly
positive Hessian direction, since $(d-1)\tau(1)>1$.
It is therefore not a local maximum, leaving exactly
$r_0$ and $1/r_0$ as global maximizers.
Thus, for every $s\geq0$, the unique maximizer within the
plus phase has vertex statistics $(1-\alpha^\pl,\alpha^\pl)$
and edge statistics $x^\pl$. Set
\[
 \Psi(s)=\Phi_{\Bb_s}\big((1-\alpha^\pl,\alpha^\pl),x^\pl\big).
\]
The continuity of $r_s$ implies that $\Psi$ is continuous.

For the second moment, the tensor-norm identity
of~\cite[Section~5.2]{BIShard} gives the unrestricted maximum
$2\Psi(s)$ for $\Phi_{\Bb_s\otimes\Bb_s}$ with maximisers obtained by tensoring the maximisers of $\Phi_{\Bb_s}$. Analogously to Section~\ref{sec:Isinglower} we obtain that the only maximizer for the second moment $\E[(Z_{\bG,s}^\pl)^2]$ is the vector   $(\gamma^\pl,y^\pl)$ from \eqref{eq:gdyoptimalbbb}.
The moment estimates from Section~\ref{sec:Isinglower},
summed over the polynomially many admissible statistics, now give
\[
 \E[Z^\pl_{\bG,s}]\leq n^{C_0}\emm^{n\Psi(s)},
 \qquad
 \E[(Z^\pl_{\bG,s})^2]\leq n^{C_0}\emm^{2n\Psi(s)}.
\]
For the deterministic lower bound, apply the  analogous component expansion
from Section~\ref{sec:Isinglower} (see \eqref{eq:helpxhelp} and the argument thereafter) to 
$(1-\alpha^\pl,\alpha^\pl)$ and $x^\pl$, assigning spin $1$ to
every cyclic component. The final estimate gives
\[
 \frac{Z^\pl_{G,s}}{\emm^{n\Psi(s)}}
 \geq
 1-\frac{(1-\alpha^\pl)\alpha^\pl}
          {(\alpha^\pl-1/2)^2n}
       \left(1+\frac{d\tau^\pl}{1-(d-1)\tau^\pl}\right)
 \geq\frac12
\]
for all sufficiently large $n$, uniformly over $s\in[0,s_0]$.
This proves Lemma~\ref{lem:ZboundsIS}.

Finally, the statistics concentration is proved exactly as for
the hard-core model. On the compact set of joint statistics with
both spin-$1$ occupation marginals at least $1/2$,
$\Phi_{\Bb_s\otimes\Bb_s}$ has the unique maximizer
$(\gamma^\pl,y^\pl)$.
Continuity and compactness give a uniform positive gap outside
any fixed neighbourhood of this maximizer, for $s\in[0,s_0]$.
Summing the moment bound over the polynomial number of relevant vectors bounds the total expected weight
of pairs with
$\|\gamma-\gamma^\pl\|_1+\|y-y^\pl\|_1\geq\eps$ by
$n^{O(1)}\emm^{2n\Psi(s)-c_\eps n}$.
The double-planted normalizer is at least
$\frac14\emm^{2n\Psi(s)}$ by the deterministic lower bound.
Dividing and absorbing the polynomial factor gives
\[
 \Pr\left[
 \|\gammab^\pl-\gamma^\pl\|_1+
 \|\yb^\pl-y^\pl\|_1\geq\eps
 \right]\leq\emm^{-cn},
\]
uniformly for $s\in[0,s_0]$, as required.
\end{proof}

\bibliographystyle{alpha}
\bibliography{bibliography.bib}

\appendix
\section{Remaining Proofs}\label{sec:remainingproofs}
\begin{proof}[Proof of~\Cref{lem:cutwidth}]
We first show the result for trees. It is a standard fact that a tree on $n$ vertices has a centroid whose removal leaves components of size at most $n/2$. Consider the following algorithm $\textsc{CutWidth}(T)$ that takes as input a tree $T$ and returns an ordering of the vertices. 

If $T$ consists of a single vertex, then return the vertex. Otherwise, find a centroid $v$ of $T$, and let $T_1,\hdots, T_k$ be the tree components remaining after removing $v$. Output the centroid $v$ first, then $\textsc{CutWidth}(T_1),\hdots, \textsc{CutWidth}(T_k)$.
Writing $w(T)$ for the cutwidth of this ordering, we have $w(T)\leq d+\max_i w(T_i)$ and hence, by induction, $w(T)\leq d\lceil\log_2 n\rceil$, where $n=|V(T)|$. 

For a graph $C$ with tree-excess $\leq r$, apply the above to a spanning tree of $C$, the remaining edges contribute at most $r$ to any cut.
\end{proof}
\begin{proof}[Proof of \Cref{lem:anneal}]
Nonnegative spectrum gives 
$\Var_\pi(Pf)
 \leq\Var_\pi(f)-\cE_P(f,f)$.
Apply the weak Poincare inequality to $g=P^{t-1} f$, and use
$g_{\max}-g_{\min}\leq f_{\max}-f_{\min}$, to obtain
\[
 \Var_\pi(Pg)\leq (1-1/A)\Var_\pi(g)
        +(b/A)(f_{\max}-f_{\min})^2.
\]
Iterating gives
\begin{equation}\label{eq:anneal-variance-decay}
 \Var_\pi(P^t f)
 \leq\emm^{-t/A}\Var_\pi(f)+b(f_{\max}-f_{\min})^2.
\end{equation}
For $0\leq f\leq1$, reversibility and Cauchy--Schwarz imply
\[
 |\nu P^t f-\pi f|
 \leq\sqrt{\chi^2(\nu\Vert\pi)}
       \sqrt{\Var_\pi(P^t f)}.
\]
Now use $\Var_\pi(f)\leq1/4$ and take the supremum over such $f$. 

To obtain the bound on the TV-distance, write $h(x)=\sigma(x)/\pi(x)$ for $x\in \Omega$ (with $h(x)=0$ when $\pi(x)=0$).
Since $0\le h\le R$ and $\pi(h)=1$, we have
\[
 \chi^2(\sigma\Vert\pi)
 =\pi(h^2)-1
 \le R\pi(h)-1
 =R-1.
\]
Applying the second estimate with $\nu=\sigma$, we obtain
\[\|\sigma P^t-\pi\|_{\TV}\le
 \sqrt{R-1}\sqrt{\tfrac14\emm^{-t/A}+b}\le
 \sqrt{R-1}\left(\tfrac12\emm^{-t/(2A)}+\sqrt b\right)\le
 R\left(\emm^{-t/(2A)}+\sqrt b\right).\qedhere
\]
\end{proof}

\section{Proofs for the initialisation lemmas}\label{sec:initialization}
\subsection{Initialisation for Hard-Core and Ising}
\begin{leminit}
\stateleminit
\end{leminit}
\begin{proof}[Proof of \Cref{lem:init}]
Set $\eta=\varepsilon$.
Write $n=|V^\pl|=|V^\mi|$ and set for convenience $
 a=\lambda\emm^{s_0}$, $b=\lambda\emm^{-s_0}$. Let also $\mu=\mu_{G,s_0}$ and $\mu^\pl=\mu^\pl_{G,s_0}$.

We first show that the unrestricted  Glauber dynamics $(X_t^\pl)$ started from $V^\pl$
converges rapidly to the unrestricted distribution $\mu$. Couple $(X_t^\pl)$ with a stationary chain $(Y_t)$, where
$Y_0\sim\mu$, by using the same update vertex and the same
uniform random variable in each Glauber update.
Let
\[
 D_t^\pl=|(X_t^\pl\oplus Y_t)\cap V^\pl|,
 \qquad
 D_t^\mi=|(X_t^\pl\oplus Y_t)\cap V^\mi|,
 \qquad
 D_t=D_t^\pl+2dD_t^\mi.
\]
When a vertex is updated, a disagreement can remain only
if its neighbours have different occupation patterns in
the two chains. The probability of a disagreement after
the update is therefore at most the number of disagreeing
neighbours for a vertex in $V^\pl$, and at most $b$ times
that number for a vertex in $V^\mi$.
Consequently, using $b\le 1/(4d^2)$,
\begin{align*}
 \mathbb E[D_{t+1}\mid X_t^\pl,Y_t]
 &\le
 D_t-\frac{D_t^\pl+2dD_t^\mi}{2n}
       +\frac{dD_t^\mi+2d^2bD_t^\pl}{2n}\leq
 \left(1-\frac{1}{4n}\right)D_t.
\end{align*}
Moreover, under $\mu$, every vertex in $V^\mi$ is occupied
with probability at most $b/(1+b)\le b$.
Since $X_0^\pl=V^\pl$, this gives
\[
 \mathbb E[D_0]
 \le n+2dnb
 \le n\left(1+\frac{1}{2d}\right)
 \le \frac{7n}{6}.
\]
The coupling lemma now yields
\[
 \|X_t^\pl-\mu\|_{\mathrm{TV}}
 \le \Pr[X_t^\pl\ne Y_t]
 \le \mathbb E[D_t]
 \le \frac{7n}{6}\emm^{-t/(4n)}.
\]
Taking $t=\left\lfloor 4n\log\frac{4n}{\eta}\right\rfloor$, 
we obtain $\| X_t^\pl-\mu\|_{\mathrm{TV}} \leq \eta/3$. 

We next  compare $\mu$ with $\mu^\pl$ its plus-phase restriction.
For $I\notin\mathcal I^\pl$, we have
$|I\cap V^\pl|<|I\cap V^\mi|$, and hence
\[
 w_{G,s_0}(I)
 =\lambda^{|I|}
   \emm^{s_0(|I\cap V^\pl|-|I\cap V^\mi|)}
 \le \lambda^{|I|}, \text{ so } \sum_{I\notin\mathcal I^\pl}w_{G,s_0}(I)
 \leq (1+\lambda)^{2n}.
\]
On the other hand, the contribution of $V^\pl$ alone gives $Z_{G,s_0}\ge a^n
 \ge 10^n(1+\lambda)^{2n}$. It follows that
\[
 \|\mu-\mu^\pl_{G,s_0}\|_{\mathrm{TV}}
 =\mu(\mathcal I_G\setminus\mathcal I^\pl_G)
 \le 10^{-n}
 \le \frac13\emm^{-n/2}
 \le \eta/3.
\]

Finally, let $(Z_t)$ be the restricted chain started
from $Z_0=V^\pl$, coupled to $(X^\pl_t)$ and $(Y_t)$ using identical updates.
Recall the bipartite order $I\preceq J$ when 
$I\cap V^\pl\subseteq J\cap V^\pl$ and $ I\cap V^\mi\supseteq J\cap V^\mi$.  Since  $Y_0\preceq X_0^\pl$ and Glauber updates preserve the ordering we have $Y_t\preceq X_t^\pl$ for all $t\geq 0$. In particular, $Y_t\in \cI^\pl_G$ implies that $X^\pl_t\in \cI^\pl_G$. Using the stationarity of $Y_t$, we therefore have 
\[
 \Pr[X_t^\pl\notin\cI_G^\pl]
 \leq\Pr[Y_t\notin\cI_G^\pl]
 =\mu(\mathcal I_G\setminus\mathcal I^\pl_G)
 \leq10^{-n}.
\]
The chains $Z_t$ and $X^\pl_t$ agree until $X^\pl$ first
leaves $\cI_G^\pl$. Thus, by a union bound over $t$ steps,
\[
 \|Z_t-X_t^\pl\|_{\TV}
 \leq  t10^{-n}\leq \eta/3.
\]
Combining the three total-variation estimates gives
\[
 \|Z_t-\mu^\pl_{G,s_0}\|_{\TV}
 \leq
 \|Z_t-X_t^\pl\|_{\TV}
 +\|X_t^\pl-\mu\|_{\TV}
 +\|\mu-\mu^\pl_{G,s_0}\|_{\TV}
 \leq \eta, 
\]
completing the proof.
\end{proof}

\begin{leminitIS}
\stateleminitIS
\end{leminitIS}
\begin{proof}
At any vertex, the conditional probability of a minus spin is at
most $\emm^{\beta d-s_0}\leq1/(2d)$. Thus the common-update coupling
contracts expected Hamming distance by $1-1/(2n)$, and the unrestricted
chain from all-plus is within $n\emm^{-t/(2n)}$ of $\nu_{G,s_0}$.
Comparing with the all-plus weight gives
$\nu_{G,s_0}(M\leq0)\leq2^n\emm^{-s_0n/2}\leq\emm^{-3n}$.
By monotonicity, the chain from all-plus dominates a stationary chain,
so its probability of leaving the phase by time $t$ is at most
$t\emm^{-3n}$. Coupling to the restricted chain until this first exit
and then conditioning the stationary law on the phase gives total
error at most $n\emm^{-t/(2n)}+(t+1)\emm^{-3n}$.
\end{proof}

\subsection{Initialisation for Random-Cluster}
\begin{leminitRC}
\stateleminitRC
\end{leminitRC}
\begin{proof}
Whp $G$ is an edge expander and is locally treelike, i.e., there are constants $\alpha,\kappa>0$ and an integer
$L$ such that for all nonempty vertex subsets $U$ with $|U|\leq n/2$ it holds that $|\partial U|\geq\alpha|U|$ and every ball of radius $\kappa\log n$ can be made
into a tree by removing at most $L$ edges.
Fix such a graph.

We first verify the percolation estimate needed for
the spatial-mixing argument.
Let $F$ contain each edge independently with probability
$p$. There are at most $d^{2k}$ connected vertex sets
of size $k$ containing a specified vertex $v$.
Consequently,
\[
 \Pr\left[
 \begin{array}{c}
 \exists\,U\ni v:\ G[U]\text{ connected},\
 \ell\leq|U|\leq n/2,\\
 F\cap\partial U=\emptyset
 \end{array}
 \right]
 \leq
 \sum_{k\geq\ell}d^{2k}(1-p)^{\alpha k}
 \leq \emm^{-c(p)\ell},
\]
for $p$ sufficiently close to $1$, where
$c(p)\to\infty$ as $p\uparrow1$.
This is precisely the exponentially strong
supercriticality condition of
\cite[Definition~1.4]{BG25}. Throughout this proof, references to~\cite{BG25} use the numbering of arXiv:2304.03182.

The spatial-mixing and local-mixing argument of
\cite[Theorem~1.6 and Remark~1.7]{BG25} gives an
$O(n^{5/4})$ mixing bound uniformly over
$\beta\geq\beta_0$, for a sufficiently large constant
$\beta_0$. The spatial-mixing argument works verbatim, we briefly justify the uniformity for the local-mixing argument. Write $p=1-\emm^{-\beta}$ and
$\widehat p=p/(p+q(1-p))$, and initially assume
only that $p\geq1/2$.

Consider a local chain on a treelike ball of radius
$h$ with $m\leq d^{h+1}$ edges. Working in
continuous time with rate-one edge clocks, the
canonical-path and comparison bounds in
\cite[Lemmas~5.10 and~6.10]{BG25} give
$\operatorname{gap}^{-1}
 \leq
 \frac{C_1\emm^{A_1h}}
      {\min\{\widehat p,1-p\}}$, where $C_1,A_1$ depend only on $d,q$ and the fixed
treelikeness and boundary-wiring constant. 
If $1-p\geq(16qm)^{-1}$, then
$\widehat p\geq1/(q+1)$ and
$\log(1/\pi_{\min})=O_q(m\log(m+1))$.
The standard spectral-gap mixing bound therefore gives
$t_{\mathrm{mix}}(1/4)\leq C_2\emm^{A_2h}$,
uniformly in this range. If $1-p<(16qm)^{-1}$, a simpler argument applies.
For any initial state, the probability that a given
edge is absent at time $t$ is at most
$\emm^{-t}+q(1-p)$: either it has not been updated,
or its last update leaves it absent.
The stationary absence probability is also at most
$q(1-p)$. Comparing both distributions with the
all-in configuration therefore bounds their
total-variation distance at time $t=\log(16m)$ by  $m\emm^{-t}+2mq(1-p)\leq3/16$. 
Thus the local mixing bound
$C_2\emm^{A_2h}$ is uniform for all $p\geq1/2$.

Choose $h=\varepsilon_0\log n$ within the available
treelike radius, with $\varepsilon_0>0$ sufficiently
small that amplification to local error $n^{-5}$
takes $O(n^{1/4})$ continuous time. We then choose
$\beta_0$ sufficiently large for the spatial-mixing
and burn-in estimates of \cite{BG25}. Their
censoring argument yields the uniform
$O(n^{5/4})$ discrete-time mixing bound.
Remark~1.7 allows every real $q\geq1$. By standard amplification, the unrestricted chain
$(X_t)$ at any $\beta\geq\beta_0$, started from $E$,
satisfies
\[
 \big\|X_t-\pi_{G,\beta}\big\|_{\TV}\leq\eta/2
\]
for $t=\lceil Cn^{5/4}\log(1/\eta)\rceil$ and
sufficiently large $C$, independently of $\beta$.

It remains to account for the phase restriction.
Every conditional edge-occupation probability under
$\pi_{G,\beta_0}$ is at least $p_*=\frac{\emm^{\beta_0}-1}
           {\emm^{\beta_0}-1+q}$.
Increase $\beta_0$ if needed so that $p_*>\mathfrak m_c$.
Stochastic domination by independent
$p_*$-percolation and a Chernoff bound give $ \pi_{G,\beta}(\cE_G^\mi)\leq \pi_{G,\beta_0}(\cE_G^\mi)\leq\emm^{-c_0n}$
for some constant $c_0>0$. Couple $(X_t)$ above a stationary unrestricted chain
using the monotone coupling. Since $\cE_G^\pl$ is
increasing, we have $
 \Pr[X_j\notin\cE_G^\pl]\leq\emm^{-c_0n}$ for all $j\geq 0$. The restricted chain $(\widehat X_t)$, also started
from $E$, agrees with $(X_t)$ until its first exit
from $\cE_G^\pl$. Hence
\[
 \begin{aligned}
 \big\|\widehat X_t
             -\pi^\pl_{G,\beta}\big\|_{\TV}
 &\leq
 \big\|X_t-\pi_{G,\beta}\big\|_{\TV}
 +(t+1)\emm^{-c_0n}\leq\eta,
 \end{aligned}
\]
for sufficiently small $c>0$ and all sufficiently
large $n$, as required.

We next obtain the partition-function estimate (possibly by 
increasing the constant $\beta_0$ further as needed).
Write $B=\emm^{\beta}$, $Q=q-1$ and $M=|E|$.
We use the polymer construction of~\cite{BIS1},
expressed through the active/inactive lifting so that
it applies also to noninteger $q$. A polymer is a nonempty vertex set $U$ such that
$G[U]$ is connected and $|U|\leq n/2$.
Two polymers are compatible if they are disjoint
and no edge of $G$ joins them.
Assign each $U$ weight
\[
 w(U)=
 B^{-|E[U]|-|\partial U|}
 \sum_{A\subseteq E[U]}
       (B-1)^{|A|}Q^{\kappa(U,A)},
\]
where $\kappa(U,A)$ counts the components of $(U,A)$,
including isolated vertices. Since $Q>1$ and $G$ has edge expansion $\alpha$, we have
\begin{equation}\label{eq:initRC-polymer-weight}
 w(U)\leq Q^{|U|}B^{-|\partial U|}
 \leq \emm^{-(\alpha\beta-\log Q)|U|}\leq 
 \emm^{-(\alpha\beta_0-\log Q)|U|}.
\end{equation}

Let $ \Xi_G=\sum_{\Gamma\text{ compatible}}
              \prod_{U\in\Gamma}w(U)$. We first relate $\Xi_G$ to the desired partition
function. Expanding the polymer weights shows that $B^M\Xi_G$ is the total lifted RC weight of pairs $(F,S)$
for which every component of $G[V\setminus S]$
has at most $n/2$ vertices.
Indeed, the polymers are precisely these components;
edges crossing from $S$ to $V\setminus S$ are absent,
and summing over the active edges contributes
$B^{|E[S]|}$. If $(V,F)$ has a component larger than $n/2$,
the allowed labelings are exactly those making that
component active. Their total weight is therefore
$1/q$ times the RC weight of $F$.
Consequently,
\[
 \left|qB^M\Xi_G-Z_{G,\beta}\right|
 \leq q\,Z_{G,\beta}\,
       \pi_{G,\beta}(|C_1(F)|\leq n/2).
\]
To bound the probability on the right, observe that
when every component has size at most $n/2$,
some union of components has size between $n/4$
and $n/2$ and has no occupied boundary edges.
Under independent edge percolation with retention
probability $1-B^{-1}$, a union bound and expansion
bound this event by $2^nB^{-\alpha n/4}$.
Since every RC component factor is at most $q^n$
and $Z_{G,\beta}\geq qB^M$, it follows that
\[
 \pi_{G,\beta}(|C_1(F)|\leq n/2)
 \leq q^{n-1}2^nB^{-\alpha n/4}
 \leq \emm^{-c_1n}
\]
for sufficiently large $\beta_0$ and some $c_1>0$.
Together with the exponentially small bound on
$\pi_{G,\beta_0}(\cE_G^\mi)$ proved above, this gives for some $c_2>0$ that
\begin{equation}\label{eq:initRC-polymer-partition}
 qB^M\Xi_G
 =Z^\pl_{G,\beta}
       \bigl(1+O(\emm^{-c_2n})\bigr).
\end{equation}

It remains to approximate $\Xi_G$ sufficiently quickly.
Fix a constant $a>0$, to be chosen below.
There are at most $d^{2k}$ connected sets of size $k$
containing a specified vertex. Hence
\eqref{eq:initRC-polymer-weight} gives
\[
 \sum_{U':\,\operatorname{dist}_G(U,U')\leq1}
       w(U')\emm^{(a+1)|U'|}
 \leq
 (d+1)|U|
 \sum_{k\geq1}
 d^{2k}
 \emm^{-(\alpha\beta_0-\log Q-a-1)k}
 \leq |U|,
\]
provided $\beta_0$ is sufficiently large.
This is the Koteck\'y--Preiss condition with decay
function $g(U)=a|U|$. Moreover, $w(U)$ can be computed in
$\exp(O(d|U|))$ time by enumerating the subsets of
$E[U]$.
The algorithmic cluster-expansion bound
of~\cite[Theorem~8]{BIS1} therefore computes
$\widehat\Xi_G$ with
\[
 \left|\log\widehat\Xi_G-\log\Xi_G\right|
 \leq \tfrac14 n^{-J}
\]
in time $ n\bigl(4n^{J+1}\bigr)^{O((d+\log d)/a)}$.
Choose $a$ sufficiently large in terms of $d,J$,
and then choose $\beta_0$ sufficiently large.
The running time is then $\widetilde O(n^{5/4})$.
Finally, set
$\widehat Z^\pl_{G,\beta}=qB^M\widehat\Xi_G$.
By \eqref{eq:initRC-polymer-partition}, its relative
error is at most $n^{-J}$ for all sufficiently
large $n$, completing the proof.
\end{proof}

\section{Remaining Technical Proofs}\label{sec:remainingproofsb}
\begin{proof}[Proof of \Cref{lem:alphabeta}]
Set $
 x=\frac{\alpha}{1-\alpha-\beta}$ and $y=\frac{\beta}{1-\alpha-\beta}$, then
the system of equations is equivalent to
\[
 x(1+y)^{d-1}=\lambda\emm^s,
 \qquad
 y(1+x)^{d-1}=\lambda\emm^{-s}.
\]
In particular, setting $f(z)=z(1+z)^{d-1}$, we have
$f(x)f(y)=\lambda^2$. Since $f$ is strictly increasing on $[0,\infty)$,
consider $\omega=f^{-1}(\lambda)$; since $\lambda>\lambda_c(d)=f(\frac{1}{d-2})$, we conclude that $\omega> \frac{1}{d-2}$. 
Now, for $x\geq \omega$, define
\[
 y(x)=f^{-1}\!\left(\frac{\lambda^2}{f(x)}\right),
 \qquad
 g(x)=\log\frac{x(1+y(x))^{d-1}}{\lambda}.
\]
Solutions with $\alpha>\beta$ therefore correspond
exactly to solutions of $g(x)=s$ with $x>\omega$.

Note that the function $y(x)$ is strictly decreasing, with
$y(\omega)=\omega$ and $y(x)\to0$ as $x\to\infty$, so $y(x)<x$ for all $x>\omega$.
Direct differentiation further gives
\begin{equation}\label{eq:alphabeta-derivatives}
 xg'(x)=
 \frac{1-\chi(x)}{1+(d-1)\frac{y(x)}{1+y(x)}},
 \mbox{ where }  \chi(x)=(d-1)^2\frac{xy(x)}{(1+x)(1+y(x))}.
\end{equation}
We further have that 
 $\chi'(x) =
 \frac{d\chi(x)(y(x)-x)}
      {x(1+x)(1+dy(x))}$, so from $y(x)<x$ we conclude that $\chi$ is strictly decreasing on $(\omega,\infty)$. We have  $\chi(\omega)=\big(\frac{(d-1)\omega}{1+\omega}\big)^2>1$ and  $\chi(x)\to0$ as $x\to\infty$.
so there is a unique $x^*>\omega$ with
$\chi(x^*)=1$.
By \eqref{eq:alphabeta-derivatives}, $g$ is strictly
decreasing on $(\omega,x^*)$ and strictly increasing on
$(x^*,\infty)$.

Since $g(\omega)=0$ and
$g(x)\geq\log(x/\lambda)\to\infty$, there is a unique
$x_{\pl}>x^*>\omega$ with $g(x_\pl)=0$. 
For every $s\geq0$, there is therefore exactly one
$x(s)\geq x_\pl$ with $g(x)=s$, and no solution in $(\omega,x_\pl)$.
This proves existence and uniqueness of the required
solution.

It remains to obtain uniform bounds.
As $s$ increases, at the corresponding solution $x=x(s)$ increases
and $y=y(x(s))$ decreases. Hence
\[
 \alpha(s)-\beta(s)
 =\frac{x(s)-y(x(s))}{1+x(s)+y(x(s))}
 \geq \alpha(0)-\beta(0)>0.
\]
Furthermore,
\[
 (d-1)^2
 \frac{\alpha(s)\beta(s)}
      {(1-\alpha(s))(1-\beta(s))}
 =\chi(x(s))
 \leq \chi(x_\pl)<1.
\]
Thus, we obtain the lemma by setting  $\zeta=\frac12\min\left\{
 \alpha(0)-\beta(0),\,
 1-\chi(x_\pl)
 \right\}>0$.
\end{proof}

\begin{proof}[Proof of Lemma~\ref{lem:RC-edge-statistics}]
Write $Q=q-1$.
For the ordered phase, recall that
$a^\pl=\frac{t^{d-1}}{t^{d-1}+Q}$ and $\alpha^\pl=\frac{t^{d}}{t^{d}+Q}$,
where $t>1$ is the largest solution of $B=1+\frac{(t-1)(t^{d-1}+Q)}{t^{d-1}-t}$. Recall also $x^\pl=\frac{1}{X^\pl}\Big[\begin{smallmatrix}\hat B(1-a^\pl)^2 &  a^\pl(1-a^\pl)\\ a^\pl(1-a^\pl)  & B(a^\pl)^2\end{smallmatrix}\Big]$ where $X^\pl=\hat B(1-a^\pl)^2+2a^\pl(1-a^\pl)+B(a^\pl)^2$.

With $S(t)=\frac{t^{d-1}-1}{t-1}=1+t+\cdots+t^{d-2}$, we therefore obtain that 
\begin{equation}\label{eq:RC-retention-identities}
\scalebox{1.1}[1.1]{\mbox{$p_0 \frac{x_{00}^\pl}{1-\alpha^\pl}=\frac{1}{S(t)}$},
 \quad
 \mbox{$p_1 \frac{x_{11}^\pl}{\alpha^\pl}=\frac{t^{d-2}}{S(t)}$},
 \quad
 \mbox{$p \frac{x_{00}^\pl x_{11}^\pl}{\alpha^\pl(1-\alpha^\pl)} =
 \frac{t^{d-2}(t+Q)}{S(t)(t^{d-1}+Q)}$.}}
\end{equation}
We have
$d-1<S(t)<(d-1)t^{d-2}$ for $t>1$, so $p_0 \frac{x_{00}^\pl}{1-\alpha^\pl}<\frac{1}{d-1}$ and $p_1 \frac{x_{11}^\pl}{\alpha^\pl}>\frac{1}{d-1}$. Moreover
\begin{align*}
\scalebox{1.1}[1.1]{\mbox{$p \frac{x_{00}^\pl x_{11}^\pl}{\alpha^\pl(1-\alpha^\pl)}-\frac{1}{d-1}=-\frac{t^{d-1}\big(S(t)-(d-1)\big)-Q\big((d-1)t^{d-2}-S(t)\big)}{(d-1)S(t)(t^{d-1}+Q)}$}}
\end{align*}
At $\beta=\beta_c$, we have $t=t_c$ where
$t_c=Q^{2/d}$. For $\beta\geq \beta_c$, we have $t\geq t_c$ (using that $t$ is the largest solution), so $Q\leq t^{d/2}$. We can therefore lower bound the numerator in the last expression by
\[t^{d-1}\big(S(t)-(d-1)\big)-t^{d/2}\big((d-1)t^{d-2}-S(t)\big)=\big(t^{d-1}+t^{d/2}\big)
       \big(S(t)-(d-1)t^{(d-2)/2}\big)>0 \]
whereas the final inequality follows from $S(t)=1+t+\hdots+t^{d-2}>(d-1)t^{(d-2)/2}$ for $t>1$ (using for example AM-GM).
By continuity, these strict inequalities extend to a small interval below
$\beta_c$. 

For the disordered phase, substitution of
$a^\mi=\alpha^\mi=1/q$ gives
\begin{equation}\label{eq:RC-disordered-retention}
\scalebox{1.1}[1.1]{\mbox{$p_0 \frac{x_{00}^\mi}{1-\alpha^\mi}=p_1 \frac{x_{11}^\mi}{\alpha^\mi}=p \frac{x_{00}^\mi x_{11}^\mi}{\alpha^\mi(1-\alpha^\mi)} =\frac{B-1}{B+Q}<\frac{1}{d-1}$}}
\end{equation}
Since \eqref{eq:RC-disordered-retention} is increasing
in $\beta$, the inequality holds uniformly
for $\beta\leq\beta_c+\delta$ when $\delta>0$ is small.

For the edge-densities, we have the expressions 
\[\mathfrak{m}^\mi(\beta)=\frac{(B-1)A^\mi}{1+(B-1)A^\mi}=\frac{B-1}{B+q-1}\text{ and }\mathfrak{m}^\pl(\beta)=\frac{(B-1)A^\pl}{1+(B-1)A^\pl}\] 
where $A^\plm=(a^\plm)^2+\frac{(1-a^{\plm})^2}{q-1}$. Since $B>1$ and $q>2$, it follows that $\mathfrak{m}^\pl(\beta)>\mathfrak{m}^\mi(\beta)$ iff $A^\pl>A^\mi$ iff $a^\pl>1/q$; the latter is true for all $\beta$ where $a^\pl$ is defined (i.e., $\beta>\beta_u$). We also have that $\mathfrak{m}^\mi(\beta)$ is trivially increasing as $\beta$ (and $B$) increases, and similarly for $\mathfrak{m}^\pl(\beta)$ we have that both $a^\pl$ and $A^\pl$ increase as $\beta$ increases.
The strict gap at $\beta_c$ and continuity allow us to
choose $\theta,\delta>0$ such that
\[
 \mathfrak{m}^\pl(\beta_c-\delta)\geq \mathfrak{m}_c+4\theta,
 \qquad
 \mathfrak{m}^\mi(\beta_c+\delta)\leq \mathfrak{m}_c-4\theta.
\]
Monotonicity gives the claimed bounds throughout the
respective parameter ranges.
\end{proof}
\begin{proof}[Proof of Lemma~\ref{lem:first-moment-bipartite}]
There are
$\binom{n}{n\alphab}\binom{n}{n\betab}$
vertex configurations with the prescribed vertex  statistics.
Fix one such configuration.
There are $dn\alpha_i$ half-edges incident to vertices
of colour $i$ in $V^\pl$, and $dn\beta_j$ half-edges
incident to vertices of colour $j$ in $V^\mi$.
The number of half-edge matchings having exactly
$dnx_{ij}$ edges between colour $i$ in $V^\pl$
and colour $j$ in $V^\mi$ is
\[
\frac{
\prod_i(dn\alpha_i)!\prod_j(dn\beta_j)!
}{
\prod_{i,j}(dnx_{ij})!
}.
\]
Indeed, we first partition the half-edges on each side
according to the colour of their partner, and then
match the resulting groups.
There are $(dn)!$ equally likely matchings in total,
and every matching with these edge statistics has
configuration weight
$\prod_{i,j:B_{ij}>0}B_{ij}^{dnx_{ij}}$.
Consequently,
\begin{equation}\label{eq:first-moment-exact}
\mathbb{E}\big[Z_{\bG}^{\alphab,\betab,\xb}\big]
=
\binom{n}{n\alphab}\binom{n}{n\betab}
\frac{\binom{dn}{dn\xb}}
{\binom{dn}{dn\alphab}\binom{dn}{dn\betab}}
\prod_{i,j:B_{ij}>0}B_{ij}^{dnx_{ij}}.
\end{equation}
For a probability vector $\mathbf{p}\in \mathbb{R}^t_{\geq 0}$ we have the standard multinomial estimate
\[
(N+1)^{-t}\emm^{NH(\mathbf{p})}
\leq \binom{N}{N\mathbf{p}}
\leq \emm^{NH(\mathbf{p})}.
\]
Applying this bound to the multinomials in \eqref{eq:first-moment-exact}, we obtain (with room to spare)
\begin{equation}\label{eq:first-moment-table-bounds}
\begin{aligned}
(dn)^{-2q^3}\leq \frac{\mathbb{E}\big[Z_{\bG}^{\alphab,\betab,\xb}\big]}{\emm^{n\Phi(\alphab,\betab,\xb)}}\leq (dn)^{2q^3}
\end{aligned}
\end{equation}

For the rounding assertion, consider arbitrary real
$(\hat\alphab,\hat\betab,\hat\xb)\in\mathcal M$.
We round $n\hat\xb$ to an integral matrix, rounding
each non-integral entry and each row and column sum to its floor
or ceiling, while preserving the total sum $n$ (such a rounding exists by integrality of network
flows). We let $n\xb$ denote the resulting integral matrix and $n\alphab,n\betab$ be its row and column  marginals. Then, we have $\big\|(\hat\alphab,\hat\betab,\hat\xb)-(\alphab,\betab,\xb)\big\|_\infty
\leq \frac{1}{n}$. For $f(x)=-x \log x$ and $\delta\in(0,1)$, we have the continuity bound $|f(x)-f(y)|\leq \delta \log \frac{1}{\delta}+\delta$  for $x,y\in [0,1]$ with $|x-y|\leq \delta$. Therefore,
\[
\left|H(\widehat\xb)-H(\xb)\right|
\leq \frac{4q^2}{n}\log n.
\]
Let $b_{\min}$ be the smallest positive entry of $\Bb$.
Since $\hat\xb$ and $\xb$ both have total mass
one, we also have
\begin{align*}
\bigg|
\sum_{i,j:B_{ij}>0}
(\hat x_{ij}-x_{ij})\log B_{ij}
\bigg|
=
\bigg|
\sum_{i,j:B_{ij}>0}
(\hat x_{ij}-x_{ij})
\log\frac{B_{ij}}{b_{\min}}
\bigg|
\leq \frac{q^2}{n}\log R.
\end{align*}
It follows that
\[
n\Phi(\alphab,\betab,\xb)
\geq
n\Phi(\hat\alphab,\hat \betab,\hat \xb)
-5dq^2\log(nR).
\]
Using the lower bound in \eqref{eq:first-moment-table-bounds} for $\E[Z_{\bG}^{\alphab,\betab,\xb}]$ yields the statement of  the lemma (with room to spare for the stated constants).
\end{proof}

\section{Proof  of the percolation Lemma~\ref{lem:percsub}}

 Here we prove the percolation bounds in Lemma~\ref{lem:percsub}.  We import some relevant results from \cite{coja2023}. 
 For $d\geq 3$ and $p\in (\frac{1}{d-1},1)$, let $\phi=\phi(p)\in (0,1)$ be the probability that a branching process with offspring distribution $\mathrm{Bin}(d-1,p)$ dies out, i.e., $\phi(p)\in (0,1)$ is  the (unique) solution of $\phi =(p \phi+1-p)^{d-1}$. Define $\chi=\chi(p), \psi=\psi(p)$ from 
\begin{equation}\label{eq:kappachipsi}
\chi=1-(p \phi+1-p)^{d},\quad \psi=\tfrac{1}{2}dp(1-\phi^2).
\end{equation}

\begin{lemma}[{\cite[Lemmas 5.3 \& 5.5]{coja2023}}]\label{lem:giantperc}
Fix integer  $d\geq 3$, and  reals $p\in (\frac{1}{d-1},1)$ and $\delta>0$. 

 Let $H\sim\CM_d(n,m)$, where $m=ndp/2$. Then, with probability $1-\emm^{-\Omega(n)}$, it holds that 
\[|C_1(H)|=(\chi\pm \delta) n\mbox{ and } |E(C_1(H))|=(\psi\pm \delta) n,\] where
$\chi=\chi(p),\psi=\psi(p)$ are as in \eqref{eq:kappachipsi}.  Moreover, it holds that $\frac{2(dp/2-\psi)}{d(1-\chi)}< \tfrac{1}{d-1}$.
\end{lemma}

\begin{proof}[Proof of Lemma~\ref{lem:percsub}]
We may assume that $d N $ is even: otherwise, apply the even case on $N+1$ vertices and condition the additional vertex to be isolated, which has a probability bounded below by a positive constant $C>0$. We can then use the bounds for $N+1$ multiplied by $1/C$.

So we focus on $d N$ even, where we can  generate  $H\sim \mathrm{CM}_{d}(N,M)$  by first sampling
$G\sim\CM_d(N,dN/2)$ and then performing a percolation step that keeps a random subset
of exactly $M$ edges.

We consider first the subcritical case $(d-1)p\leq1-\eta$. The bound on the size of the components follows from a standard exploration argument and it works actually for any $d$-regular graph. To bound the size of the component containing $v$, we can examine which edges remain in a BFS manner; if the size of the component is $\leq \ell\leq \eps N$ for some $\eps>0$ then we have revealed $\leq d\ell $ edges, so the next edge  has probability at most $\frac{M}{(dN)/2-d\ell}\leq \frac{p}{1-2\eps}$ of being retained. After the root, each vertex has at most $d-1$ edges to examine. By choosing $\eps>0$ a small enough constant, we can ensure $(d-1)\frac{p}{1-2\eps}$ is less than $1-\eta/2$, so the process is subcritical and we have the standard exploration bound $\Pr[|C(v)|\geq\ell]\leq K\emm^{-\kappa\ell}$ (for $\ell>\eps N$, apply the exploration bound at
$\eps N$ and decrease $\kappa$). For the tree-excess, assuming the currently explored component has size at most $\ell$, each edge examination creates a
collision with a previously discovered vertex with
conditional probability at most $K\ell/N$.
Each collision increases the tree excess by one.
Thus, we obtain the bound
\[
 \Pr\big[|C(v)|<\ell,\
          \mathtt{exc}(C(v))\geq r\big]
 \leq
 \binom{d\ell}{r}
 \left(\frac{K\ell}{N}\right)^r
 \leq
 \left(\frac{K\ell^2}{N}\right)^r.
\]
Taking a union bound over the $N$ possible roots proves
Item~2. For $\ell>N/4$, its right-hand side is at least
one after increasing $K$.

To bound the generating function of the number $W=W(H)$ of cyclic components, we will show shortly that for an arbitrary integer $k\geq0$ it holds that
\begin{equation}\label{eq:Wmoments}
 \mathbb E\binom{W}{k}
 \leq
 \frac{2}{k!}
 \bigg(
   \sum_{\ell\geq1}\frac{((d-1)p)^\ell}{2\ell}
 \bigg)^k
 \leq
 \frac{2}{k!}
 \bigg(\frac12\log\frac1\eta\bigg)^k,
\end{equation}
Then, the binomial expansion of
$z^W=(1+(z-1))^W$ and the series  $\emm^x=\sum_{k}\frac{x^{k}}{k!}$ yields the desired 
\[\mathbb E[z^W]
 =\sum_{k\geq0}(z-1)^k\mathbb E\binom{W}{k}
 \leq
 2\emm^{\frac{1}{2}\log\frac1\eta(z-1)}.\]
To show the bound in \eqref{eq:Wmoments}, we  upper-bound $W$ by  the number of \emph{vertex disjoint} cycles in $H$. We first estimate the number of such disjoint cycles in $G$. Consider arbitrary integers $\ell_1,\ldots,\ell_k\geq1$,
with $L=\ell_1+\cdots+\ell_k\leq N$. The expected number of ordered tuples
of $k$ vertex-disjoint cycles with length $\ell_1,\ell_2,\hdots,\ell_k$ in $G$ is given by 
\begin{equation}\label{eq:Gcycles}
\frac{[d(d-1)]^L\prod_{j=0}^{L-1}(N-j)}
 {\prod_{i=1}^k(2\ell_i)
  \prod_{j=0}^{L-1}(dN-2j-1)}\leq 2\prod_{i=1}^k\frac{(d-1)^{\ell_i}}{2\ell_i}.
  \end{equation}
   Indeed,  we choose and order the $L$ distinct vertices in $\prod_{j=0}^{L-1}(N-j)$ ways,
choose two ordered half-edges at each vertex in $(d(d-1))^L$ ways, and divide
by the $2\ell_i$ possible starting points and orientations
of each cycle. The probability that the resulting $L$
prescribed pairs occur is
$\prod_{j=0}^{L-1}(dN-2j-1)^{-1}$. The inequality follows by observing that $\prod_{j=0}^{L-1}
 \frac{d(N-j)}{dN-2j-1}
 \leq \frac{dN}{dN-1}\leq2$,

Now consider the percolation step of keeping $M$ uniformly chosen edges of $G$,
obtaining $H$. A given tuple uses $L$ distinct edges,
so when $L\leq M$ its probability of surviving is $\prod^{L-1}_{j=0}\frac{M-j}{dN/2-j}\leq p^L$ (and 0 otherwise). 
Consequently, the expected number of surviving ordered
tuples is at most $2\prod_{i=1}^k
 \frac{((d-1)p)^{\ell_i}}{2\ell_i}$. Since $\binom{W}{k}$ is bounded by the number of such unordered tuples, dividing by $k!$ (and relaxing the constraint $\ell_1+\cdots+\ell_k\leq M$), we conclude the upper bound
 \[ \mathbb E\binom{W}{k}
 \leq \frac{2}{k!}\sum_{\ell_1,\hdots,\ell_k\geq 1}\prod_{i=1}^k
 \frac{((d-1)p)^{\ell_i}}{2\ell_i},\]
 implying the first inequality in \eqref{eq:Wmoments}. The second inequality in \eqref{eq:Wmoments} follows from $(d-1)p\leq 1-\eta$ and the series $\log(1-x)=-\sum_{\ell\geq 1} x^\ell/\ell$.

For the supercritical case, the proof is completely analogous after conditioning on the size of the giant, but we first need a cruder bound to handle the generating function when the giant-component has atypical size. We start with this crude bound first. 

Let $A=\frac{1}{4}\log_{d-1}N$. Write $W=W_0+W_1$ where $W_0$ are the cyclic components with size  $\leq A$, and $W_1$ the components with size $>A$. We have the deterministic bound $W_1\leq N/A$. For the smaller cyclic components, repeat the preceding
cycle count, restricting each cycle length to at most
$A$ and using $p\leq1$. Then $\mathbb E\binom{W_0}{k}\leq 2N^{k/4}/k!$, so we obtain  uniformly for $1\leq z\leq \emm^{\sqrt{\log N}}$,
\begin{equation}\label{eq:crudebound}
 \mathbb E[z^{2W}]\leq z^{2N/A} \mathbb E[z^{2W_0}]
 \leq
 Kz^{2N/A}
 \emm^{Kz^2N^{1/4}}
 =\emm^{o(N)}.
\end{equation}

Now, to the main part of the argument. Suppose that $(d-1)p\geq1+\eta$,  with $p$
bounded away from $1$.
Lemma~\ref{lem:giantperc} (and the inequality therein)
ensures 
that outside the giant-component the edge density is subcritical. In particular,  except with probability $\emm^{-\Omega(N)}$, we have $N-|C_1(H)|=\Theta(N)$ and $\frac{2(M-|E(C_1(H))|)}
      {d(N-|C_1(H)|)}
 \leq\frac{1-\kappa}{d-1}$.
Let $\cE$ be the event that the giant component satisfies these bounds.

Condition on the vertex set and the complete
half-edge matching of $C_1(H)$.
The remaining graph has distribution
$\CM_d\bigl(N-|C_1(H)|,\,
            M-|E(C_1(H))|\bigr)$
conditioned not to have a component larger than the
revealed component. On $\cE$, the remaining graph is subcritical so by the bound above the  conditioning event has probability
$1-\emm^{-\Omega(N)}$ and hence  increases any nonnegative expectation
by at most a factor of two. Applying Item~2 to the remaining graph and adding the
exceptional probability from
Lemma~\ref{lem:giantperc} gives Item~3.
Likewise, the subcritical generating-function bound
applies to its cyclic components.
The removed component contributes at most an additional
factor $z$, and $z\leq\emm^{z-1}$. We therefore obtain
\[
 \mathbb E[z^W]\leq  \mathbb E[z^W\mathbf{1}_{\cE}]+\mathbb E[z^W\mathbf{1}_{\neg\cE}]\leq K\emm^{K(z-1)}+\sqrt{\mathbb E[z^{2W}]\, \Pr(\neg\cE)}\leq 2K\emm^{K(z-1)}.
\]
where in the last inequality we used the crude bound \eqref{eq:crudebound}. This finishes the proof.
\end{proof}

\end{document}